\documentclass[12pt,a4paper]{article} 
\pdfoutput=1
\usepackage{jheppub} 
\usepackage{graphicx} 
\usepackage{amsmath} 
\usepackage{amssymb} 
\usepackage{slashed} 
\usepackage{bigstrut} 
\usepackage{mathtools} 
\usepackage{multirow} 
\usepackage[utf8]{inputenc}
\usepackage{todonotes}
\title{Prompt neutrino fluxes in the atmosphere with PROSA parton distribution functions} 

\author[]{{\large
    PROSA Collaboration:}\\ \noindent
  }
\author[a]{\!\!M.V.~Garzelli,} 
\author[a]{S.~Moch,} 
\author[b]{O.~Zenaiev,}
\author[c]{A.~Cooper-Sarkar,}
\author[b]{A.~Geiser,}
\author[b]{K.~Lipka,}
\author[a]{R.~Placakyte,}
\author[a]{G. Sigl} 

\affiliation[a]{II.~Institute for Theoretical Physics, Hamburg University
\\
Luruper Chaussee 149, D--22761 Hamburg, Germany}
\affiliation[b]{DESY, Hamburg \\
Notkestrasse 85, D--22607 Hamburg, Germany}
\affiliation[c]{University of Oxford, Department of Physics\\ 
Keble Road, Oxford OX1 3RH, UK}

\emailAdd{maria.vittoria.garzelli@desy.de}

\abstract{
Effects on atmospheric prompt neutrino fluxes of present uncertainties
affecting the nucleon composition are studied by using the PROSA fit to parton distribution functions (PDFs). 
The PROSA fit extends the precision of the PDFs to low~$x$, which is the
kinematic region of relevance for high-energy neutrino production, by
taking into account LHCb data on charm and bottom hadroproduction.
In the range of neutrino energies explored by present Very Large Volume Neutrino Telescopes, it is found that
PDF uncertainties are far smaller with respect to those due to
renormalization and factorization scale variation and to assumptions on the cosmic ray composition, which at present dominate and limit our knowledge of prompt neutrino fluxes.  
A discussion is presented on how these uncertainties affect the expected number 
of atmospheric prompt neutrino events in the analysis of high-energy events characterized by interaction vertices fully contained within the instrumented volume of the detector, performed by the IceCube collaboration. 
}  
 

\preprint{DESY 16-152}
\begin{document} 
\maketitle 

\section{Introduction}
\label{sec:intro}
The production of leptons due to the interaction of cosmic rays 
with the Earth's atmosphere introduces a background in the searches
for astrophysical neutrinos performed by means of Very Large Volume Neutrino
Telescopes (VLV$\nu$Ts), like IceCube and KM3NeT~\cite{Aartsen:2013jdh,Adrian-Martinez:2016fdl}. 
Although at least part of this background can be eliminated by using veto experimental techniques,
accurate predictions for atmospheric lepton fluxes are of crucial
importance both to refine or further develop these techniques and to get a
precise estimate of the actual spectrum of astrophysical neutrinos (the
signal), in terms of its slope and normalization, after subtraction of the
background. This may allow to understand the origin of a\-stro\-phy\-si\-cal
neutrinos, distinguishing between different potential galactic and
extragalactic sources. The importance of this issue relies on the fact that
neutrinos may travel to the Earth from very distant sources, subject only to
weak interactions, undeflected by cosmic magnetic fields, thus providing
 direct information on the sources and regions of the sky where they are
produced and where cosmic rays are probably ac\-ce\-le\-ra\-ted. In this
respect neutrinos are unique carriers of information, complementing those 
obtained by the detection of other messengers (charged cosmic rays, gamma rays,
gravitational waves....) in the multimessenger approach to astroparticle physics, 
allowing to reach a deeper understanding of our Universe. Furthermore, an
accurate knowledge of the atmospheric neutrino spectrum, at least for energies
up to a few TeV, is fundamental for the precise determination of the
parameters
governing the oscillations of
atmospheric neutrinos and for understanding the neutrino mass hierarchy~\cite{Franco:2013in}. 
Nowadays this research can be performed by making use of extensions
of VLV$\nu$Ts, like the DeepCore instrumentation and its extension PINGU in
the inner core of IceCube~\cite{Yanez:2016ojt}, or the ORCA project in KM3NeT~\cite{Katz:2014tta}, 
complementing more traditional searches for atmospheric neutrino oscillations.  

Atmospheric leptons are typically separated into two components, according to
the decaying hadrons from which they originate: the conventional component is
created by the decay of light mesons, i.e. charged pions and kaons, whereas
the prompt component is generated by the decay of heavier mesons and baryons,
in\-clu\-ding charm or bottom valence quarks. In particular, the prompt component is
expected to become larger than the conventional component at energies larger than $\sim 6^{\, +12}_{\, -3}\, \times \, 10^{5}$~GeV~\cite{Garzelli:2015psa}.  
At present, uncertainties affect the shape and normalization of both the
conventional and prompt contributions. 
As for the conventional component, uncertainties of about 10\% - 20\% are quoted 
in the literature~\cite{Honda:2006qj}. 
Uncertainties on the prompt component are larger, due to the poor
knowledge of the charm hadroproduction process~\cite{Garzelli:2015psa,Accardi:2016ndt}. 

In this paper, we investigate how the uncertainties on nucleon composition in
terms of parton distribution functions (PDFs) affect prompt neutrino fluxes,
by making use of the PROSA PDF fit~\cite{Zenaiev:2015rfa}, 
and we compare this uncertainty to that 
from QCD and astrophysical origin. 
Current PDF fits are based on data collected at the HERA $ep$ collider, 
in fixed-target Deep Inelastic Scattering (DIS) and Drell-Yan experiments and in some cases also make use of $p\bar{p}$ data from Tevatron and $pp$ data from the Large Hadron Collider (LHC) at central rapidities. 
Due to the limited phase-space coverage in Bjorken-$x$ by these experiments 
all the data allow to constrain PDFs only in the range $x \gtrsim 10^{-4}$.
The PROSA fit was the first exploiting the potential of the LHCb data in
constraining gluon PDFs at low Bjorken-$x$ values, encouraging global fits to include these data as well. A first study in this direction was subsequently done in Ref.~\cite{Gauld:2015yia}, including, a-posteriori, information from the LHCb charm data at $\sqrt{s}$~=~7~TeV, on top of the NNPDF3.0 NLO fit, by using the Bayesian reweighting method, leading to the so-called NNPDF3.0 + LHCb set.~\footnote{Very recently, during the completion of our paper, the study of Ref.~\cite{Gauld:2015yia} was extended further in Ref.~\cite{Gauld:2016kpd} by using additional selected LHCb charm data at $\sqrt{s}$ = 13 and 5 TeV~\cite{Aaij:2015bpa, Aaij:2016jht} in a restricted kinematic region.}. As detailed in Ref.~\cite{Gauld:2015yia}, several approximations were applied within the heavy flavour scheme used for the extraction of these PDFs.
The PROSA analysis~\cite{Zenaiev:2015rfa}, on the other hand, contains a full PDF fit in a fully self-consistent scheme without approximations, and also includes LHCb data on beauty hadroproduction~\cite{Aaij:2013noa}. These developments have allowed to constrain gluon PDFs  for
$x \lesssim 10^{-4}$, a regime that is important to study in view 
of high-energy applications such as discussed in this work. 
In fact, VLV$\nu$Ts have so far detected neutrinos with deposited energies up to a few PeV in the laboratory frame, 
i.e. $E_{\nu,\, lab} \sim \mbox{(few)} \times 10^{6}$~GeV.
In this context it is important to note that
the production of neutrinos at a given laboratory energy $E_{\nu,\,lab}$ is affected by collisions of cosmic rays with nuclei in the Earth's atmosphere at energies $E_{p,\,lab}$ even larger by a factor of order ${\cal O}(10 - 1000)$. Further, $E_{p,\, lab} \sim 10^{8}$~GeV corresponds to $E_{pp, \,cm} \sim 13.7$~TeV, close to the present LHC center-of-mass energy. 
Plans exist to extend the accessible neutrino energy range even further, 
by increasing the active volume of ice/water probed by the optical module instrumentation. 
With the IceCube-Gen2 program~\cite{Aartsen:2014njl}, 
an extension from an instrumented fiducial volume of $\sim 1$ $\mbox{km}^3$ to a larger one, of $\sim 10$ $\mbox{km}^3$, has been proposed,
which claims the potential to deliver significant samples of neutrinos with laboratory energies in the EeV range, 
i.e. increased by a factor of  $\sim$ 1000. 
In nucleon-nucleon collisions such high-energy neutrinos 
may arise from collision energies $E_{pp, \,cm}$ up to a few hundred TeV.

So far IceCube has detected several tens of leptonic events with a cha\-rac\-te\-ri\-stic topology of tracks or showers,
with interaction vertices fully contained within the instrumented volume of the detector. This feature, in addition to a minimum charge/energy requirement, define a category of so-called high-energy-starting events (HESE), in relation to their specific signatures\footnote{IceCube is also able to record with higher statistics and analyze partially contained track or cascade events. Those events starting outside the active volume of the detector and entering it are not included in the HESE sample.}. 
Three of those HESE events populate the high-energy tail in the $\mathcal{O}$(PeV) region of neutrino deposited energies in the laboratory frame.
As a practical application of our study we consider the case of the IceCube HESE analyses~\cite{Aartsen:2013jdh, Aartsen:2014gkd, Aartsen:2015zva},
and show how the theoretical uncertainties affect the number of expected events 
from prompt neutrinos as a function of the energy deposited in the detector.  
We also compare our predictions to the upper limit on the prompt ($\nu_\mu$ + $\bar{\nu}_\mu$) flux recently obtained by the IceCube collaboration in a complementary analysis, restricted to $\sim$ 350000 up-going events from the Northern hemisphere, with neutrino interaction vertices either outside or inside the instrumented volume~\cite{Aartsen:2016xlq}. 

The paper is organized as follows: a short review of the PROSA PDF fit and of
the details of the QCD framework we adopt is presented in Section~\ref{sec:theory}. 
Predictions for prompt neutrino fluxes and their uncertainties are discussed in Section~\ref{sec:fluxes}, 
whereas comparisons with the data collected in the aforementioned IceCube analyses are provided in
Section~\ref{sec:icecube}, followed by our conclusions in Section~\ref{sec:conclu}. 
Appendix~A contains details of the QCD predictions 
for charm hadroproduction compared with the LHCb experimental data taken in
collisions at center-of-mass energy $\sqrt{s}=7$ and 13~TeV.
\section{Charm hadroproduction in QCD and Parton Distribution Functions}
\label{sec:theory}

The hadroproduction of heavy-quark pairs has received a lot of attention since
the time when early colliders were built.
The attention focused on the production of charm-anticharm and bottom-antibottom pairs~\cite{Lourenco:2006vw,Behnke:2015qja} already well before the advent of the high-energy colliders Tevatron and LHC.
The charm and bottom masses $m_c$ and $m_b$ are large enough with respect to $\Lambda_{QCD}$, 
so that the use of perturbative QCD is justified to describe 
the hard partonic scatterings giving rise to these quarks.
However, these masses are far smaller than present day collider energies.  
Thus, processes involving charm and bottom hadroproduction at high energies
are typically multiscale processes for which it is worth to investigate the
role of logarithms of the ratios of the different scales involved. 
The effect of resummation of different kinds of logarithms has been studied in 
some specific cases (e.g. resummation up to a certain accuracy
of threshold logarithms and of transverse momentum dependent logarithms $(p_T/m_q)$ at high $p_T$), while there are other resummations which have not yet been performed.
NLO QCD corrections to charm hadroproduction are available
since the eighties~\cite{Nason:1987xz, Mangano:1991jk}, and have been further
implemented in Monte Carlo event genera\-tors matching NLO QCD corrections to
parton shower (starting from pio\-nee\-ring work of~\cite{Frixione:2003ei}
and~\cite{Frixione:2007nw}), which are routinely used in the analysis of LHC data. 
NLO electroweak corrections are available in the case of top-antitop quark hadroproduction and, consistently with the naive expectations, they have been found to be far
smaller than the QCD ones, see e.g., \cite{Uwer:2016htd,Pagani:2016pbr}.
In case of $c\bar{c}$ and $b\bar{b}$ hadroproduction the electroweak corrections are expected to lie well within the NLO QCD scale uncertainty band.
Thus we neglect electroweak corrections in the present work.
On the other hand, NNLO QCD predictions are already available at least for the
total cross-section~\cite{Garzelli:2015psa,Mangano:2016jyj,Accardi:2016ndt}.
Extensions of the first differential studies on $t\bar{t}$ pairs 
at NNLO~\cite{Czakon:2015owf, Czakon:2016dgf} to the case of $c\bar{c}$ pairs still require further effort, 
particularly to take into account the small value of the charm mass, $m_c < 2$~GeV,
which may lead to stability issues in the fully numerical methods used so far for
investigating heavy-quark hadroproduction at NNLO.  
The calculation of charm-pair hadroproduction in this paper follows Ref.~\cite{Garzelli:2015psa} 
and uses recent results from perturbative QCD together with estimates on various sources of uncertainties.
In particular, we consider NLO QCD corrections matched to parton showers 
(with their standard logarithmic accuracy) as implemented in the \texttt{POWHEGBOX}~\cite{Alioli:2010xd} + \texttt{PYTHIA6}~\cite{Sjostrand:2006za} framework. Hadronization is also
taken into account by means of \texttt{PYTHIA6}, with non-perturbative
parameters regulated according to one of the most recent Perugia tunes~\cite{Skands:2010ak}. 
The {\texttt{POWHEGBOX}} framework includes
hard-scattering matrix-elements for charm hadroproduction according to
Ref.~\cite{Mangano:1991jk, Frixione:2007nw}. The charm is considered massive throughout the calculation of the scattering amplitudes performed in the fixed flavour number scheme ($n_f $~=~3). Consistently, we adopt 3-flavour PDFs.
{\texttt{POWHEGBOX}} is capable of producing events in the Les Houches format, including up to a first resolved radiation emission. These events are subsequently further showered by the $p_T$-ordered parton shower algorithms included in {\texttt{PYTHIA6}}.
The {\texttt{POWHEGBOX}} framework was also used in Ref.~\cite{Gauld:2015kvh} with {\texttt{PYTHIA8}}~\cite{Sjostrand:2014zea} instead of {\texttt{PYTHIA6}}. 

Alternatively, it is possible to use approaches where predictions at the
parton level are directly convoluted with
phenomenological non-perturbative fragmentation functions (FFs) for the
transformation of partons into charmed hadrons. The latter choice was adopted in
the computation of prompt neutrino fluxes by Ref.~\cite{Bhattacharya:2015jpa,
  Bhattacharya:2016jce}. It was also adopted when performing the original
PROSA fit for PDFs, in the computation of theoretical predictions for charmed-
and bottomed- meson hadroproduction which were compared with LHCb experimental
data. In particular, data for $D^\pm$, $D_s^\pm$, $\Lambda_c^\pm$, $D^{*\pm}$, 
$D^0$ and $\bar{D}^0$ hadroproduction~\cite{Aaij:2013mga} 
in $pp$ collisions at $\sqrt{s} = 7$~TeV 
in different transverse momentum and rapidity bins in the interval 
$0 < p_T < 8$~GeV and $2 < y < 4.5$ and 
data for $B^+$, $B^0$ and $B_s^0$ meson production~\cite{Aaij:2013noa} with 
$0 < p_T < 40$~GeV in the same rapidity range and at the same center-of-mass energy 
were used in the fit, as detailed in Ref.~\cite{Zenaiev:2015rfa}.
The performances of the PROSA PDF fit in association with our \texttt{POWHEGBOX}~+~\texttt{PYTHIA6} 
setup in comparison with these data at both $\sqrt{s} = 7$ and 13~TeV are shown 
in Appendix A.
We emphasize here that the PROSA PDFs used in this paper were derived by fitting the ratios of LHCb rapidity distributions in each fixed $p_T$ bin.
It turns out that 
we obtain absolute $p_T$ distributions consistent with LHCb experimental data within QCD uncertainties. We consider this fact as a proof of the robustness of the PROSA fit. This consideration remains valid both if we produce theory predictions for differential cross-sections by the NLO QCD computation interfaced with FFs adopted in the PROSA fit, with fragmentation fractions in agreement with the most recent measurements~\cite{Lisovyi:2015uqa}, and if we use for the same purpose the approach of this paper, matching NLO QCD matrix elements with parton shower and hadronization, pointing to the consistency of the two theoretical approaches.
\section{Predictions for prompt neutrino fluxes and their uncertainties}
\label{sec:fluxes}

The collision of primary cosmic rays with the Earth's atmosphere
lead to secondary particles, which during their propagation towards
the Earth surface may in turn reinteract or decay.
This mechanism, characterized by an interplay between in\-te\-rac\-tions and
decays, leads to the production of fluxes of different kinds of particles,
including neutrinos, and is described by cascade equations,
which define a set of coupled differential equations 
regulating particle flux evolution through an air column of slant depth $X$,
representing the amount of matter traversed downward along the direction of the
particle that initiated the cascade. These equations admit approximate solutions with the help of the so-called $Z$-moment approach initially proposed in Refs.~\cite{Gaisser:1990vg, Lipari:1993hd} (see also Refs.~\cite{Bhattacharya:2015jpa, Garzelli:2015psa}).

The differential distributions for the hadroproduction of $c\bar{c}$ quark pairs
are one of the essential ingredients to the solution of
those equations, yielding the prompt neutrino fluxes.
The necessary $Z$-moments, e.g., $Z_{p\, h_c}$ for charm hadroproduction in the atmosphere, are then obtained from the differential distributions $d\sigma/d x_E$ for the process $NN \rightarrow c\bar{c} \rightarrow h_c + X$, 
for all lowest-lying charmed hadrons $h_c$ = ($D^\pm$, $D^0$, $\bar{D^0}$, $D_s^\pm$
and $\Lambda_c^\pm$), with $x_E = E_{h,\, {lab}}/E_{N,\, {lab}}$, 
by an integration over the entire kinematically accessible range.
To that end we assume that the interaction of cosmic rays with nuclei in the atmosphere can be approximated by the superposition of nucleon-nucleon ($N\,N$) interactions. 

Those moments $Z_{p\, h_c}$ are combined together with the $Z$-moments for $h_c$ decays into neutrinos
$Z_{h_c\, \nu_\ell}$, those for $h_c$ regeneration, $Z_{h_c \, h_c}$, and those for nucleon regeneration, $Z_{N \, N}$, 
see for instance Refs.~\cite{Lipari:1993hd, Pasquali:1998ji,Garzelli:2015psa}. 
In this work, all moments $Z_{p\, h_c}$ were computed using as input the normalized variant of the PROSA PDF fit~\cite{Zenaiev:2015rfa}, with central values for the factorization and renormalization scales as $\mu_R = \mu_F = \mu_0 = \sqrt{p_{T,c}^2 + m_c^2}$. 
The scales $\mu_R$ and $\mu_F$ were allowed to vary independently in the range 
$\mu_0/2 < \mu_R, \mu_F < 2 \mu_0$, with exclusion of the combinations 
$(1/2, 2)\mu_0$ and $(2, 1/2)\mu_0$, see, e.g. Ref.~\cite{Cacciari:2012ny}, 
and the charm pole mass was fixed to $m_c = 1.40 \pm 0.15$~GeV, as in Ref.~\cite{Garzelli:2015psa}. 
All other moments $Z_{h_c\, \nu_\ell}$, $Z_{h_c \, h_c}$ and $Z_{N \, N}$ were computed as in Ref.~\cite{Garzelli:2015psa} (see also Ref.~\cite{Bhattacharya:2015jpa, Bhattacharya:2016jce}).

\begin{figure}[h]
\begin{center}
\includegraphics[width=0.54\textwidth]{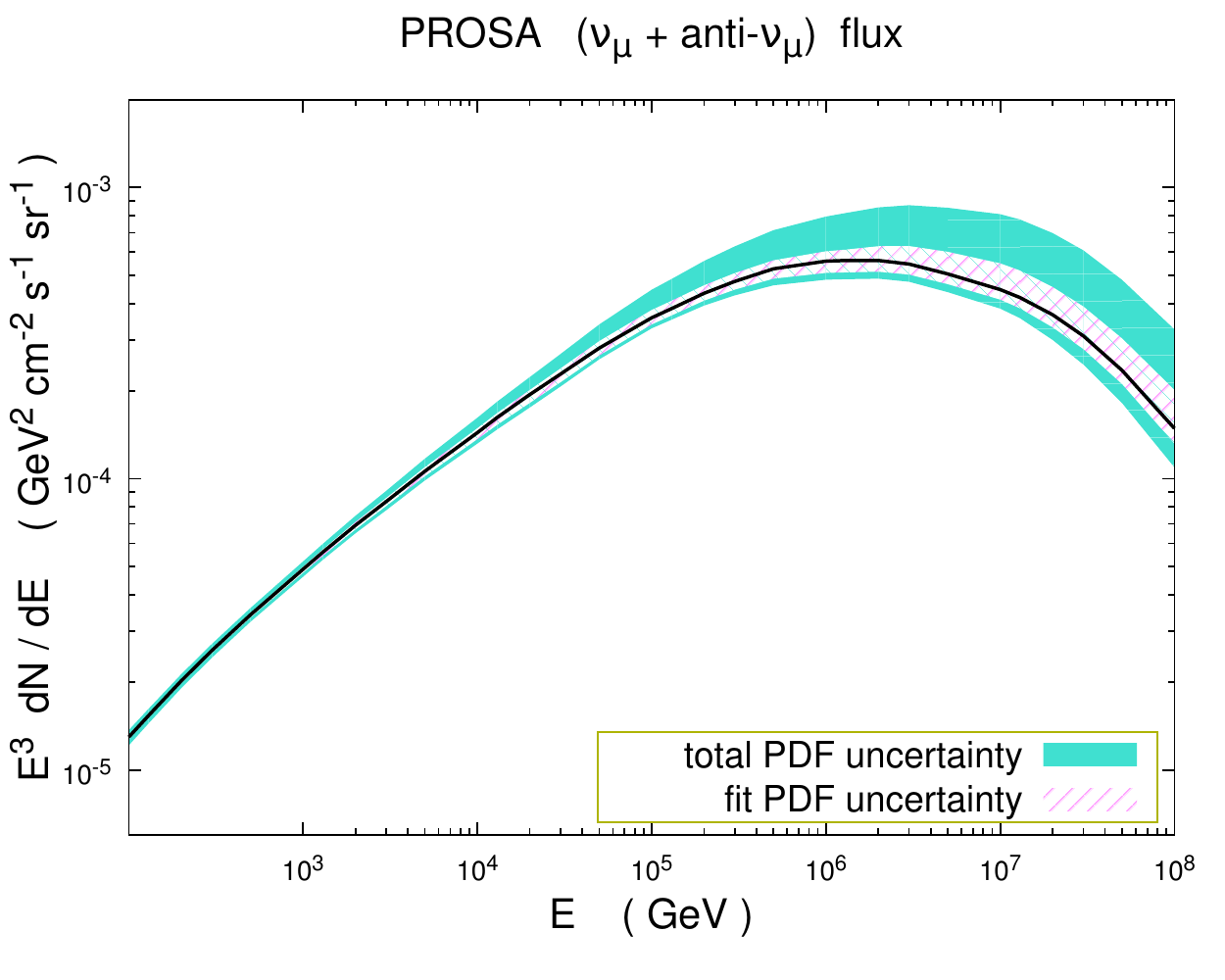}
\includegraphics[width=0.54\textwidth]{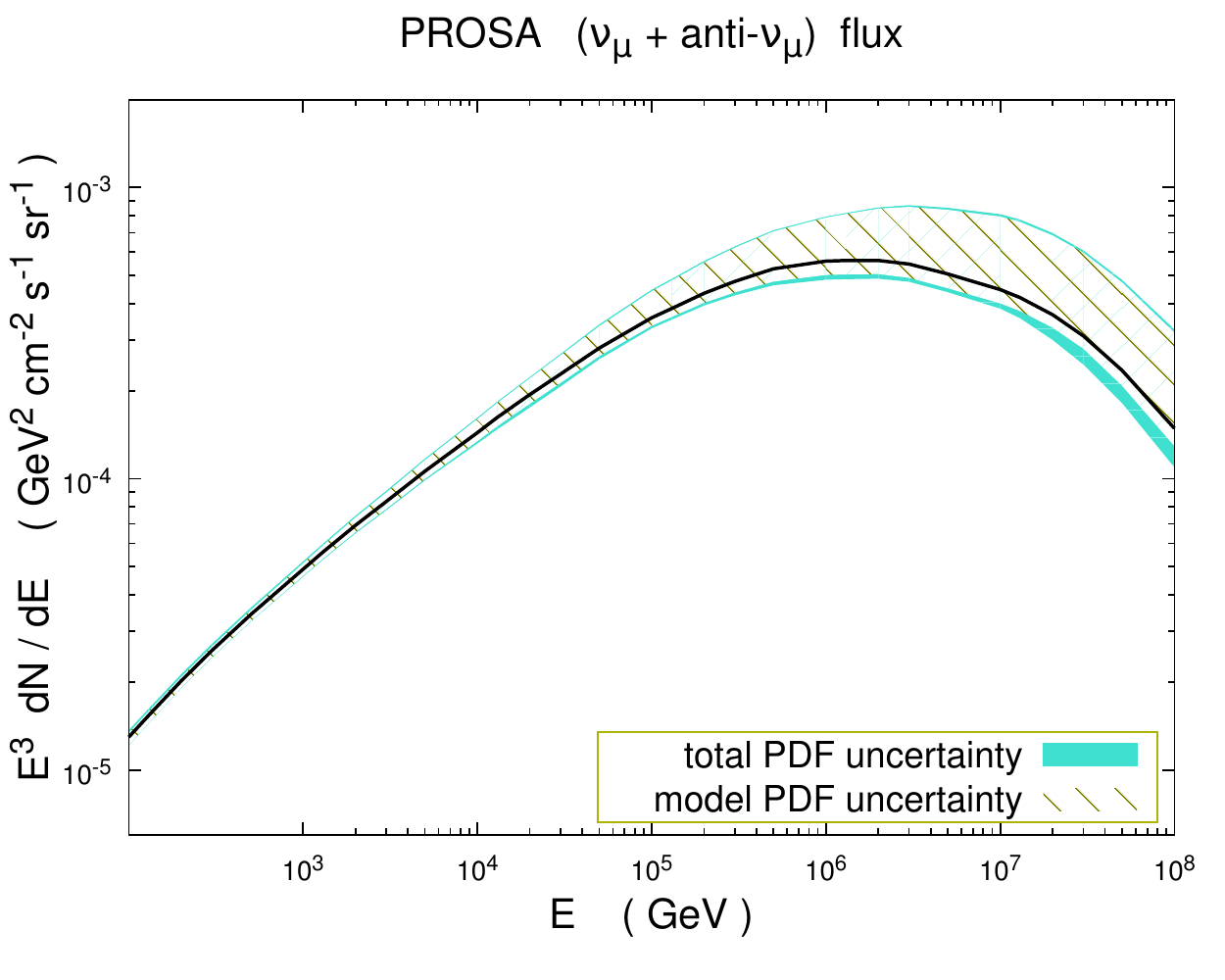}
\includegraphics[width=0.54\textwidth]{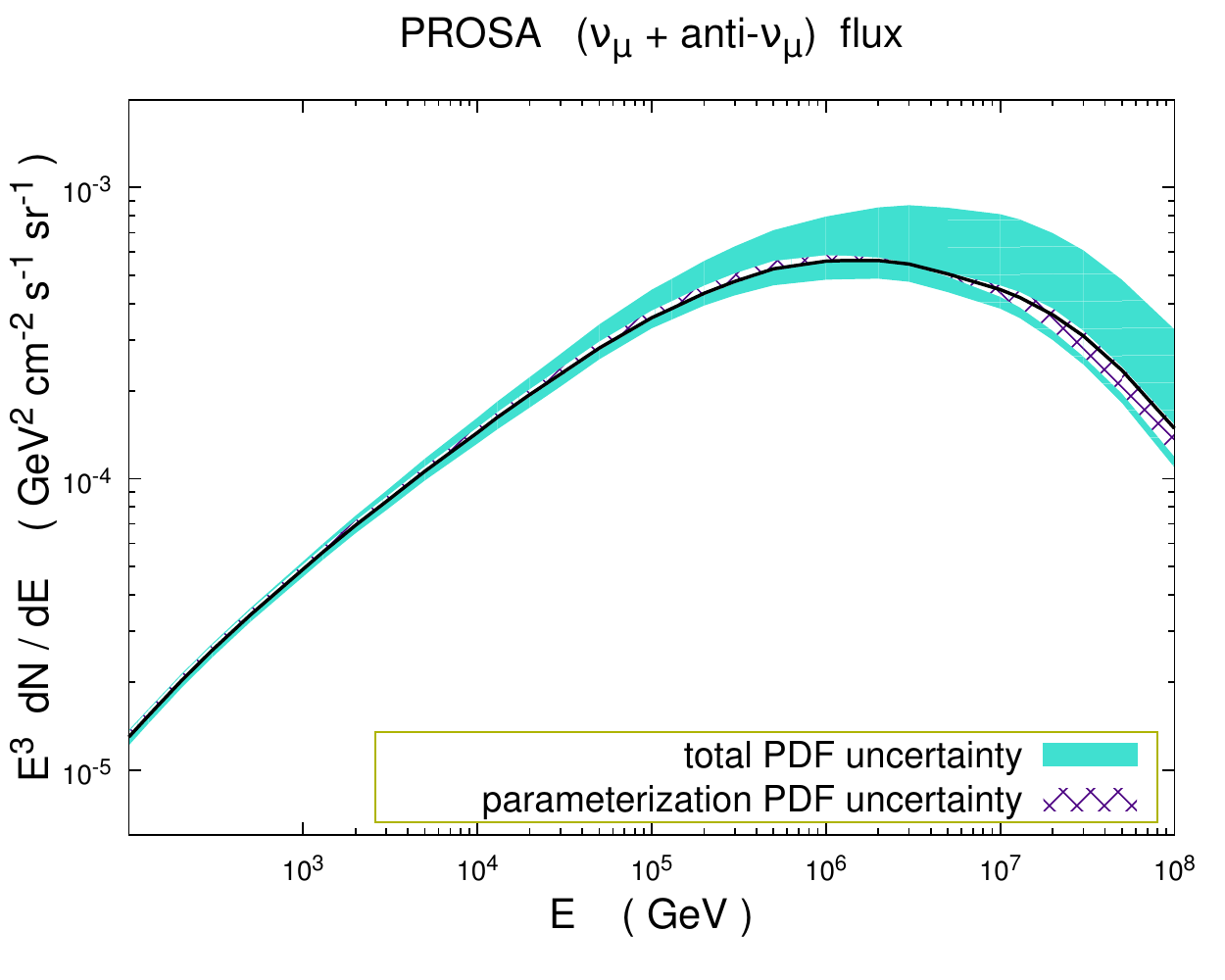}
\end{center}
\vspace*{-7mm}
\caption{\label{fig:pdfunctoflux} 
  PROSA PDF uncertainties on the prompt 
  ($\nu_\mu + \bar{\nu}_\mu$) atmospheric flux as a function of
  the neutrino energy $E_{\nu,lab}$: the contribution due to (fit + experimental), model and parameterization PDF uncertainties are shown in separate panels,
  respectively, and compared to the total PDF uncertainty (blue band).
  A broken power-law all-nucleon spectrum for the cosmic ray flux impinging on
  the Earth atmosphere is used as input, cf. Ref.~\cite{Garzelli:2015psa}.
} 
\end{figure}

\begin{figure}[ht!]
\begin{center}
\includegraphics[width=0.67\textwidth]{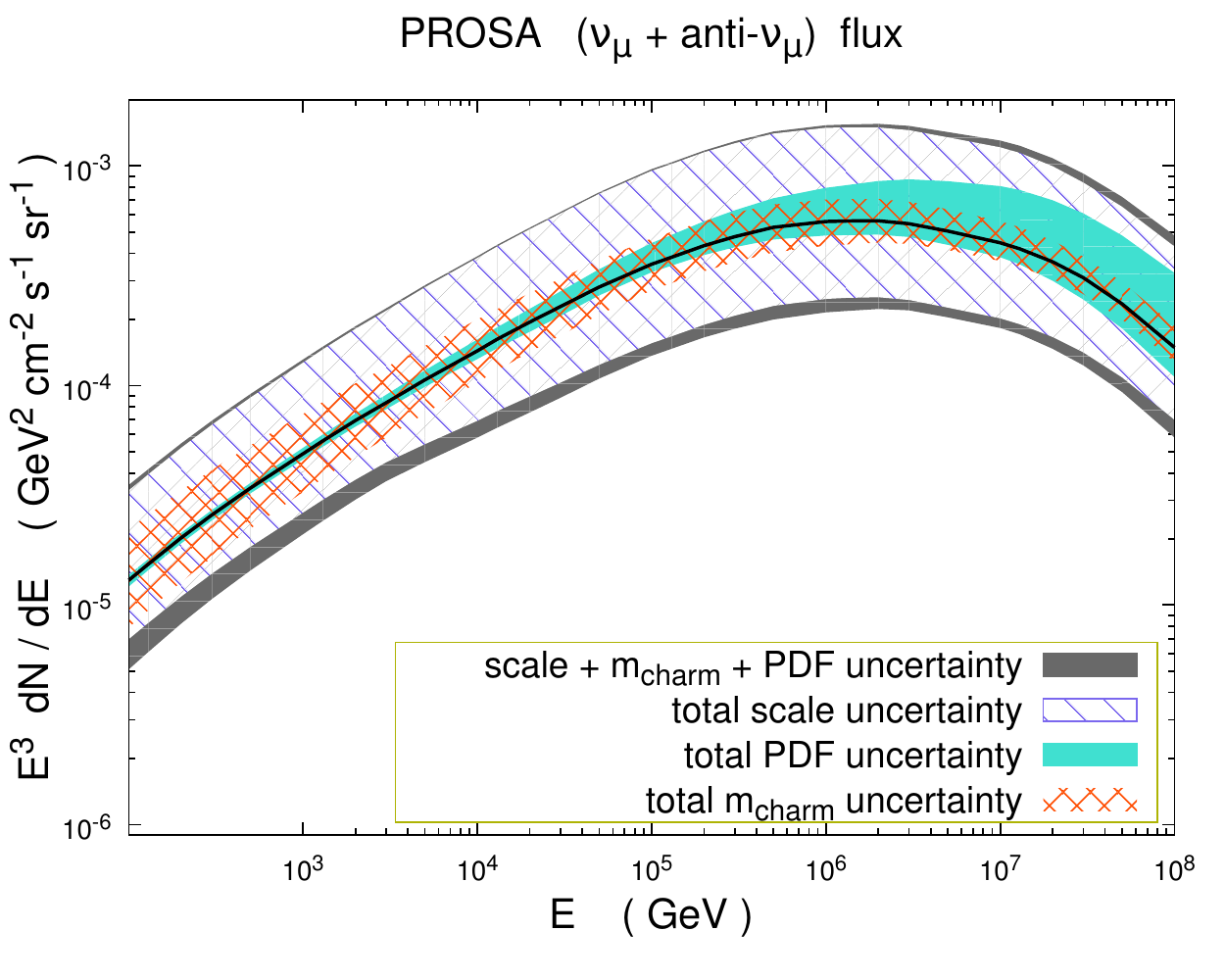}
\end{center}
\vspace*{-5mm}
\caption{\label{fig:unctoflux} 
  Central prediction for the prompt 
  ($\nu_\mu + \bar{\nu}_\mu$) flux together with its QCD uncertainties as a function of
  the neutrino energy $E_{\nu,lab}$. The uncertainty contributions due to
  $\mu_R$ and $\mu_F$ scale variation around $\mu_0$, $m_{c}$ 
   and the PDF eigenvalues within the PROSA fit, are shown
  separately by bands of different styles and colors, together with their combination in quadrature.  The same broken power-law all-nucleon spectrum for the cosmic ray flux as in Fig.~\ref{fig:pdfunctoflux}  is used as input.
} 
\end{figure}

\begin{figure}[ht!]
\begin{center}
\includegraphics[width=0.49\textwidth]{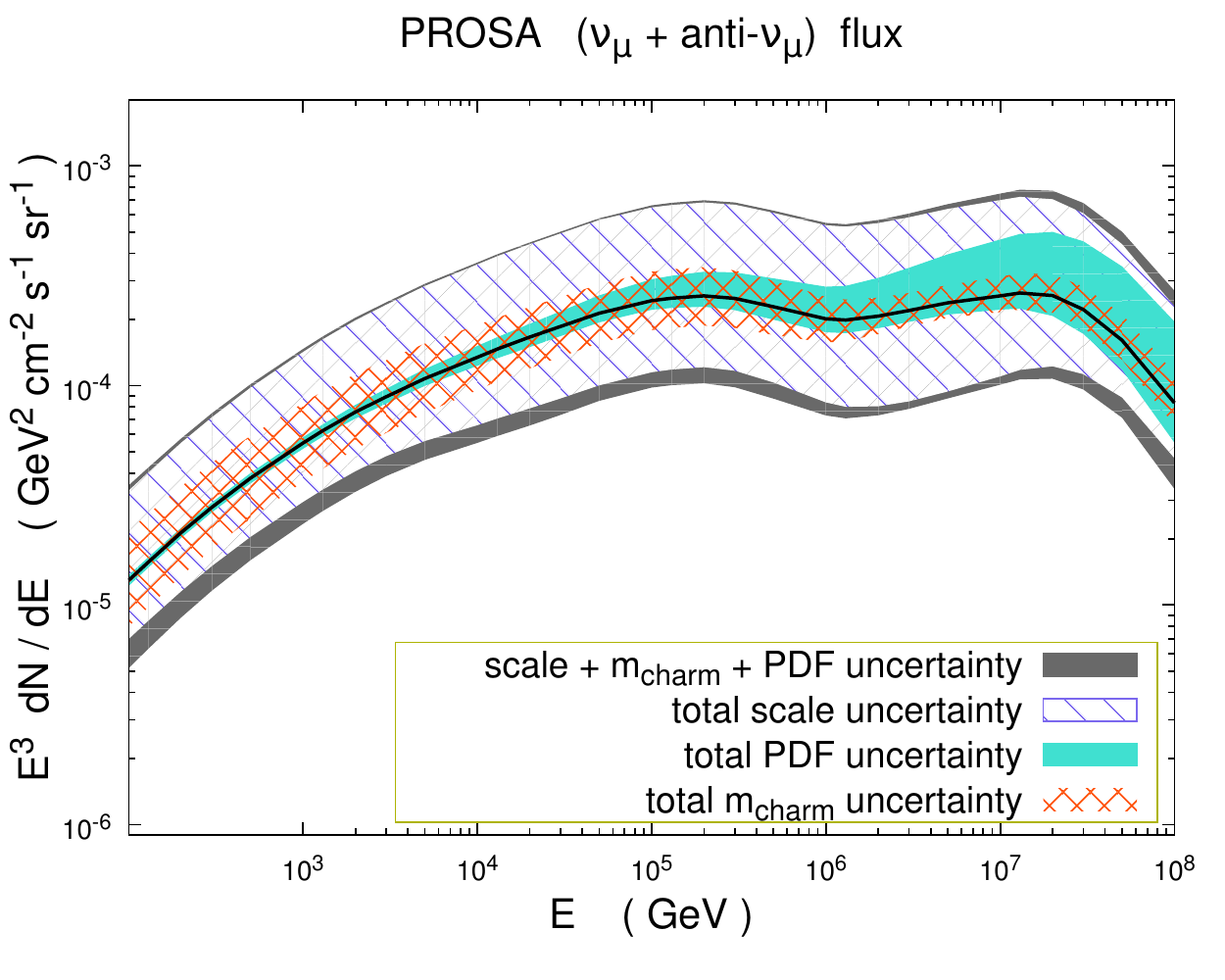}
\includegraphics[width=0.49\textwidth]{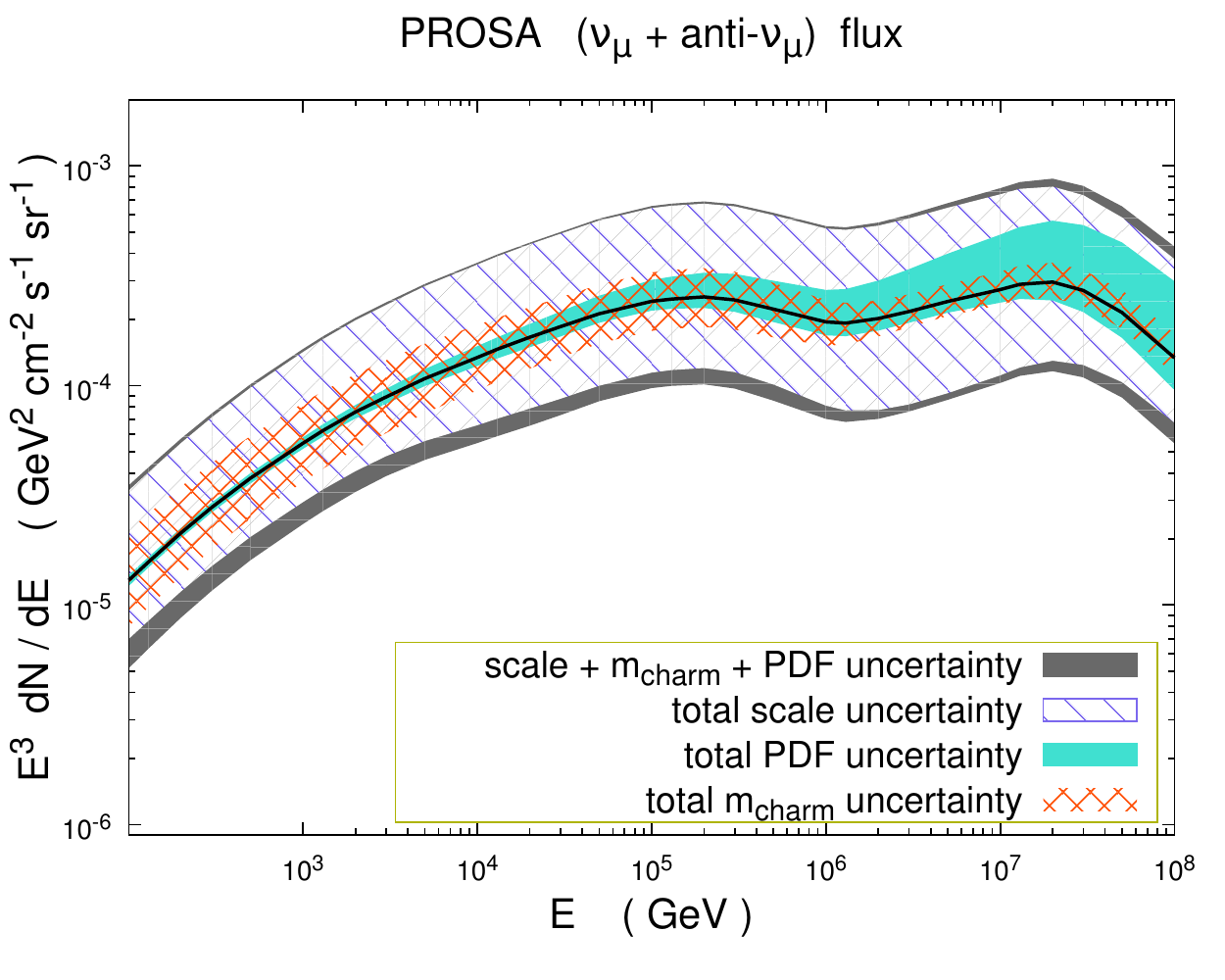}
\includegraphics[width=0.49\textwidth]{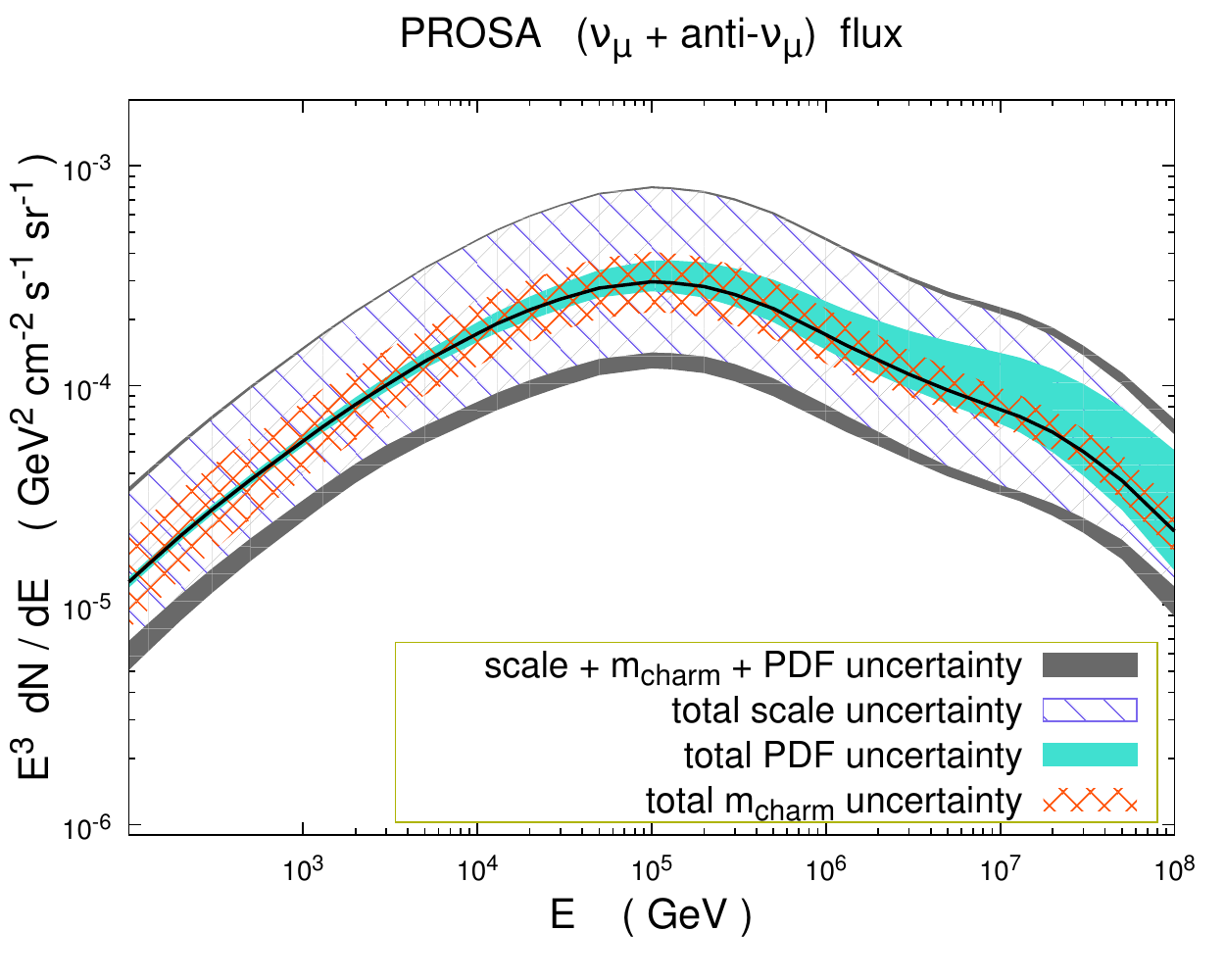}
\includegraphics[width=0.49\textwidth]{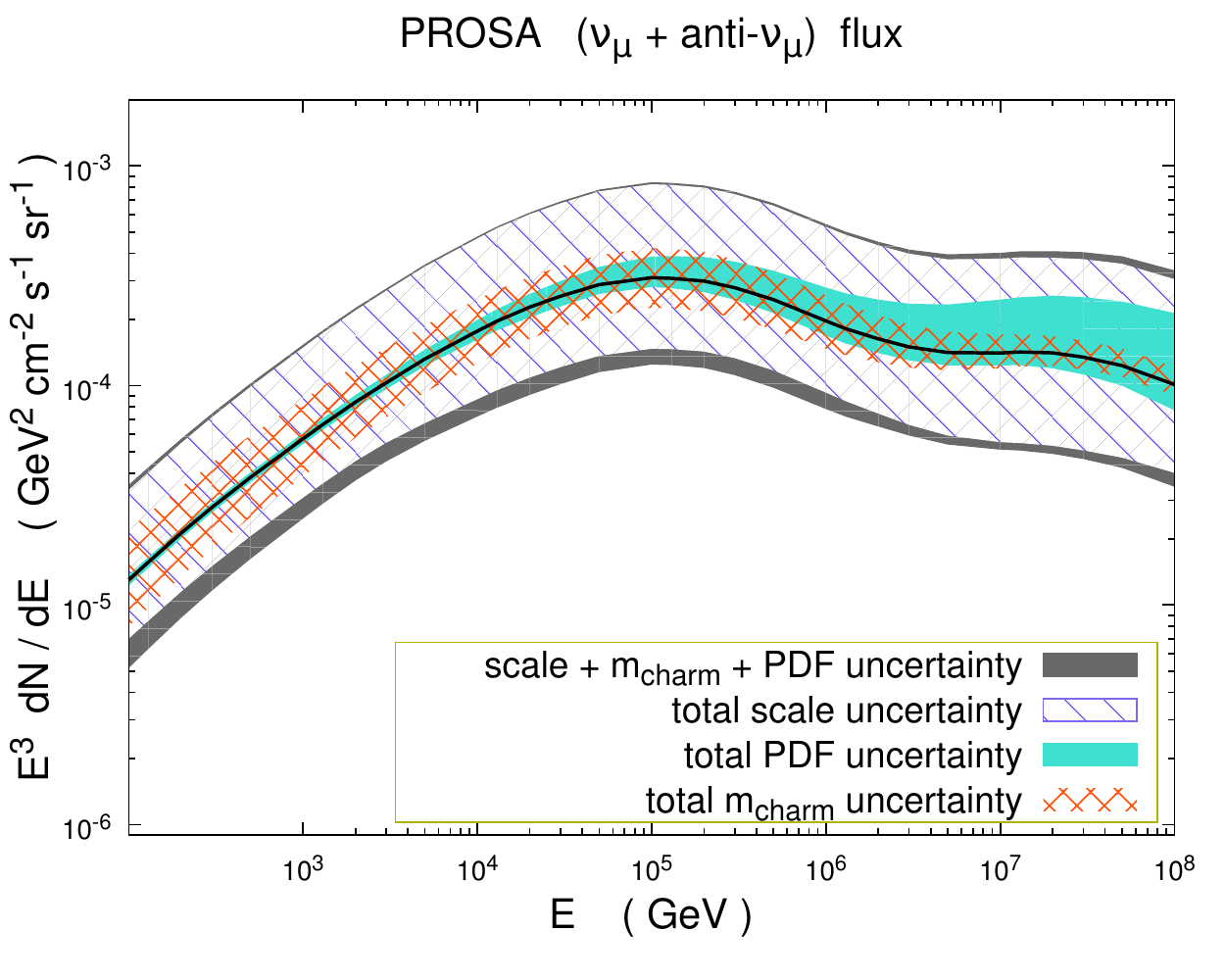}
\end{center}
\vspace*{-5mm}
\caption{\label{fig:unctoflux2} 
  Same as Fig.~\ref{fig:unctoflux}, using
  as input different more realistic cosmic ray all-nucleon spectra. The upper
  two panels correspond to the %Gaisser 2012
  GST fit described in Ref.~\cite{Gaisser:2013bla,Stanev:2014mla},
  whereas the lower two panels correspond to the %Gaisser 2014
  Gaisser H3a and H3p fit of Ref.~\cite{Gaisser:2013bla, Gaisser:2011cc}.
  Both fits have two variants:
  one corresponding to a heavy composition at the highest energies (left upper panel: GST-3, left lower panel: H3a),  and a second one including a dominating light proton component instead (right upper panel: GST-4, left lower panel: H3p).} 
\end{figure}

\begin{table}[b]
  \begin{center}
\begin{tabular}{|l|c|c|}
\hline    
PDF uncertainty component & Number of variations & Uncertainty source\\
\hline
fit + exp. stat. + exp. syst. & 30 & data uncertainties \\
model & 16 & $\alpha_s$, $f_s$, $Q^2_{min}$, $\mu_R$, $\mu_F$, $\alpha_K$\\
parametrization & 4 & $Q_0^2$, $D_{u_v}$, $D_{\bar{U}}$, $D_{\bar{D}}$ \\
\hline    
\end{tabular}
\caption{\label{tab:prosapdf} 
  Summary of the sources of uncertainty included in the PROSA PDF fit, 
  together with the list of the corresponding uncertainty sources.} 
\end{center}
  \end{table}

The resulting predictions for the ($\nu_\mu + \bar{\nu}_\mu$) fluxes 
are presented in Figs.~\ref{fig:pdfunctoflux},~\ref{fig:unctoflux} and~\ref{fig:unctoflux2}.
In particular, the central predictions were computed using as input the central
set of the PROSA PDFs, whereas PDF uncertainties were computed using all
different PROSA PDF variations.  
The latter consist of three components, as described in detail in
Ref.~\cite{Zenaiev:2015rfa}: fit uncertainties originating from experimental
uncertainties of the measurements, model uncertainties 
and parametrization uncertainties, cf. Tab.~\ref{tab:prosapdf}.
Model uncertainties arise from the variations of the model assumptions, 
such as the value of the strong coupling constant $\alpha_s(M_Z)$, 
the strangeness fraction $f_s$ in the PDF fit, 
the minimum virtuality cut $Q^2_{min}$ on the $ep$ DIS data entering the fit, 
the choices of the renormalization and factorization scales, $\mu_R$ and $\mu_F$, and the parameters $\alpha_K$ of the charm and beauty fragmentation functions~\cite{Kartvelishvili:1977pi}.
Parametrization uncertainties are assessed by varying the functional form of the PDFs 
using additional parameters $D_{u_v}$, $D_{\bar{U}}$, $D_{\bar{D}}$ 
as described in Refs.~\cite{Zenaiev:2015rfa,Aaron:2009aa} at the starting
scale $Q_0^2$ of the QCD evolution, as well as the value of the starting scale. 

\begin{figure}[ht]
\begin{center}
\includegraphics[width=0.49\textwidth]{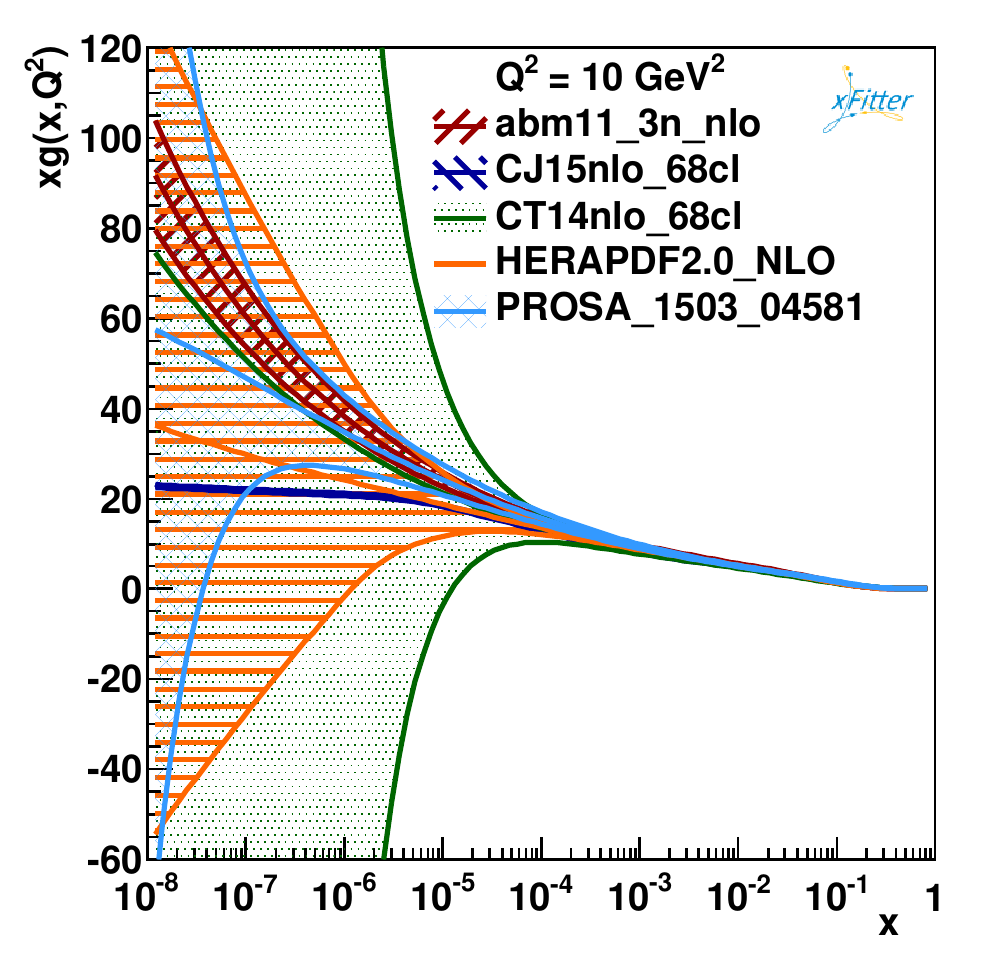}
\includegraphics[width=0.49\textwidth]{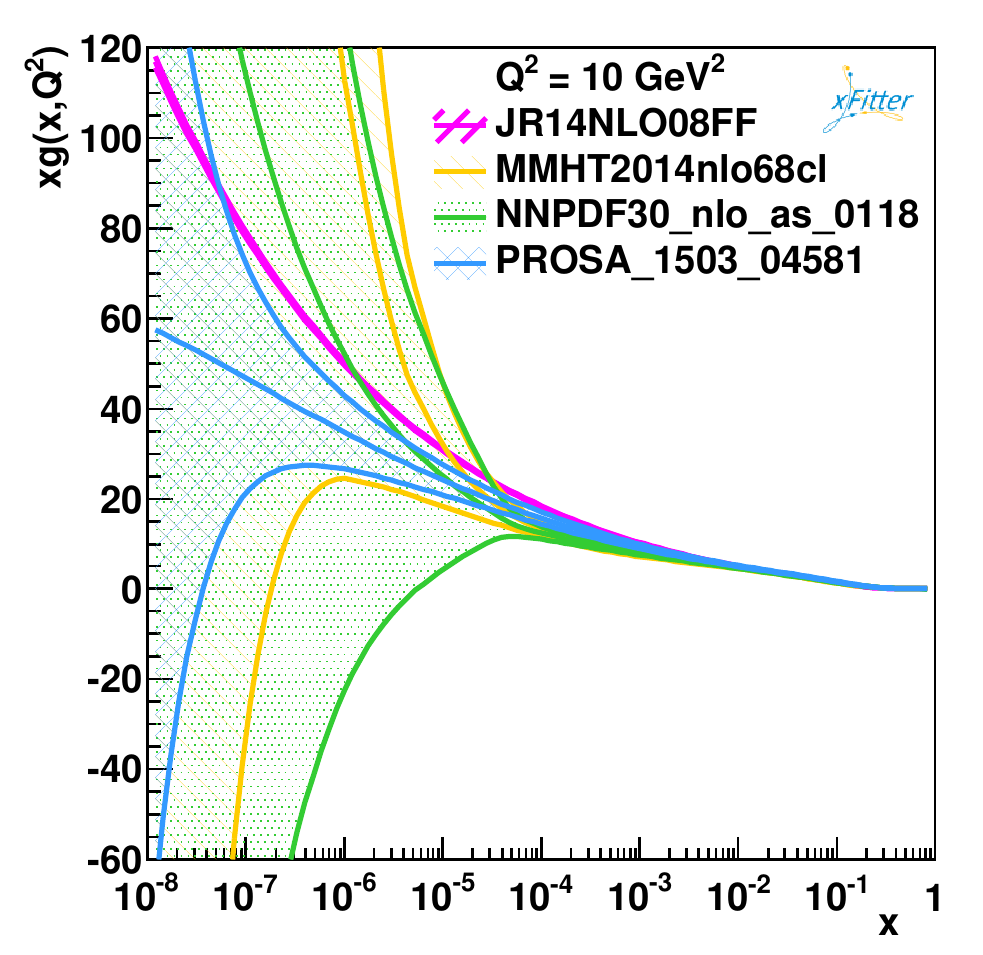}
\end{center}
\vspace*{-5mm}
\caption{\label{fig:prosagluon} 
  The PROSA gluon distribution compared to ABM11~\cite{Alekhin:2012ig}, 
  CJ15~\cite{Accardi:2016qay}, CT14~\cite{Dulat:2015mca}, HERAPDF2.0~\cite{Abramowicz:2015mha} (left) and 
  JR14~\cite{Jimenez-Delgado:2014twa}, MMHT2014~\cite{Harland-Lang:2014zoa} and NNPDF3.0~\cite{Ball:2014uwa} (right) 
  at NLO at the scale of $10$ GeV$^2$. 
  The plots are obtained using the \texttt{Xfitter} program~\cite{Alekhin:2014irh,HERAFitter}. Note that the ABM11, PROSA and JR14 PDFs employ the fixed flavour number scheme (with $n_f$ = 3) in the fit of DIS data, whereas CT14, CJ15, HERAPDF2.0, NNPDF and MMHT use different implementations of the variable flavour number scheme, so the latter distributions should only be compared qualitatively to the former ones.}
\end{figure}

The fit uncertainties (referred to as experimental uncertainties in Ref.~\cite{Zenaiev:2015rfa}) were provided as 30 eigenvectors arising from 13
fitted PDF parameters and charm and beauty masses left free in the fit.  
The model uncertainties were determined as positive and negative variations of each model parameter and were technically provided as 16 eigenvectors. 
The parametrization uncertainties were assessed with 4 individual variations,
and the total parametrization uncertainty was built as an envelope of the
maximal differences between these variations and the central value. 
The total PDF uncertainties were obtained by adding fit, model and parametrization uncertainties in quadrature. 
In Fig.~\ref{fig:prosagluon} the PROSA gluon distribution is compared to the results of other PDF groups~\cite{Alekhin:2012ig,Accardi:2016qay,
Dulat:2015mca,Jimenez-Delgado:2014twa,Abramowicz:2015mha,Harland-Lang:2014zoa,Ball:2014uwa}
in the relevant kinematic region of low Bjorken-$x$, at a scale of $10$~GeV$^2$. 
The use of the \texttt{LHAPDF 6.1.6} framework \cite{Buckley:2014ana} for Fig.~\ref{fig:prosagluon} ensures the extrapolation of the PDFs to the low-$x$ region beyond the kinematic range provided and fitted by the individual groups~\cite{Alekhin:2012ig,Accardi:2016qay, Dulat:2015mca,Jimenez-Delgado:2014twa,Abramowicz:2015mha,Harland-Lang:2014zoa,Ball:2014uwa}.

The separate contribution of each of the uncertainty sources  
in the PROSA PDFs listed in Tab.~\ref{tab:prosapdf} is shown
in Fig.~\ref{fig:pdfunctoflux} together with the total PROSA PDF
uncertainty and using a broken power-law all-nucleon spectrum as input for the cosmic ray flux, cf. Ref.~\cite{Garzelli:2015psa}.
The total PDF uncertainty turns out to be dominated by the model
uncertainties, which in turn for the gluon distribution in the region 
$x < 10^{-4}$ are influenced by theoretical uncertainties on heavy-flavor
hadroproduction, mostly arising from renormalization and factorization scale
variations.  
However, due to the use of the normalized heavy-flavor LHCb cross sections in
the PROSA fit, these theoretical uncertainties are strongly reduced, since
variations of the renormalization and factorization scales as well as of the
fragmentation parameters do not significantly affect the shape of the 
rapidity distributions for heavy-flavor production, while this shape remains sensitive to PDFs. 
For this reason, in the calculation of prompt neutrino flux the scale variations 
in the matrix elements were performed independently of the scale variations in the PDF fit.
The same applies for the value of the charm mass, 
which is well constrained in the PROSA fit by the data on DIS charm production.
The choice and uncertainity for the charm pole mass $m_c = 1.4 \pm 0.15$~GeV adopted here are compatible and have been motivated by detailed studies of the inclusive cross sections 
in Ref.~\cite{Garzelli:2015psa} computed with the very precise value for the 
charm mass reported by the Particle Data Group~\cite{Agashe:2014kda} in the $\overline{\rm MS}$ scheme.
In addition, since the scale variations have a much larger effect on the predictions than the $m_c$ value, 
possible correlations between the two can be neglected.  
The fit and parameterization PDF uncertainties turn out to have a minor role in the kinematic region of interest. 
The role of PDF uncertainties with respect to other sources of QCD
uncertainties affecting ($\nu_\mu + \bar{\nu}_\mu$) fluxes is shown in
Fig.~\ref{fig:unctoflux} again for the broken power-law primary cosmic-ray input spectrum as in Fig.~\ref{fig:pdfunctoflux}.
It is evident that, at all energies, the dominant source of QCD uncertainty is
represented by factorization and renormalization scale variations. 
On the other hand, the role of charm mass and PDF uncertainties is 
complementary: the former dominate over the latter at lower energies, 
whereas at higher energies the behavior is the opposite, with PDF uncertainties
increasing due to the increasing number of collisions occurring in an asymmetric situation when one of the partons inside the nucleon is characterized by low $x$ and the other one by high $x$. 
In this respect we would like to note that the present PROSA fit extends down to $x \sim 10^{-6}$, while below this value the gluon distribution is not directly constrained by any data and should be considered as an extrapolation which relies on assumptions for the parametrization of the PDFs.\footnote{On the other hand, $pp$ collisions at present LHC energies, involving initial state partons with $x_1$ and $x_2$ distributions, are characterized by distributions of the $min$($x_1$, $x_2$) minimum peaked around $10^{-4.5}$, with a tail extending down to $x$ $\sim$ $10^{-8}$, where the total differential cross-section $d\sigma$ / $d$ $min$($x_1$,$x_2$) is suppressed by a factor $\sim$ 50 with respect to its maximum.} 
In particular,
for neutrino energies $E_{lab,\,\nu}$ around~1~and~2 PeV, corresponding to the leptonic events with highest energy observed so far by IceCube, the PDF uncertainties look to be already far better constrained than the scale ones, and amount to
(+42\%, -13.5\%) and (+52\%, -13.5\%), respectively, for the power-law cosmic
ray spectrum. These values decrease by few percent when considering more
realistic input cosmic ray spectra. They can be compared to the charm mass uncertainties
amounting to about (+26\%, -22\%) and (+25.8\%, -20\%) at 1 and 2 PeV, respectively. 

It is plausible that, with the increasing availability of LHC experimental data 
even at higher energies, PDF uncertainties will decrease. 
Therefore, we can already conclude that, as for present day investigations 
of prompt neutrinos at neutrino telescopes, the uncertainty related to PDFs 
does not form a bottleneck.
Of course, if neutrino telescopes will be able to discover events at higher energies, 
either thanks to the extension of their fiducial volume, or simply by accumulating 
much more statistics over the years, 
the uncertainty due to the PDF variation may become a more critical issue. 

Another important input to cascade equations is represented by primary cosmic ray
fluxes, i.e. the energy spectra of cosmic rays on top of the Earth atmosphere,
as a function of their mass number. At energies above a PeV, an important aspect, at least presently, is related to our superficial knowledge of the composition of cosmic ray fluxes~\cite{Olive:2016xmw}.
Here the problem is that at the highest energies, cosmic ray spectra
cannot be measured directly by satellites or balloon-born experiments,
because the flux of cosmic rays decreases too rapidly. 
It is presently impossible to build an instrument capable of measuring a small,
but non negligible, flux because the detector surface and exposure time necessary are too big for current capabilities. Therefore balloon-born instruments or those in satellites are used at present days only for measuring cosmic ray spectra at lower energies, i.e. below the knee~\cite{Seo:2012pw}. 

On the other hand, at higher energies, cosmic ray spectra are investigated indirectly,
through extended air shower (EAS) experiments~\cite{Haungs:2003jv, Kampert:2012mx}. 
EAS experiments count the rate of leptons of different origin ($e$ and $\mu$) reaching the array of detectors on the Earth's surface and 
compare it with the data on the maximum development of the EAS electromagnetic component seen by fluorescence telescopes pointing to the
upper layer of the Earth's atmosphere.  
Last studies in this direction, ex\-ploi\-ting correlations between different EAS observables, seem to point to a CR spectrum characterized by a mixed composition in the energy region around the dip/ankle~\cite{Aab:2016htd}, which tends to become heavier at $E_{lab}$ $\sim$ a few $10^{19}$~eV. However, not all questions are solved, in particular in the comparison of the experimental data with the expectations from the Monte Carlo generators used for simulating the formation and development of EAS, 
characterized by large uncertainties in their hadronic interaction models~\cite{Aab:2016hkv}. As a consequence, uncertainties on the composition of cosmic ray spectra above $E_{lab}$ = $10^{16}$ - $10^{17}$~eV are still large.
 
In order to have an idea of the effect of these uncertainties on prompt ($\nu_\mu + \bar{\nu}_\mu$) fluxes, we show in Fig.~\ref{fig:unctoflux2} prompt neutrino spectra with their QCD uncertainties, for four different recent all-nucleon cosmic ray spectra~\cite{Gaisser:2011cc,Gaisser:2013bla,Stanev:2014mla}, 
provided by the group of T. Gaisser, which were recently created 
in order to fit the measured cosmic ray all-particle spectrum data.
The transformation of an all-particle spectrum into an all-nucleon spectrum 
requires additional knowledge (or assumption) on the composition. 
We have used these same spectra in Ref.~\cite{Garzelli:2015psa}, 
where we have provided a more extensive discussion of their content. 
The resulting neutrino spectra at e\-ner\-gies~$\gtrsim$~1~PeV 
are larger if a proton component or, more generally, 
a light component in the cosmic rays is assumed and dominates over the nuclear ones at the highest energies.
The QCD uncertainties behave in a similar way in all cases, 
comparable to the case of Fig.~\ref{fig:unctoflux} already discussed, 
but at the highest $E_{lab, \nu}$ energies the uncertainties due to our poor knowledge of cosmic ray composition are becoming large, as follows from comparing one with each other the various panels of Fig.~\ref{fig:unctoflux2}.

\begin{figure}[ht]
\begin{center}
\includegraphics[width=0.49\textwidth]{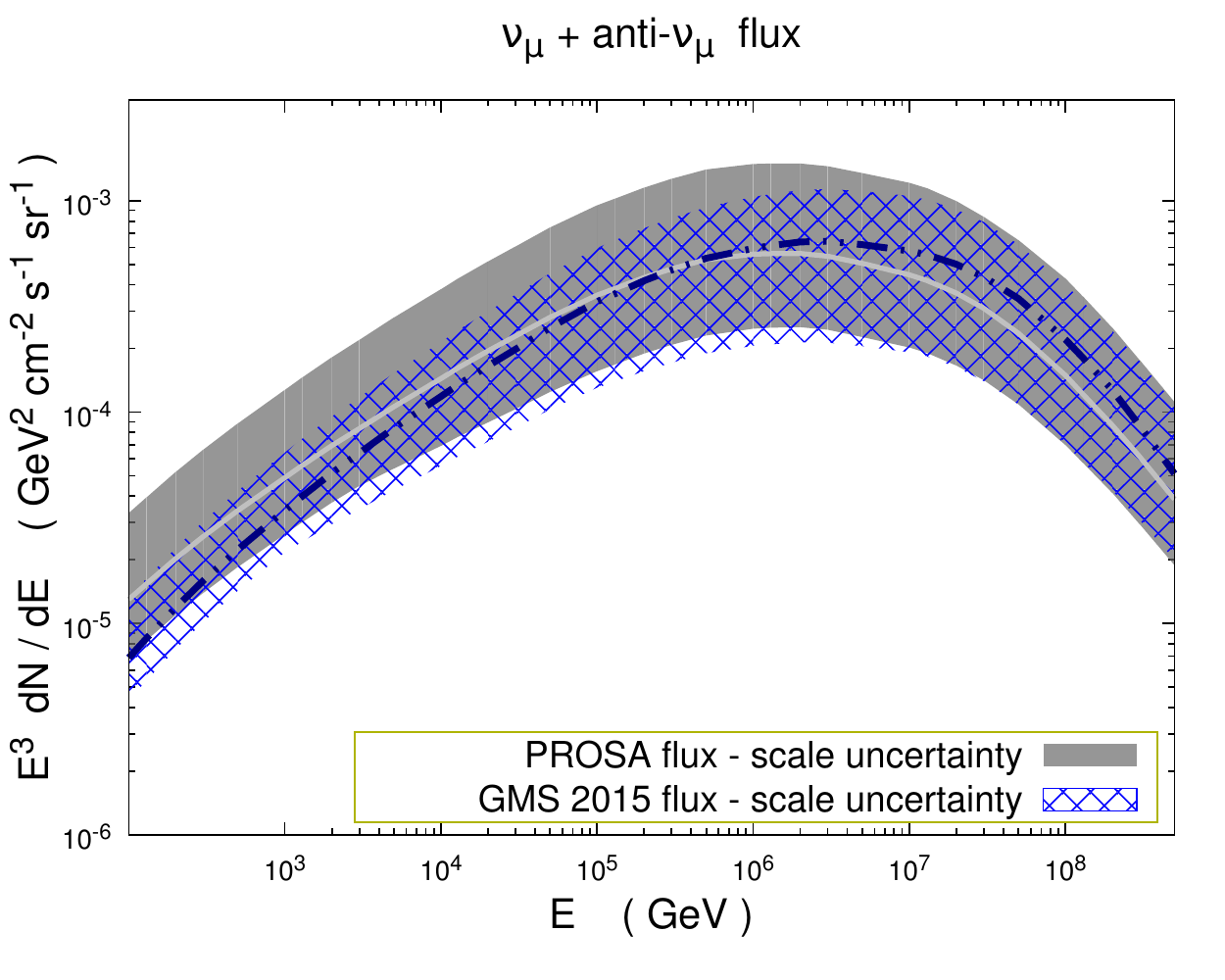}
\includegraphics[width=0.49\textwidth]{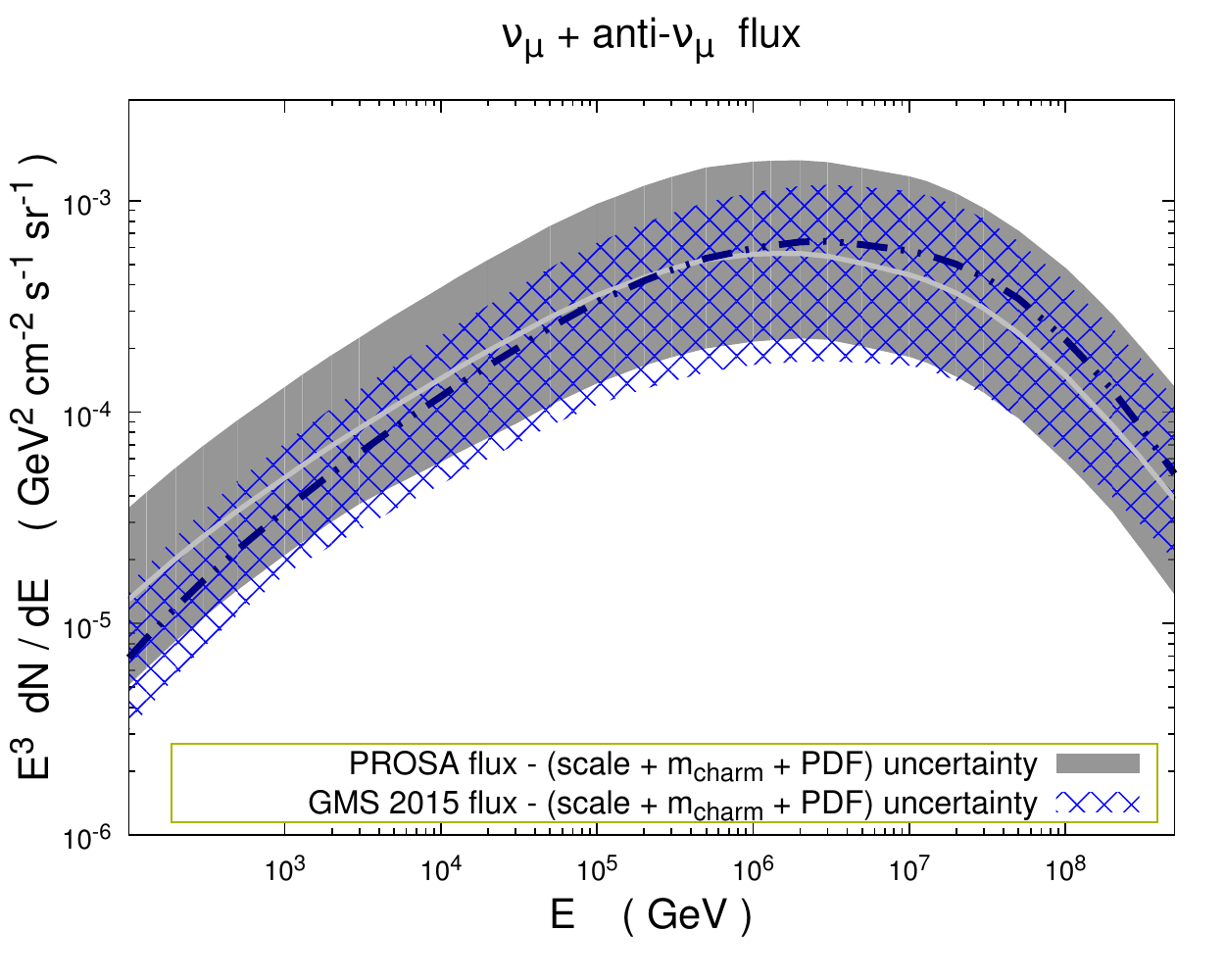}
\end{center}
\vspace*{-5mm}
\caption{\label{fig:compa_abm_prosa} 
  Prompt ($\nu_\mu + \bar{\nu}_\mu$) fluxes obtained with PROSA PDFs (solid line)  as compared to    the previous GMS 2015 predictions~\cite{Garzelli:2015psa},
  obtained using ABM11 PDFs (dash - double dotted line), shown with respective
  theory uncertainties, represented by bands of different styles. 
  Theoretical uncertainties due to scale variations are shown in
  the left panel, whereas those due to scale + $m_{c}$ + PDF variations
  are shown in the right panel. The main difference between the two fluxes
  is the use of a different central scale: 
  $\mu_0 = \sqrt{p_{T,c}^2 + m_c^2}$ for PROSA fluxes vs. 
  $\mu_0 = \sqrt{p_{T,c}^2 +4 m_c^2}$ for the GMS 2015 fluxes.
} 
\end{figure}

\begin{figure}[ht]
\begin{center}
\includegraphics[width=0.49\textwidth]{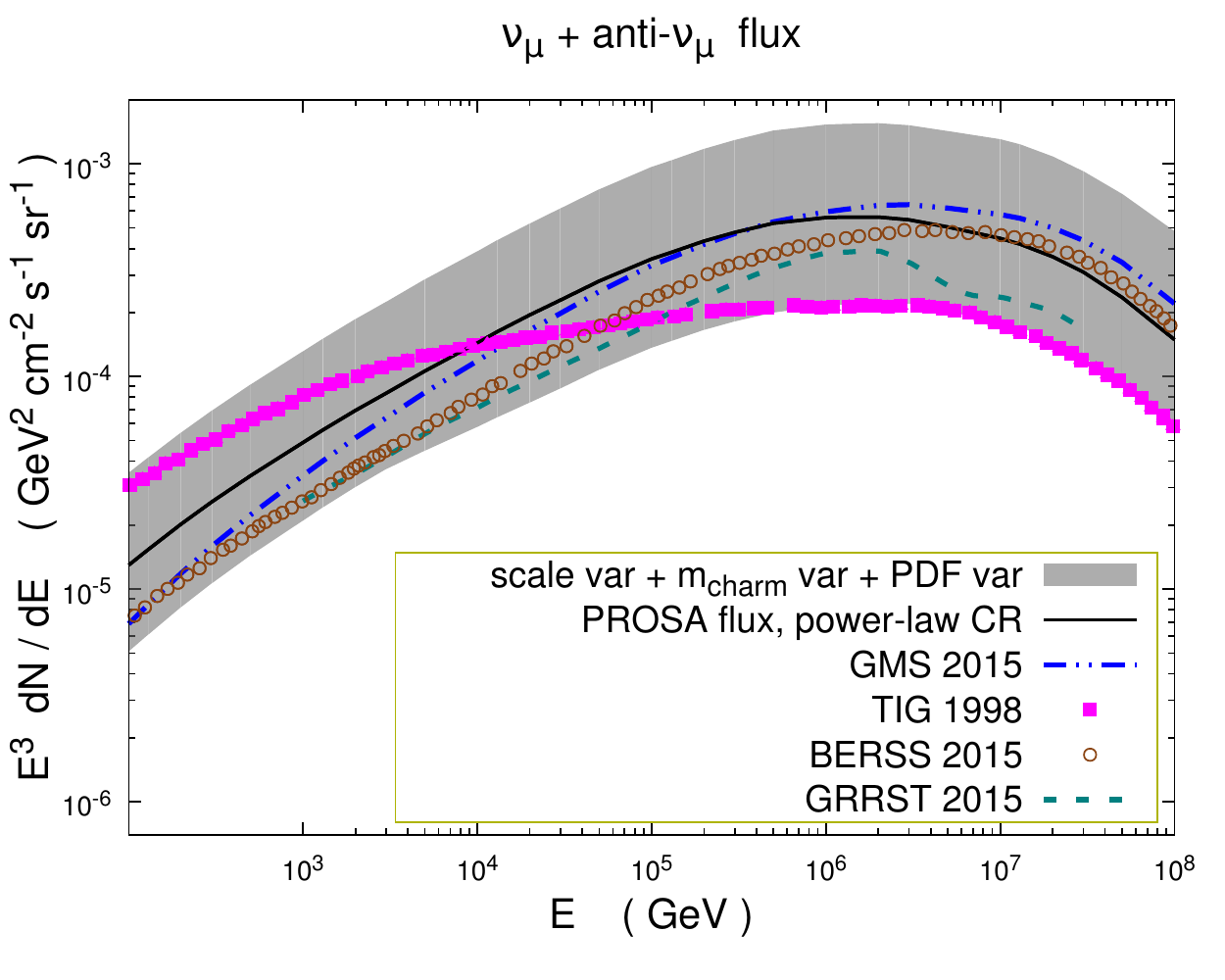}
\includegraphics[width=0.49\textwidth]{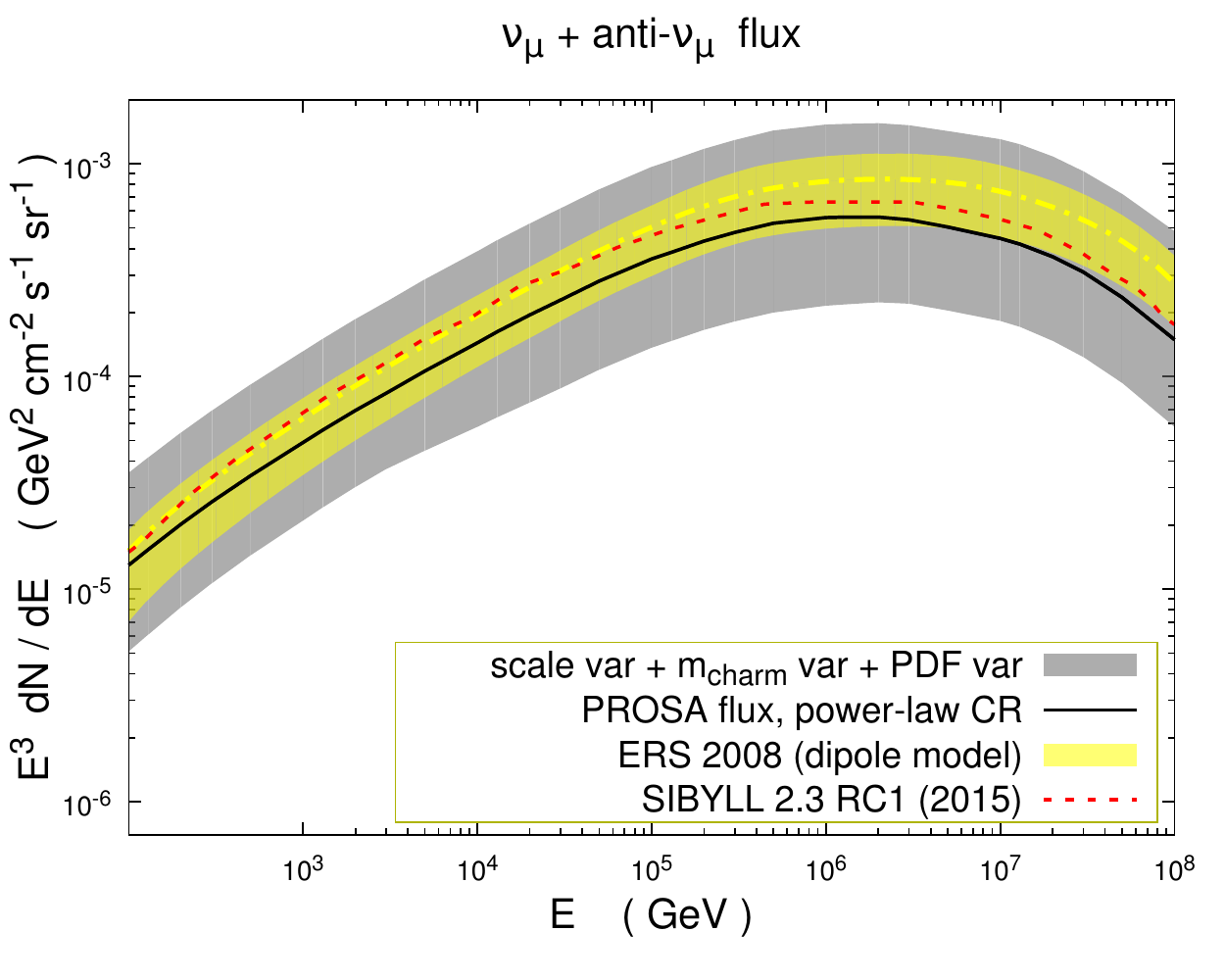}
\vspace*{-5mm}
\caption{\label{fig:fluxcompa} 
  Prompt neutrino spectrum from the PROSA PDFs with
  its uncertainties compared to other predictions from the 
  literature~\cite{Garzelli:2015psa, Gauld:2015kvh, Bhattacharya:2015jpa,
    Gondolo:1995fq, Fedynitch:2015zma, Enberg:2008te}. 
  Predictions treating charm hadroproduction 
  at parton-level by means of perturbative QCD are shown on the left. 
  On the right comparisons with predictions using the dipole model  
  and a recent version of the \texttt{SYBILL} event generator are shown.  
  The broken power-law cosmic ray spectrum is used as input in all predictions, cf. Fig.~\ref{fig:pdfunctoflux}.
} 
\end{center}
\end{figure}

It is worth to compare our results on cosmic ray fluxes using the PROSA PDFs to
those obtained in 2015 with the ABM PDFs in Ref.~\cite{Garzelli:2015psa} (labelled as GMS 2015) in Fig.~\ref{fig:compa_abm_prosa}.
The central values of the ($\nu_\mu + \bar{\nu}_\mu$) fluxes,  
together with their scale and total QCD uncertainties, 
are shown separately in the two panels of Fig.~\ref{fig:compa_abm_prosa}. 
For both fluxes the scale uncertainties dominate the total QCD uncertainty, 
as we have already observed above. 
The central GMS 2015 predictions are included in the uncertainty band of the
PROSA predictions of the present study, and viceversa, 
the PROSA central predictions are within the uncertainty band of the GMS 2015 fit. 
The differences in the shape of central predictions and of the uncertainty band 
are due to the different choice for the central scale $\mu_0 = \mu_R = \mu_F$. 
The GMS 2015 predictions were computed using $\mu_0 = \sqrt{p_{T,c}^2 + 4 m_c^2}$ 
whereas the PROSA predictions have been obtained with the more widely adopted scale choice 
$\mu_0 = \sqrt{p_{T,c}^2 + m_c^2}$. 
We have verified that when using the same scale choice for $\mu_0$, 
the predictions with ABM and PROSA PDFs have a quite similar shape. 
The scale uncertainty corresponding to the choice adopted in
this paper, $\mu_0~=~\sqrt{p_{T,c}^2 + m_c^2}$, leads to uncertainty bands larger than for GMS 2015.
Note that the scale in the GMS 2015 study was explicitly chosen to fulfill the principle of fastest convergence, 
i.e. in order to reduce the difference between NLO and NNLO predictions for
the total cross section, cf. Ref.~\cite{Garzelli:2015psa}.
In the present study we have used the scale which retains full consistency 
with the PROSA PDF fit.

Interestingly earlier predictions, evaluated at the central scale, even the leading-order ones
presented in Ref.~\cite{Gondolo:1995fq} several years ago, turn out to lie
within the uncertainty band of the PROSA predictions, as shown in Fig.~\ref{fig:fluxcompa}. 
In this plot, besides these old predictions, 
we also show more recent ones obtained by different groups in the last few years. 
Even those predictions which are obtained in approaches quite dif\-fe\-rent from ours,
e.g., on the basis of the so-called dipole model, or those obtained by a recent update of the {\texttt{SYBILL}} Monte-Carlo~\cite{Riehn:2015oba}, 
turn out to be compatible with our prediction,
cf. Fig.~\ref{fig:fluxcompa}.
This is an important result because dipole model predictions have been
extensively used by IceCube as a theory reference for prompt neutrino fluxes,
while the {\texttt{SYBILL}} Monte-Carlo is one of the event generators
extensively used in EAS simulations. 
It includes a perturbative QCD leading-order core and a soft phenomenological
component for the description of hadronic interactions.

\begin{figure}[ht]
\begin{center}
  \includegraphics[width=0.49\textwidth]{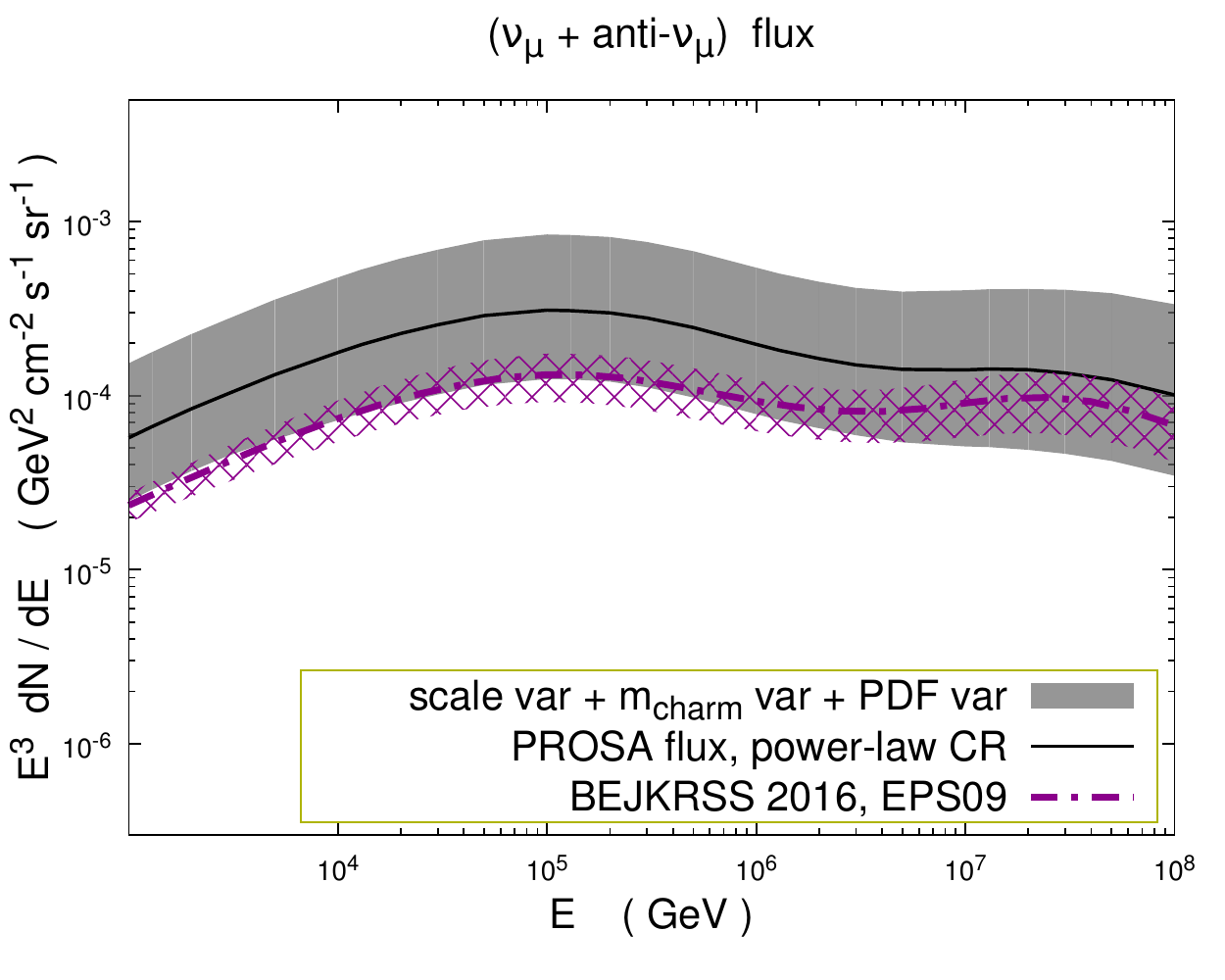}
  \includegraphics[width=0.49\textwidth]{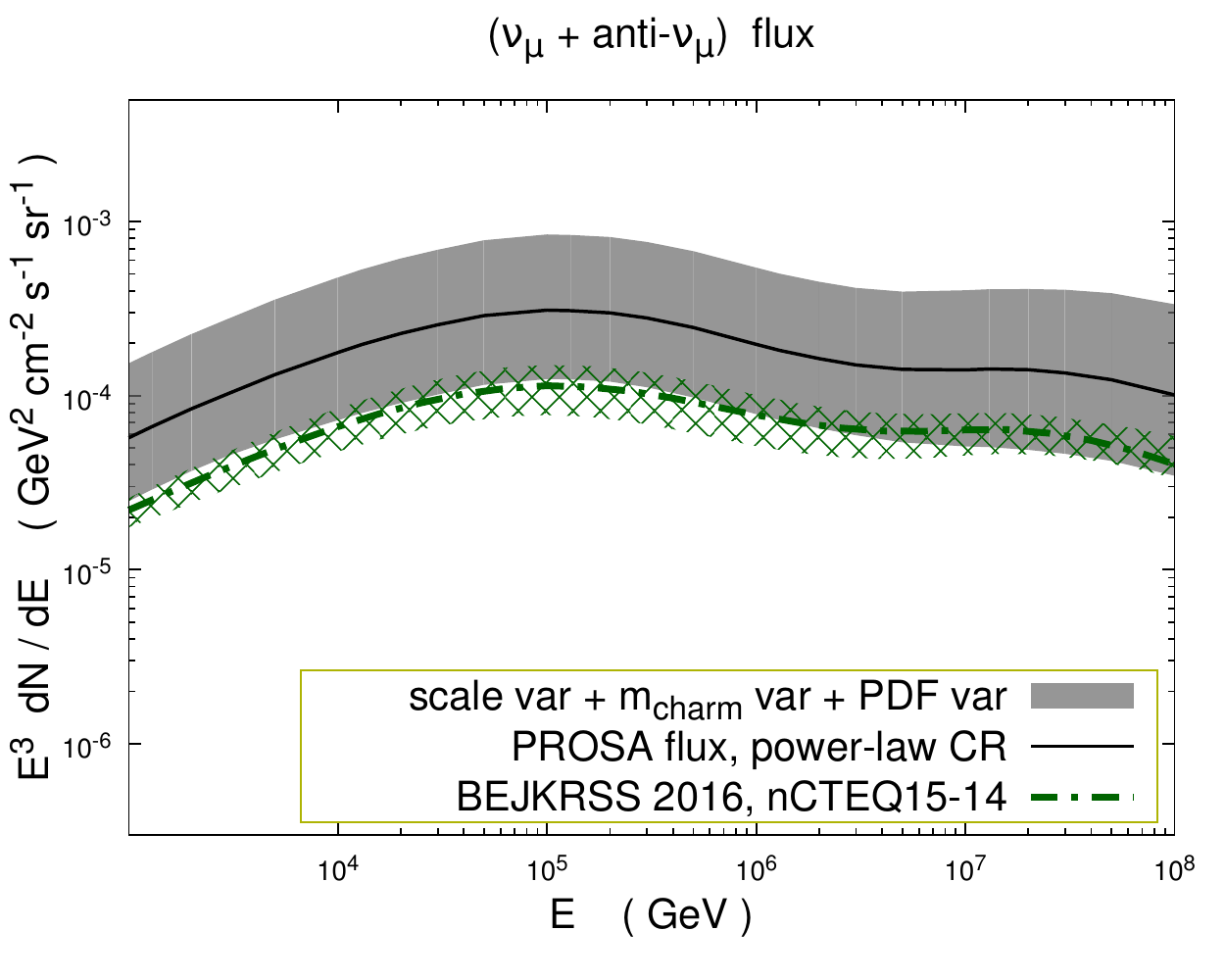}
\end{center}
\vspace*{-5mm}
\caption{\label{fig:nuclear} 
  Comparison of our predictions for ($\nu_\mu + \bar{\nu}_\mu$) fluxes with 
  the PROSA proton PDFs and the superposition approximation,
  to those of Ref.~\cite{Bhattacharya:2016jce}
  using nuclear PDFs, with their respective uncertainty bands. 
  Predictions of Ref.~\cite{Bhattacharya:2016jce} using the EPS09 nuclear PDFs 
  are shown on the left, those based on the nCTEQ15 nuclear PDFs on the right.
  Uncertainties affecting nuclear PDFs are not accounted for in these plots. 
  For consistency with Ref.~\cite{Bhattacharya:2016jce}, and differently from Fig.~\ref{fig:compa_abm_prosa} and~\ref{fig:fluxcompa},
  the cosmic ray primary flux H3p
  is used in all these
  predictions, cf. Fig.~\ref{fig:unctoflux2}.
}  
\end{figure}

Finally, we briefly address the superposition model assumption used so far, 
in which the collision of cosmic rays with air nuclei, mostly nitrogen, 
are approximated by the mere superposition of nucleon interactions, i.e.
the $N\,N$ hard scattering described within perturbative QCD.
Initial work to test the superposition approximation has recently been
presented in Ref.~\cite{Bhattacharya:2016jce}, 
where also two nuclear PDFs were used in the cross section 
computation of cosmic ray interactions with atmospheric nuclei. 
In Fig.~\ref{fig:nuclear} we compare the predictions of Ref.~\cite{Bhattacharya:2016jce} 
obtained with nuclear PDFs EPS09~\cite{Eskola:2009uj} and nCTEQ15~\cite{Kovarik:2015cma} with our own ones.
Fig.~\ref{fig:nuclear} shows that the predictions of Ref.~\cite{Bhattacharya:2016jce} using nuclear PDFs, 
although being systematically lower than ours, are still close to the lower limit of, 
or
%included
stay
within, our total QCD uncertainty band. 
At present, any uncertainties on nuclear PDFs are not included in these plots.
However, these uncertainties are actually quite large in the full $x$ range (i.e. also for $x > 10^{-4}$),  
due to the limited amount of nuclear data that can be used in the nuclear PDF
fits and the limited kinematic coverage of those data, cf. \cite{Eskola:2009uj,Kovarik:2015cma}.
In addition, model assumptions for the description of hadronic interactions involving nuclei are ingredients entering nuclear PDF fits. The uncertainty in this modellization introduces further uncertainties on the resulting PDFs. 
Thus, in order to fully address the superposition approximation,
a systematic experimental and theoretical study of collisions of nuclei at high energies, including nuclear and possibly quark-gluon plasma effects, is required.
This is clearly beyond the scope of the present study and we leave these
improvements and new developments on this issue for the future. 

\section{Uncertainties on prompt neutrino expected events in the IceCube HESE analysis and comparison of the prompt
($\nu_\mu$~+~$\bar{\nu}_\mu$) flux with the present IceCube upper limit}
\label{sec:icecube}

One of the most intriguing results of IceCube has been reported in the HESE analysis.
Over the years this analysis has collected leptonic events also known as contained-vertex events, where the incoming lepton 
has its first interaction inside the active volume of the detector, with a deposited energy in the range from tens of TeV to a few PeV.
In this analysis, 
 an excess with respect to the expected atmospheric background considered so
 far by the IceCube collaboration has been registered
 with increasing statistical evidence along the years.
The experimental data include events with shower and muon track topologies, 
coming from both the Northern and Sou\-thern hemisphere. 
In particular, results were reported for a 662-day analysis with 28 candidates
in the energy range [50 TeV - 2 PeV], corresponding to a 4.1 $\sigma$ excess
in the year 2013~\cite{Aartsen:2013jdh}. 
These results were updated in the year 2014, thanks to the 988-day analysis~\cite{Aartsen:2014gkd}, with 37 events in the energy
range [30~TeV~-~2~PeV] (5.7 $\sigma$ excess over the background), and
featuring also an ``empty'' window corresponding to the [400 TeV - 1 PeV] interval. 
A further update was presented in 2015, in the 1347-day analysis~\cite{Aartsen:2015zva},
which has collected 54 events, corresponding to a $\sim$ 7 $\sigma$
rejection of the atmospheric-only hypothesis,
and with the empty window partially filled and thus
reduced to the [$\sim$600 PeV - 1~PeV] bin. 
In these analyses, the experimental data were fitted considering two possible kinds of sources of neutrinos: the atmosphere and astrophysical ones.  
For the shape of the astrophysical signal a power-law was assumed. 
The best-fit parameter values turned out to change slightly from one
analysis to the other, although they always remained compatible with each other within the quoted uncertainties.
For the atmospheric component, the possibilities of both a conventional and a prompt
contribution were considered. 
In particular, the IceCube collaboration used
the Honda predictions~\cite{Honda:2006qj}, extended to higher energies and modified for taking into account more recent cosmic ray primary spectra with a knee component, as theoretical input for the modellization of conventional neutrino fluxes
in these analyses. A prior was used for the normalization of this contribution in the fit to IceCube experimental data, performed under the assumption that leptons of both astrophysical and atmospheric origin contribute to the total signal seen by IceCube. In the most recent versions of the analysis, part of the atmospheric background was vetoed and subtracted from the signal, by using information on the atmospheric muons detected in coincidence with neutrino events and the techniques of Ref.~\cite{Gaisser:2014bja} in the modellization of this case. As for prompt neutrinos, IceCube has used the ERS 2008 predictions~\cite{Enberg:2008te}, reweighted to a cosmic ray spectrum with a knee component, as a basis for modelling prompt neutrino fluxes. This component was included in the fit as well, and it happened its normalization fits to zero, although with a large uncertainty. As a consequence of this big uncertainty, an upper limit on the total atmospheric neutrino flux, as derived from other IceCube analyses (di\-scus\-sed in the second part of this Section), was adapted as well to the HESE analyses, in order to show how big a potential (prompt + conventional) component could look like.  

\begin{figure}[hb]
\begin{center}
  \includegraphics[width=0.75\textwidth]{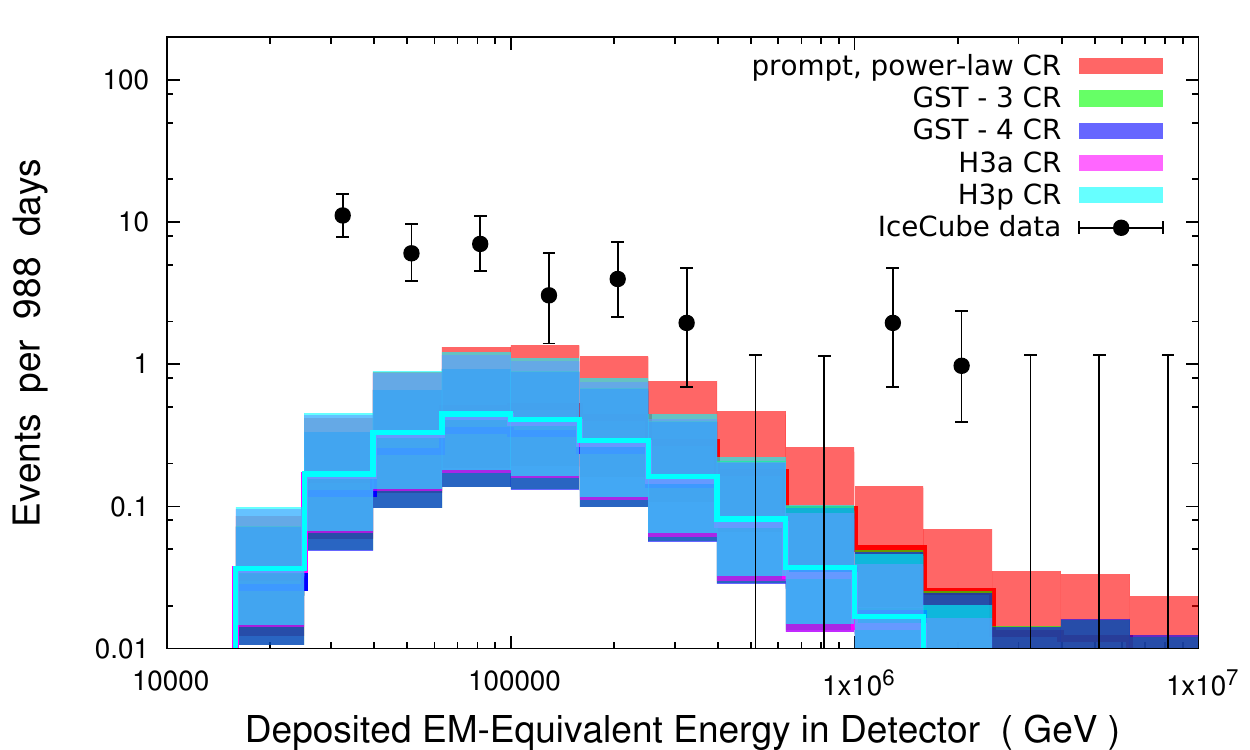}
\end{center}
\vspace*{-5mm}
\caption{\label{fig:hesecrfluxvar} 
  Predictions for the number of prompt neutrino events as a
  function of the deposited energy in the detector for the IceCube 988-day
  HESE analysis. 
  Each colored band refers to a different cosmic ray primary flux, 
  and accounts for uncertainties due to scale, charm mass and PDF variations.
  IceCube experimental data on the total number of leptons detected are also shown.
  The latter include leptons of both atmospheric and astrophysical origin. 
} 
\end{figure}

\begin{figure}[ht]
\begin{center}
  \includegraphics[width=0.75\textwidth]{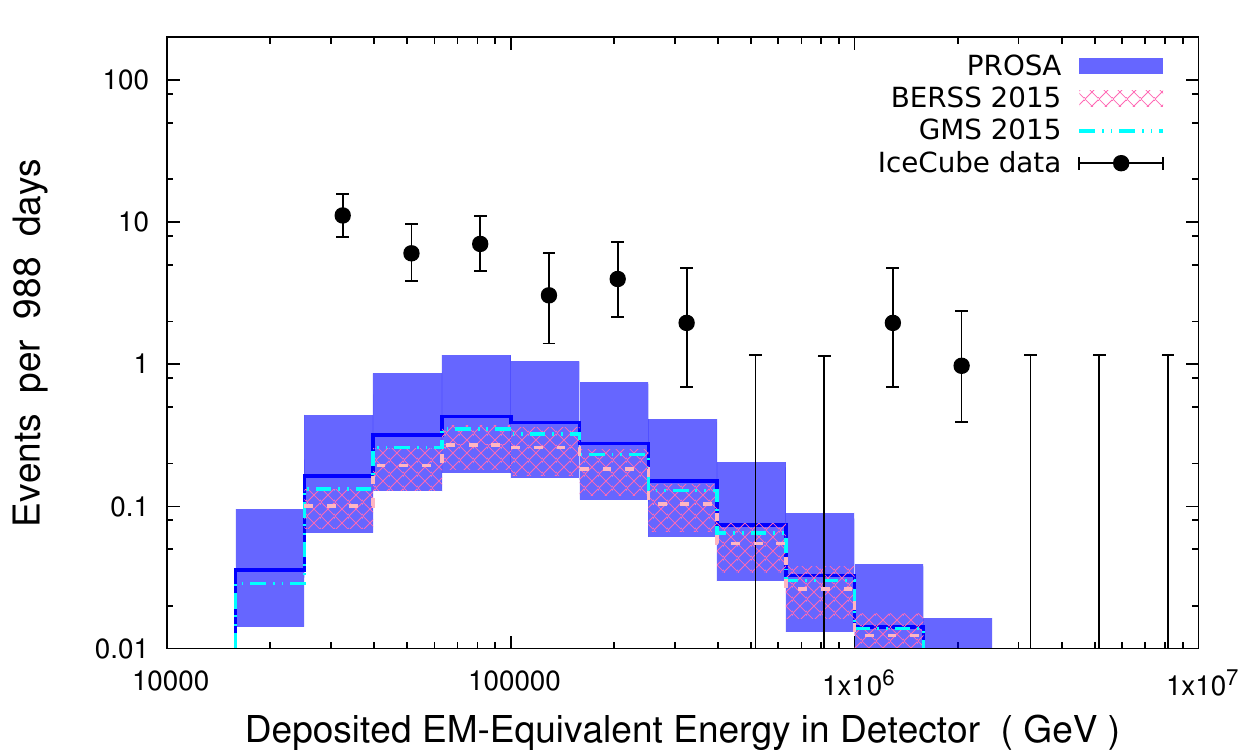}
\end{center}
\vspace*{-5mm}
\caption{\label{fig:heseprompt} 
  Predictions for the number of prompt neutrino events as a function of the deposited energy in
  the detector for the IceCube 988-day HESE analysis. The H3a cosmic ray primary flux
  is used as input. 
  Results from the PROSA calculation of this paper, shown in
  blue, with their QCD uncertainty band, are compared with the central
  predictions of GMS 2015~\cite{Garzelli:2015psa} and with the BERSS one~\cite{Bhattacharya:2015jpa}, 
  including the uncertainty band of the latter. 
  IceCube experimental data on the total number of leptons detected are also shown.
  These data include leptons of both atmospheric and astrophysical origin. 
}   
\end{figure}

\begin{figure}[ht]
\begin{center}
  \includegraphics[width=0.75\textwidth]{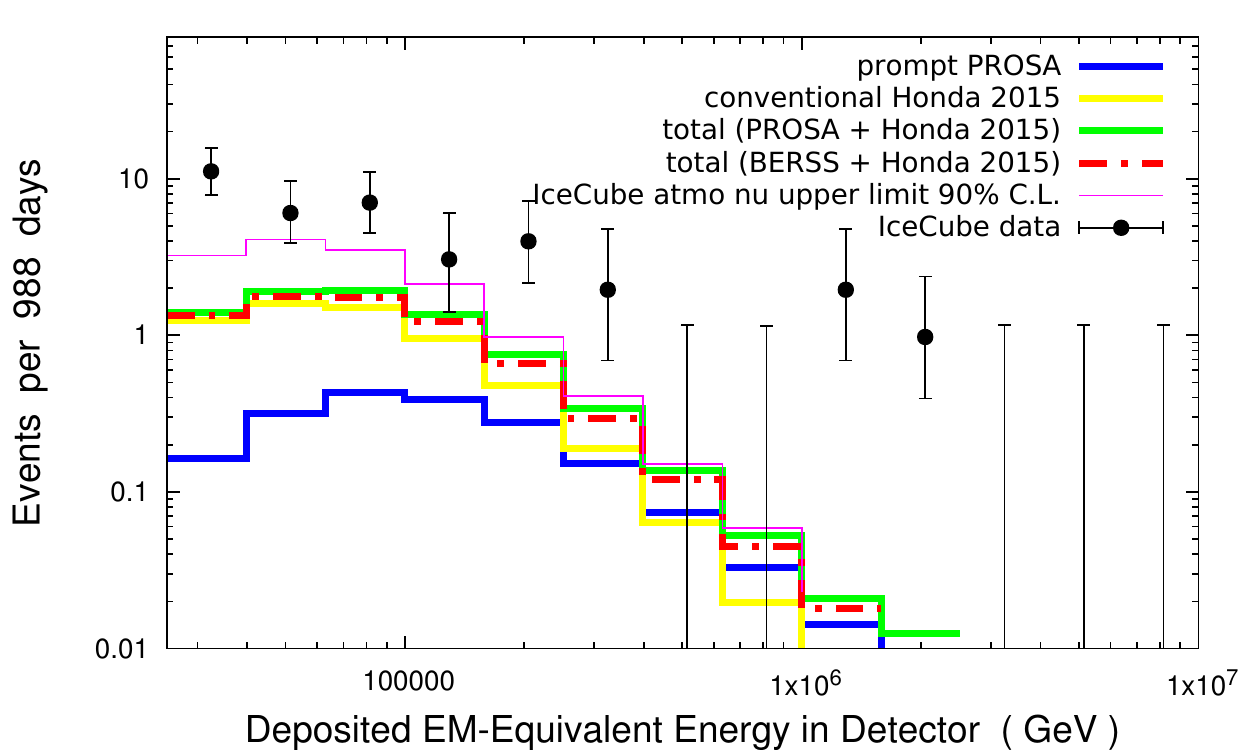}
  \includegraphics[width=0.75\textwidth]{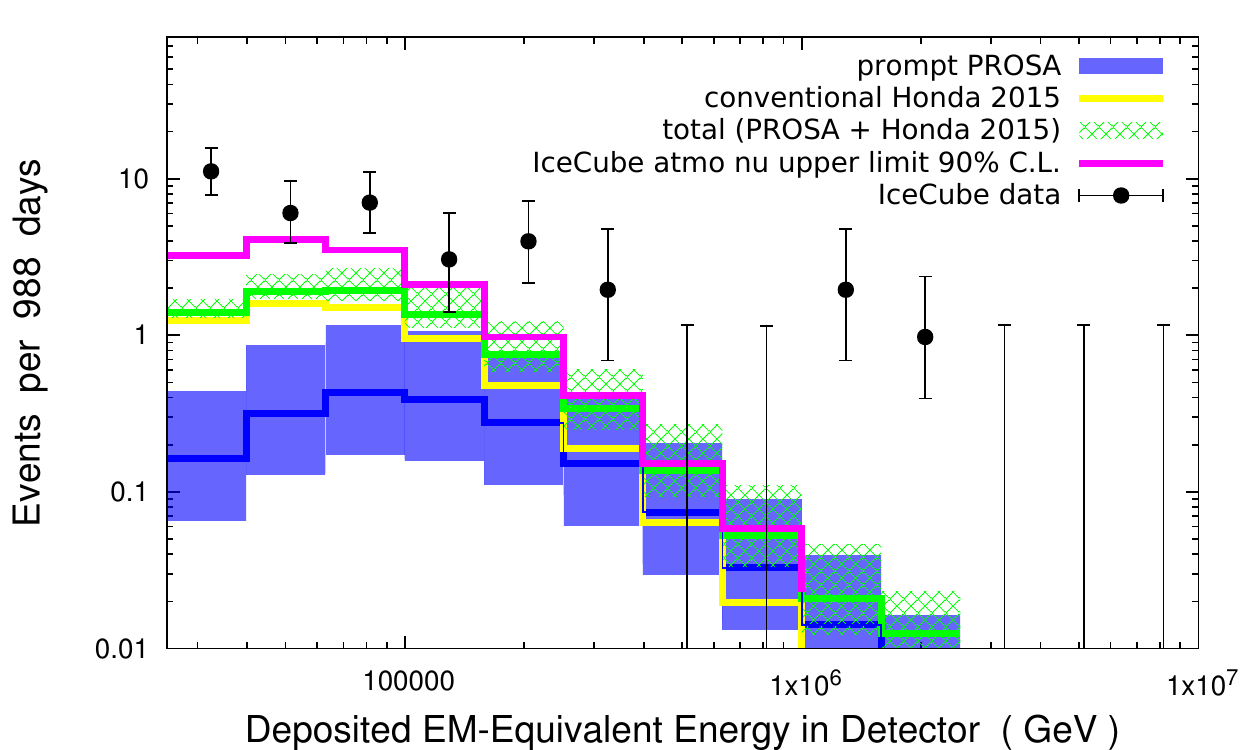}
\end{center}
\vspace*{-5mm}
\caption{\label{fig:hesetotal} 
  Predictions for the number of prompt, conventional and total
  expected atmospheric neutrino events for the IceCube 988-day HESE analysis,
  as compared to the IceCube lepton data.
The H3a cosmic ray primary flux
  is used as input. 
  On the top the 
  central predictions are shown and also compared to the total central
  predictions by BERSS~\cite{Bhattacharya:2015jpa}. 
  On the bottom the effect of uncertainties 
  on prompt neutrino contribution is propagated to an uncertainty on the total atmospheric neutrino contribution under the assumption that the conventional component does not contribute   any additional uncertainty. 
  IceCube experimental data are shown in black, and 
  the IceCube 90\% confidence level upper limit on the total atmospheric neutrino flux obtained in Ref.~\cite{Aartsen:2013eka}  and reproduced in Ref.~\cite{Aartsen:2014gkd} is shown in magenta. }
\end{figure}

\begin{figure}[ht]
\begin{center}
  \includegraphics[width=0.75\textwidth]{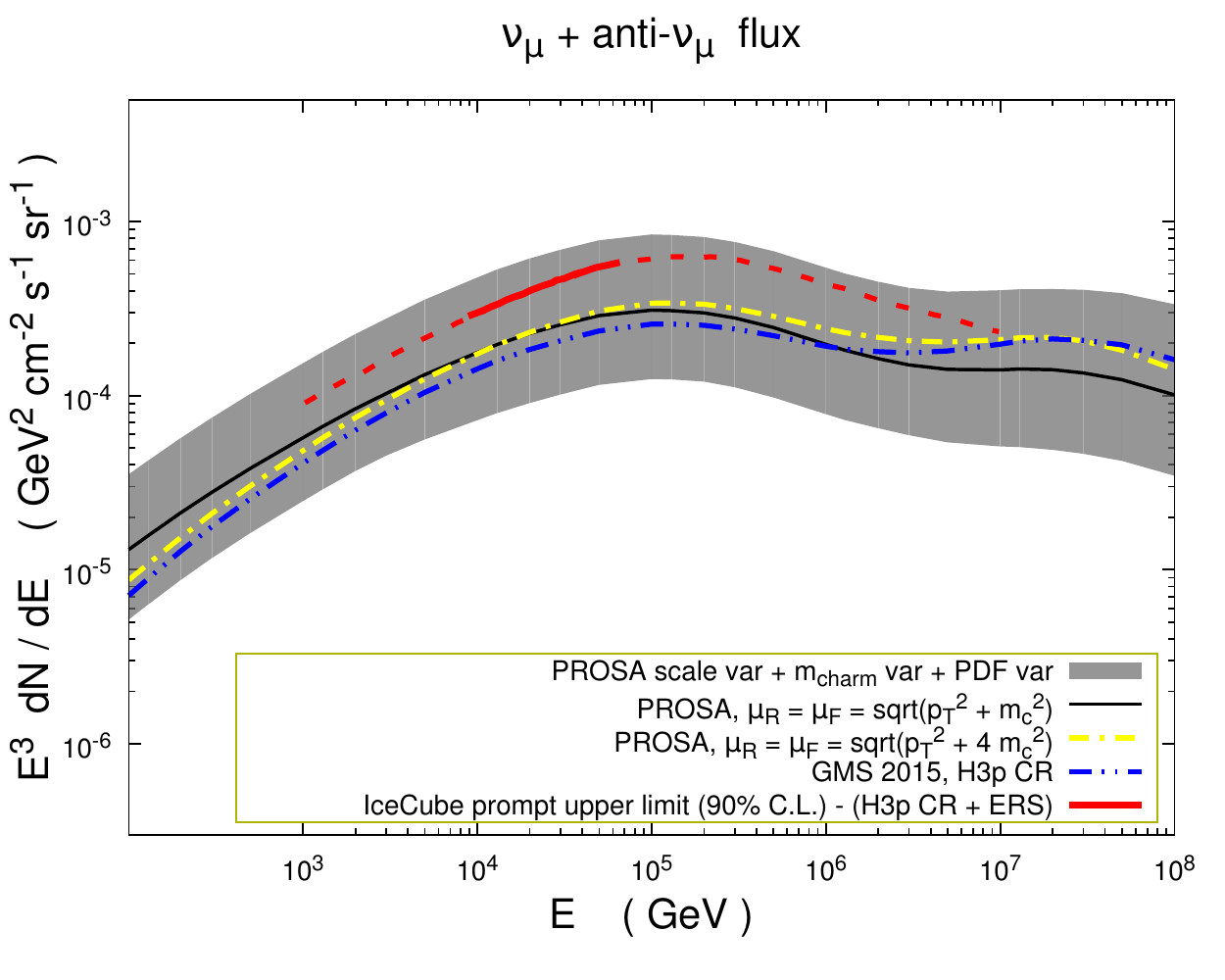}
\end{center}
\vspace*{-5mm}
\caption{\label{fig:upperlimit} 
  Comparison of the prompt ($\nu_\mu + \bar{\nu}_\mu$) flux using the PROSA PDFs with the present upper limit on prompt neutrino flux at 90\% confidence level obtained by the IceCube experiment~\cite{Aartsen:2016xlq} (solid red line) and its extrapolation (dotted red line), which adopted the ERS model~\cite{Enberg:2008te} as a basis for modelling prompt neutrinos. Central predictions using the scale $\mu_R = \mu_F = \sqrt{p_T^2 + 4 m_c^2}$ and PROSA PDFs and ABM PDFs (GMS 2015) are also shown. The limit and all predictions refer to the H3p CR flux.} 
\end{figure}

Given the model dependence understood in the description of atmospheric neutrinos by IceCube, it is instructive to investigate the effect of predictions for prompt neutrino fluxes considered in this study, on the expected number of events in the IceCube analyses. 
In particular, we consider the case of the 988-day analysis,
discussed previously also in Ref.~\cite{Bhattacharya:2015jpa}. We note that even considering the most recent IceCube 1347-day analysis gives rise to qualitatively similar results and does not alter our discussion and conclusions. 
The number of prompt neutrino events in the different energy bins reported by
IceCube, computed on the basis of PROSA predictions for prompt neutrino
fluxes, is shown in Fig.~\ref{fig:hesecrfluxvar} for the various cosmic ray spectra already presented in Section~\ref{sec:fluxes}. 
It is evident that modern CR spectra give a suppressed number of events with respect to the broken power-law spectrum (red band). 
Furthermore, for those events seen by IceCube so far 
with the maximum e\-ner\-gies deposited, limited to $\sim$ 2~PeV, the difference
between the variants of Gaisser CR spectra considered in this work 
do not introduce any dramatic changes in the number of observed events. 
The bands reported in the plot refer to QCD theoretical uncertainties stemming
from the combination of scale, charm mass and PDF uncertainties, computed as described
previously in Section~\ref{sec:fluxes}. 
Even considering these uncertainty bands, our
theoretical predictions for prompt fluxes turn out to lie below the data,
thus confirming a different origin for those most energetic IceCube events,
also shown in the plot.

A comparison between our predictions for prompt neutrino events, and
predictions computed by using different models is shown in Fig.~\ref{fig:heseprompt}, 
considering as a basis for cosmic ray primary flux the H3a model. 
In particular, predictions computed by using the GMS 2015
central flux~\cite{Garzelli:2015psa} and those reported in Ref.~\cite{Bhattacharya:2015jpa} are shown. 
These earlier predictions are compatible
with the PROSA predictions. 
Furthermore, it turns out that the uncertainty band of Ref.~\cite{Bhattacharya:2015jpa} 
is completely contained,
in all the event bins, inside the PROSA uncertainty band.
The latter is much larger because it includes a broader range of scale variations (with $\mu_R$ $\ne$ $\mu_F$) 
and also the PDF uncertainties, not considered in~\cite{Bhattacharya:2015jpa}.  

The total atmospheric flux, which can be computed by summing the conventional
and the prompt flux, is plotted in Fig.~\ref{fig:hesetotal}. 
As for the conventional flux, the Honda spectrum~\cite{Honda:2006qj} was adopted, extended to the highest energies and
reweighted to the H3a cosmic ray flux as described in Ref.~\cite{Bhattacharya:2015jpa}. 
Uncertainties on the total flux, due to uncertainties on the prompt component, are shown,
under the assumption that the conventional flux does not contribute any 
additional uncertainty~\footnote{Actually, deeply comprehensive studies
  on the uncertainties on conventional neutrino fluxes are still missing,
  especially in the high-energy region explored by VLV$\nu$Ts. 
  Thus, instead of not quoting any uncertainty, it could be more reasonable to take a more conservative attitude, and consider the possibility that this flux has an uncertainty around at least 20-30\%.}.
The IceCube upper limit on the total neutrino flux at 90\% confidence level~\cite{Aartsen:2013eka} is also shown. At the highest energies, the IceCube upper limit lies well inside our uncertainty band.
The latter limit corresponds to the result of 
a separate analysis of muonic events from the Northern hemisphere, characterized by neutrino interaction vertices which can lie both inside and outside the instrumented volume, with the Earth acting as an efficient shielding for atmospheric muons. IceCube has progressively updated this analysis in Ref.~\cite{Aartsen:2013eka, Aartsen:2015rwa, Aartsen:2016xlq}. As for the prompt neutrino component, the experimental results of this kind of analysis are presented in the form of upper limits on the prompt ($\nu_\mu$ + $\bar{\nu}_\mu$) flux~\footnote{IceCube HESE samples include events initiated by neutrinos of all possible flavours, $\nu_e$, $\nu_\mu$ and $\nu_\tau$. This has been indeed taken into account by reasonable assumptions on the ratio $\nu_e$ : $\nu_\mu$ : $\nu_\tau$ when adapting to the HESE analysis the upper limit on atmospheric neutrinos derived from the analysis of the ($\nu_\mu$ + $\bar{\nu}_\mu$) events from the Northern hemisphere.}. The comparison of the most recent IceCube estimate of this limit~\cite{Aartsen:2016xlq} with the fluxes computed in the present study and the central prediction of GMS 2015 is shown in Fig.~\ref{fig:upperlimit}.
Again, the published IceCube upper limit, although being larger than our central prompt predictions, is well inside the uncertainty band of 
the prompt ($\nu_\mu + \bar{\nu}_\mu$) flux, both over the limited range of neutrino energies $E_{lab}$ probed by present IceCube data [$8 \cdot 10^3$ -  $8 \cdot 10^4$] GeV and even when considering the extrapolation to a larger energy range [$10^3$ - $10^7$] GeV, also presented by the IceCube collaboration. This might point to a need of revising the model assumptions in the aforementioned IceCube analyses. 

In summary, Figs.~\ref{fig:hesetotal} and~\ref{fig:upperlimit} show the future potential of astrophysical measurements at VLV$\nu$Ts, especially when a higher
statistical accuracy will be reached, in complementing accelerator-based measurements by putting constraints and pro\-vi\-ding complementary information on the physics related to charm hadroproduction. 

\section{Conclusions}
\label{sec:conclu}
We have used the PROSA PDF fit to provide predictions for the flux of prompt neutrinos in the atmosphere. 
The PROSA PDFs are the first fit to include LHCb open charm and beauty data 
in order to constrain the gluon distribution inside the proton 
in regions of Bjorken-$x$ not previously covered by any other experiment.
We have shown that present PDF uncertainties on the prompt neutrino flux
increase with increasing neutrino energies. 
Moreover, in the region of interest for present day neutrino telescopes, 
which have so far detected neutrinos up to a few PeV, 
PDF uncertainties are already quite well constrained 
and are subdominant with respect to the dominating QCD uncertainties 
related to the renormalization and factorization scale variation. 
Our flux turns out to be compatible
with several previous computations of prompt fluxes, 
obtained in a variety of approaches, 
which may or may not involve a perturbative QCD description of the hard scattering. 

As a practical application, we have studied the uncertainties on the number of expected prompt neutrino events in the IceCube HESE analysis and we have compared the theoretical predictions on ($\nu_\mu$ + $\bar{\nu}_\mu$) fluxes with the IceCube upper limit published in complementary analyses, including also non-contained events.
We have found that the adoption of different assumptions for the composition of the cosmic ray primary flux has a small effect on the shape of the distribution of prompt neutrino events in the HESE analysis, at least when considering the energy range tested so far by IceCube, and that the high energy tail of the atmospheric neutrino flux has a steeper slope than the slope of IceCube events, even when including in the analysis prompt neutrino uncertainties of QCD origin. This confirms that, to explain IceCube HESE events, is necessary to add the existence of at least one additional neutrino component of non-atmospheric origin. Furthermore, we have found that the upper limit on prompt neutrino fluxes at 90\% confidence level, published by IceCube, although being model dependent, is just slightly above the central predictions for the prompt ($\nu_\mu + \bar{\nu}_\mu$) flux obtained in this study, but well inside our global QCD uncertainty band. 
This holds over the entire range of relevant neutrino energies and challenges the model assumptions on atmospheric fluxes at high-energies adopted in the IceCube analyses.

In summary, this paper has presented the first application of the PROSA PDFs to an
astrophysical problem.
This opens up the possible use of these PDFs in many other problems arising 
in the description of microscopic interactions at low Bjorken-$x$ 
at present and future high-energy $pp$ and $ep$ colliders, and in cosmic ray hadronic interactions occurring in their astrophysical sources, during propagation and in the atmosphere.

Possible extensions of the PROSA fit towards lower $x$ values using latest LHCb data on charm hadroproduction for collisions at $\sqrt{s} = 13$~TeV~\cite{Aaij:2015bpa}, left to a separate paper, could check the extrapolation of the present PROSA fit and further reduce the PDF uncertainties at low Bjorken-$x$. On the other hand, incorporating into the fit recently appeared LHCb charm data at $\sqrt{s} = 5$~TeV~\cite{Aaij:2016jht}, could provide insights on the self-consistency of the data and of their theoretical description. Likewise, the inclusion of nuclear effects in the theory description of the hadroproduction of $D$-mesons in collisions of ultra-high-energy cosmic ray nuclei with the nuclei of our atmosphere will be left for future work.
\\
\\
Our predictions for prompt neutrino fluxes with PROSA PDFs are publicly available at \texttt{https://prosa.desy.de}.

\subsection*{Acknowledgments}
We would like to thank S.~Alekhin, P.~Nadolsky, K.~Rabbertz and V.~Radescu for discussions. M.V.G. is additionally grateful to C.~Kopper, J.~Van~Santen, A.~Fedynitch, C.~James and A.~Margiotta for clarifications on data analysis at VLV$\nu$Ts and to G.~Kramer, B. Kniehl and M.~Benzke for suggestions and useful observations on the theoretical description of charmed meson hadroproduction.  
This work has been supported by the Deutsche Forschungsgemeinschaft in Sonderforschungsbereich through the Collaborative Research Center SFB 676 ``Particles, Strings and the Early Universe'', and by the Helmholtz Alliance for Astroparticle Physics (HAP) funded by the Initiative and Networking Fund of the Helmholtz Association.

%\appendix{Appendix A}
%
%
%
\section*{Appendix A: Comparisons of theoretical predictions with LHCb experimental data at $\sqrt{s}$ = 7 and 13 TeV}
\label{sec:appA}

\begin{figure}
\begin{center}
  \includegraphics[width=0.49\textwidth]{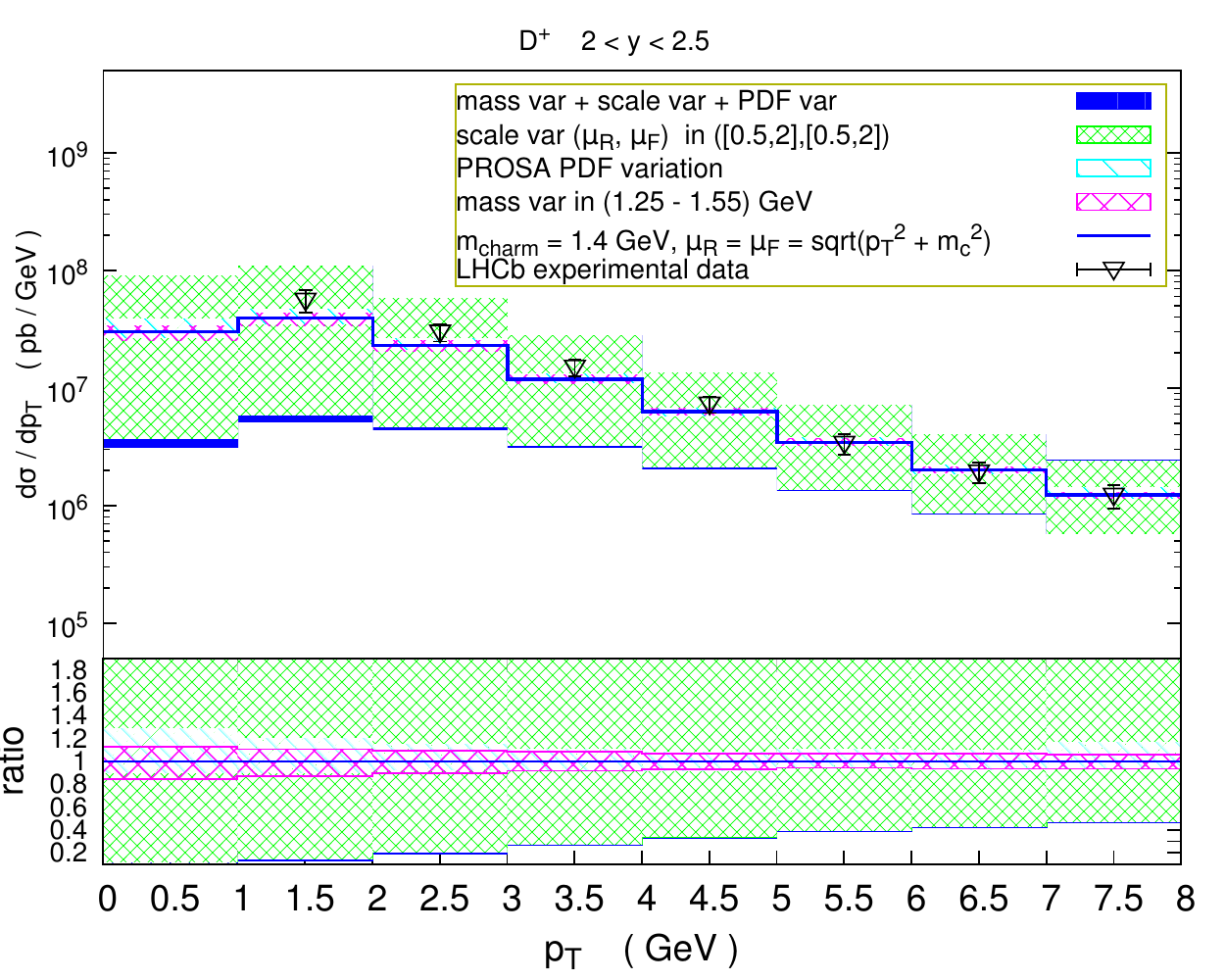}
  \includegraphics[width=0.49\textwidth]{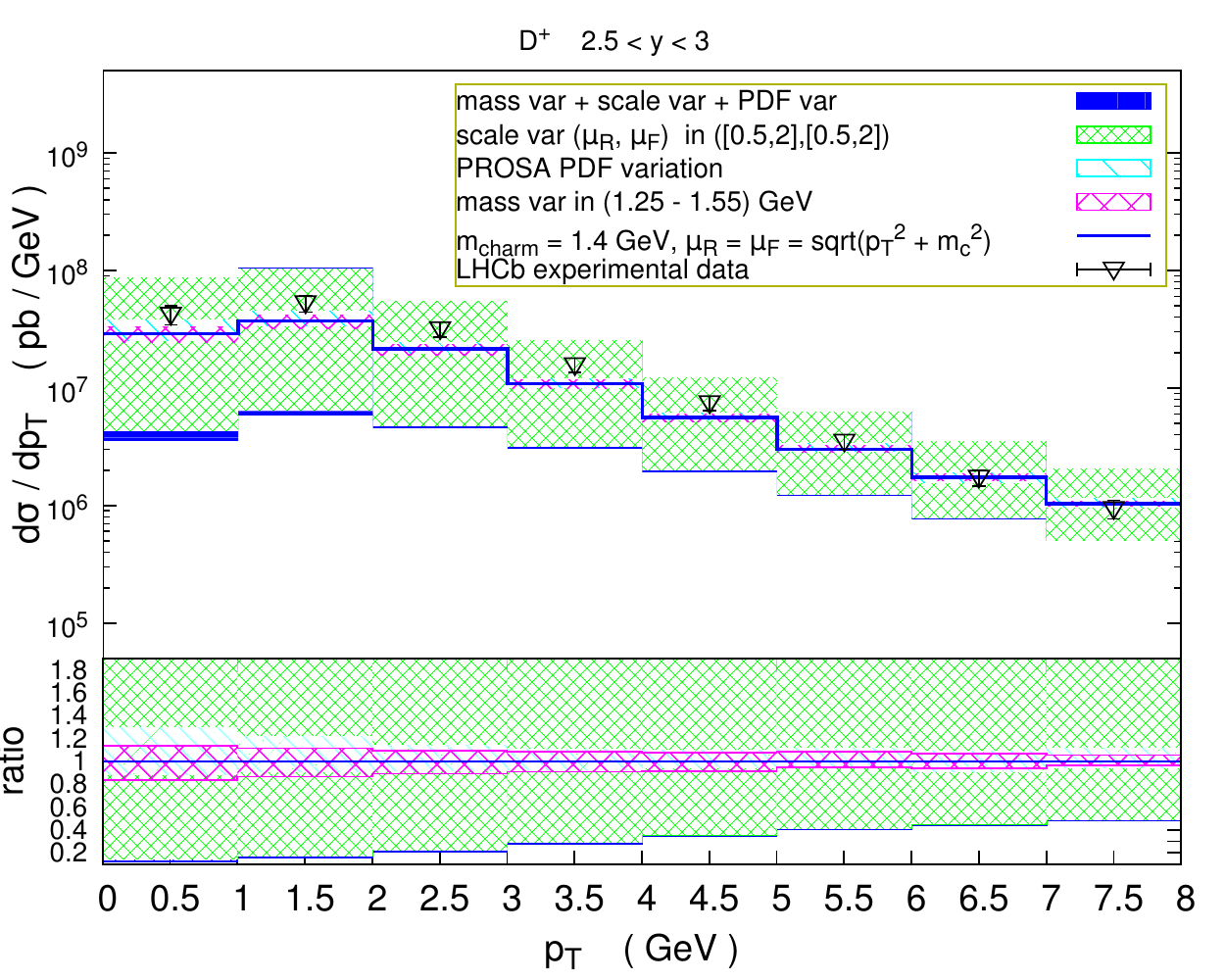}
  \includegraphics[width=0.49\textwidth]{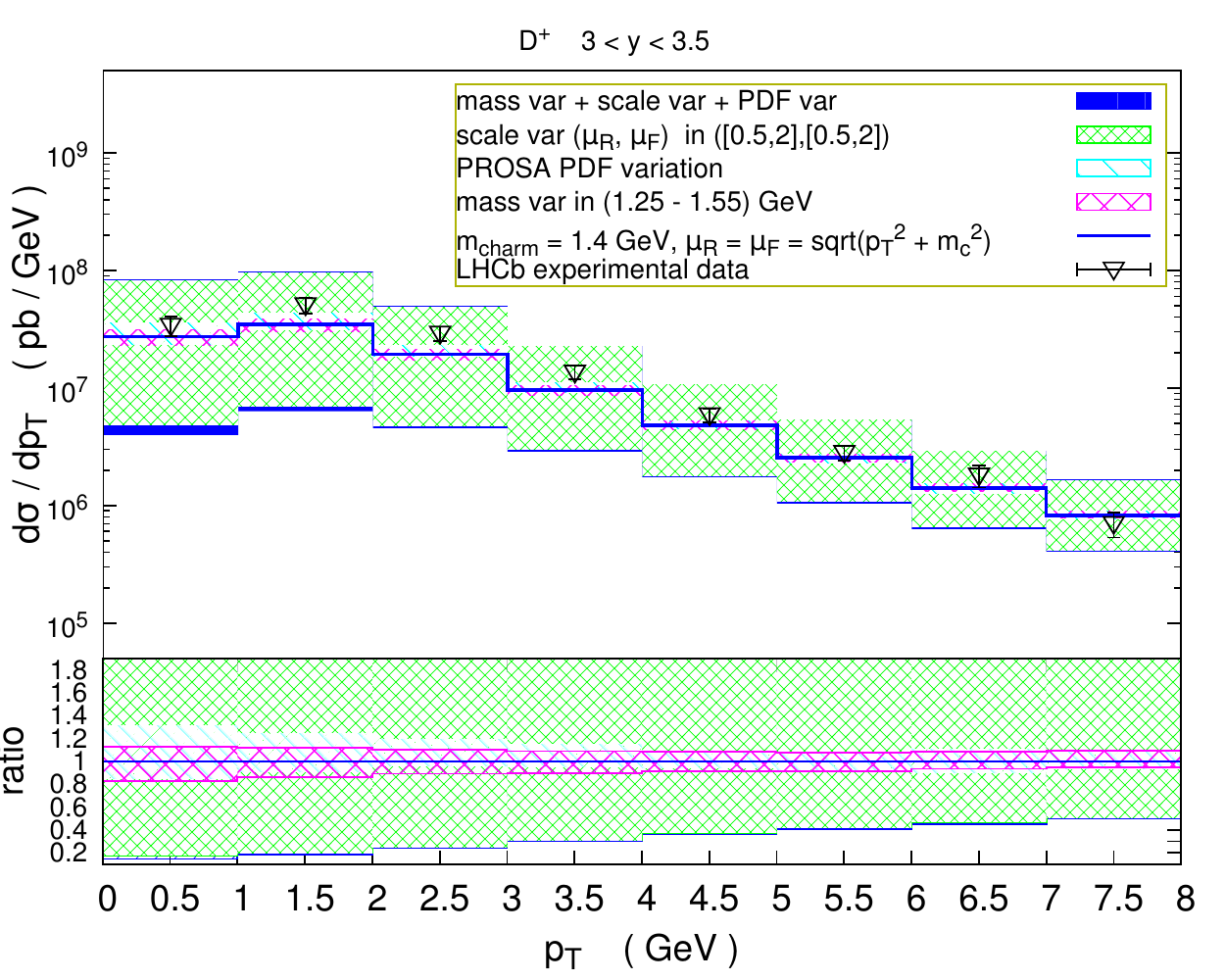}
  \includegraphics[width=0.49\textwidth]{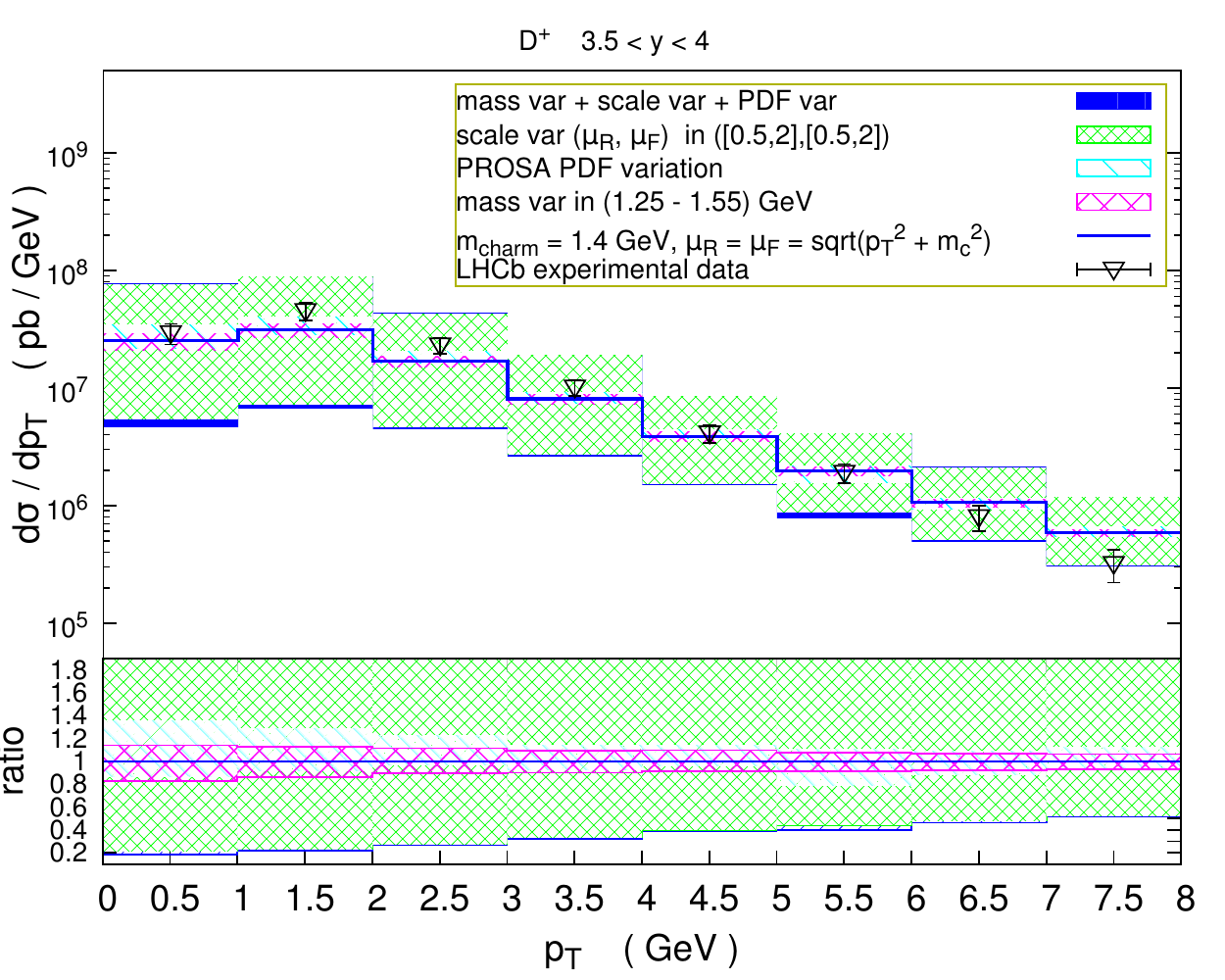}
  \includegraphics[width=0.49\textwidth]{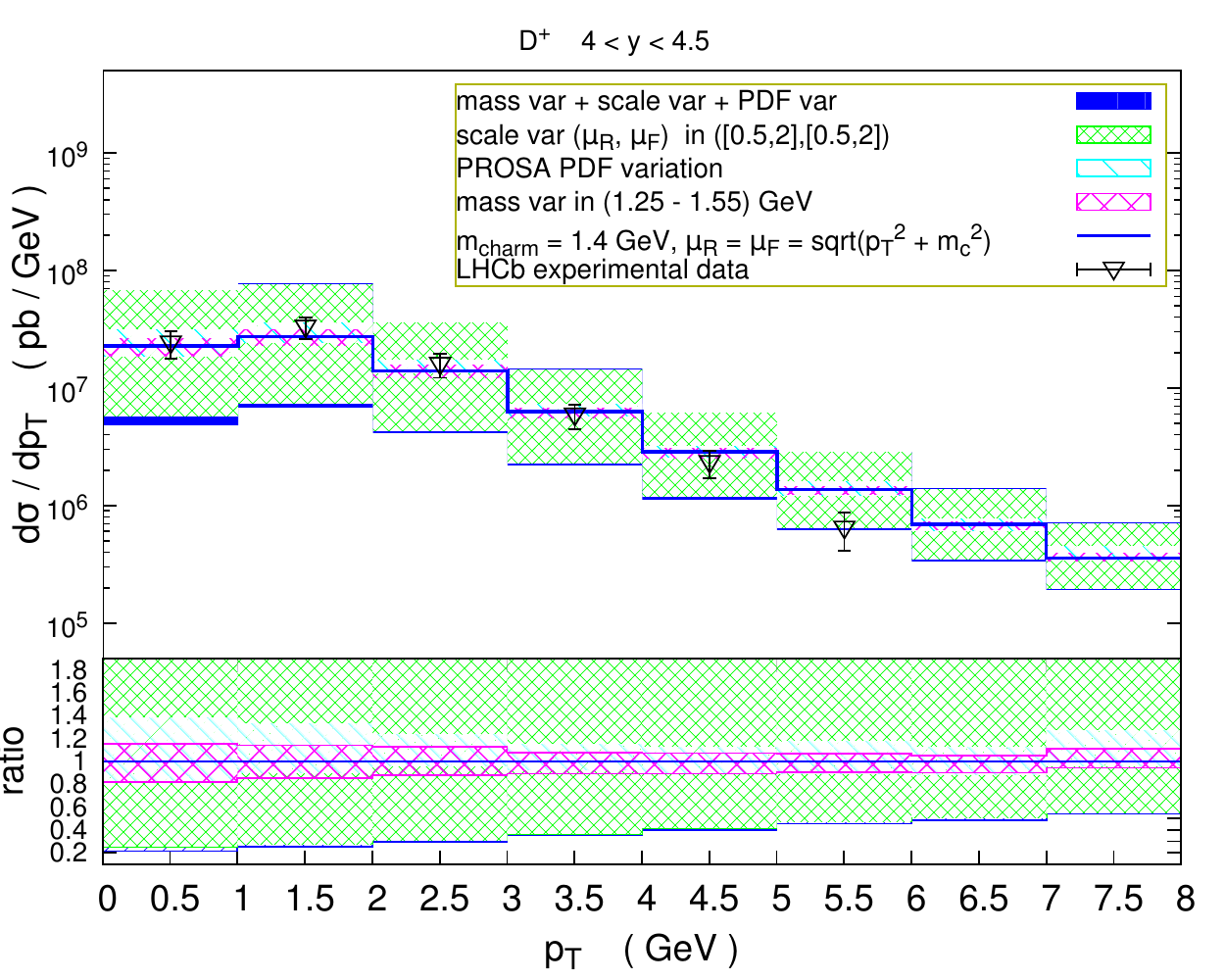}
\end{center}
\vspace*{-5mm}
\caption{\label{fig:dp} Transverse momentum distribution of $D^\pm$ mesons in $pp$ collisions at $\sqrt{s}~=~7$~TeV. \texttt{POWHEGBOX + PYTHIA 6} predictions for $\mu_0=\sqrt{p_{T,\, c}^2 + m_c^2}$, $m_c = 1.4$~GeV and with the PROSA PDFs, are compared to LHCb experimental data~\cite{Aaij:2013mga} in different rapidity bins.
  Theoretical predictions are accompanied by their uncertainty bands, 
  due to $\mu_R$ and $\mu_F$ scale variation (green), to $m_c$ (magenta) and to PROSA  PDF (light-blue hatched) variation, as described in the text. The LHCb experimental data~\cite{Aaij:2013mga}
  are shown together with their statistical and systematic uncertainties, added in quadrature. In the lower panel ratios of the uncertainties with respect to the theoretical central predictions are shown.}
\end{figure}

\begin{figure}
\begin{center}
\includegraphics[width=0.49\textwidth]{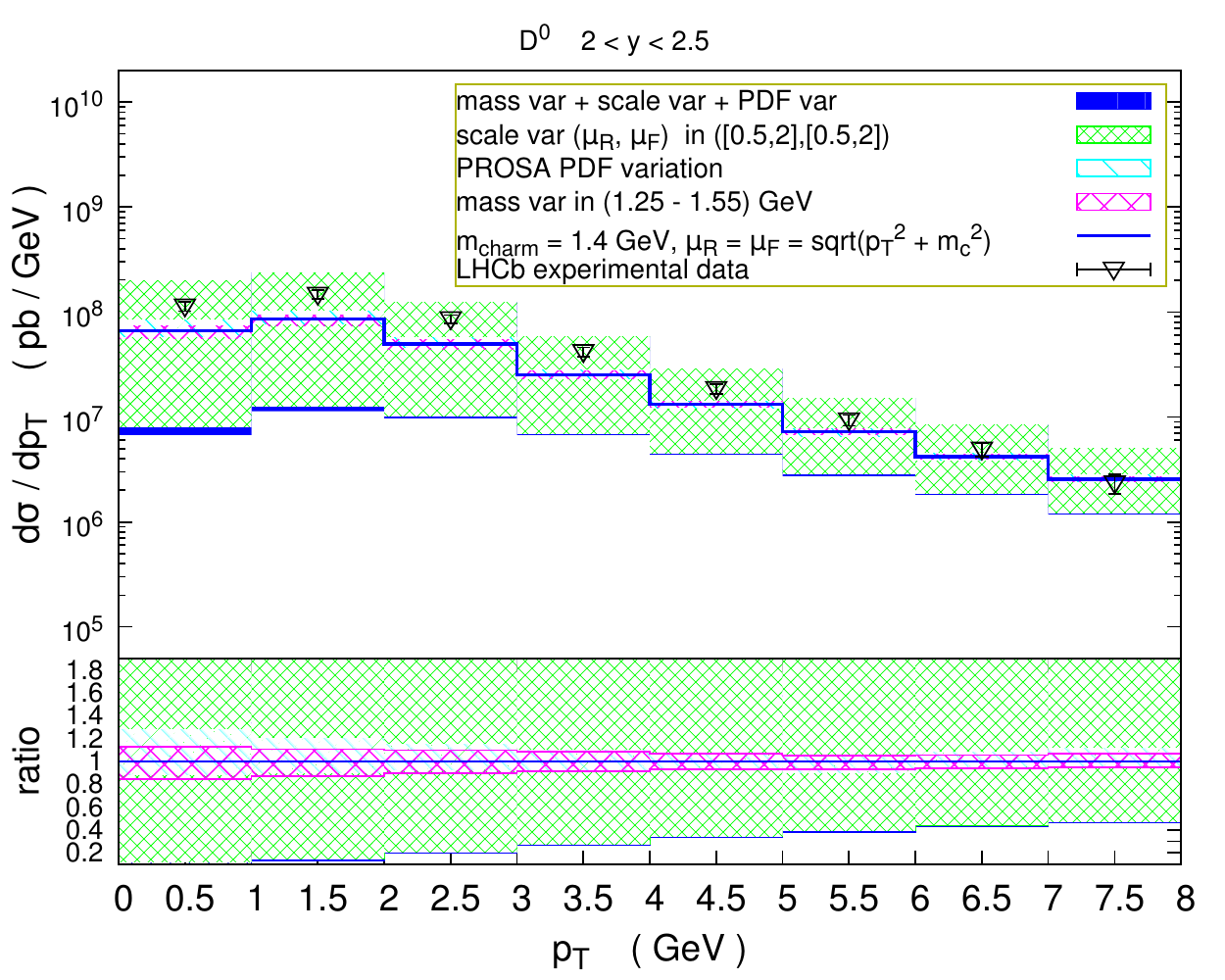}
\includegraphics[width=0.49\textwidth]{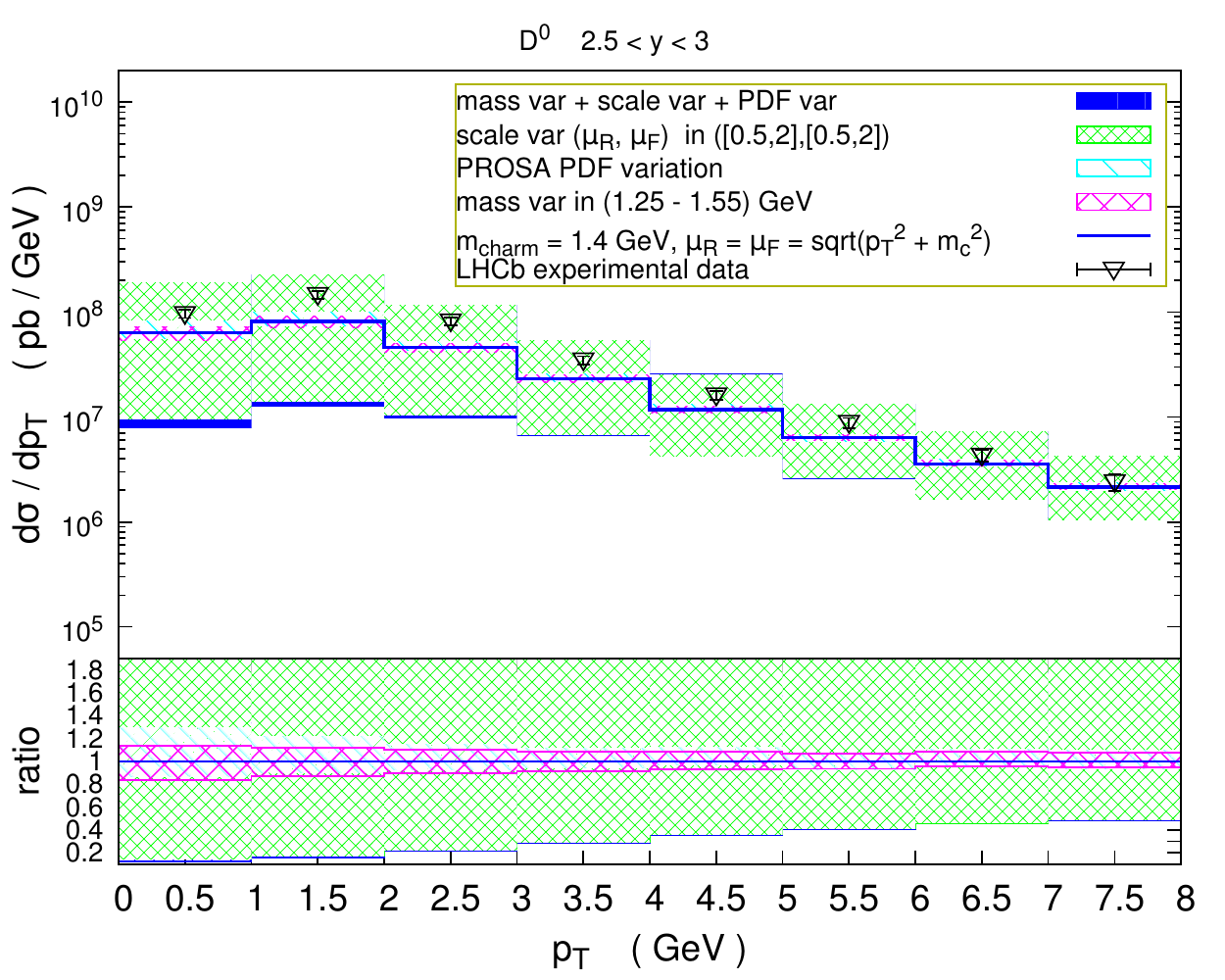}
\includegraphics[width=0.49\textwidth]{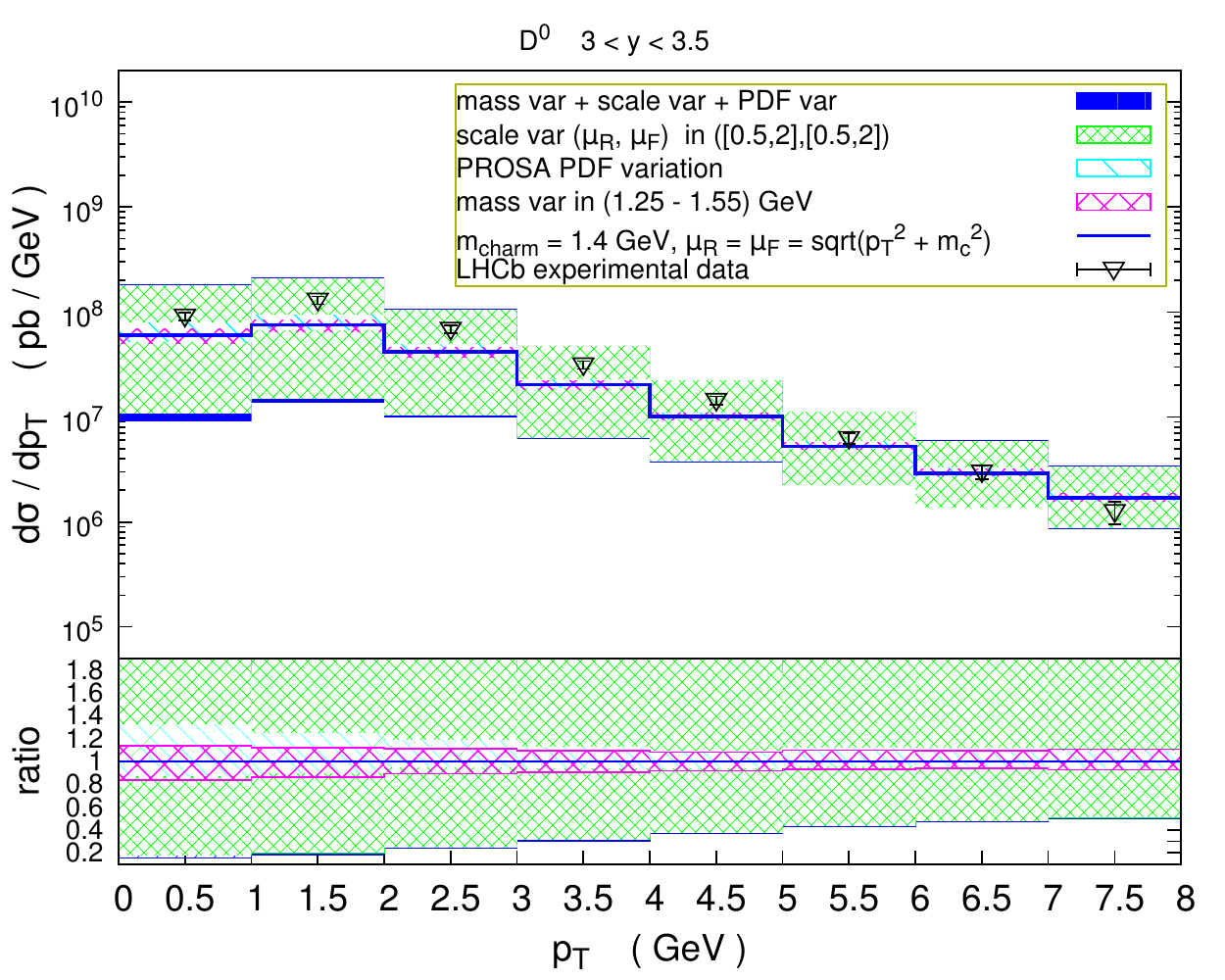}
\includegraphics[width=0.49\textwidth]{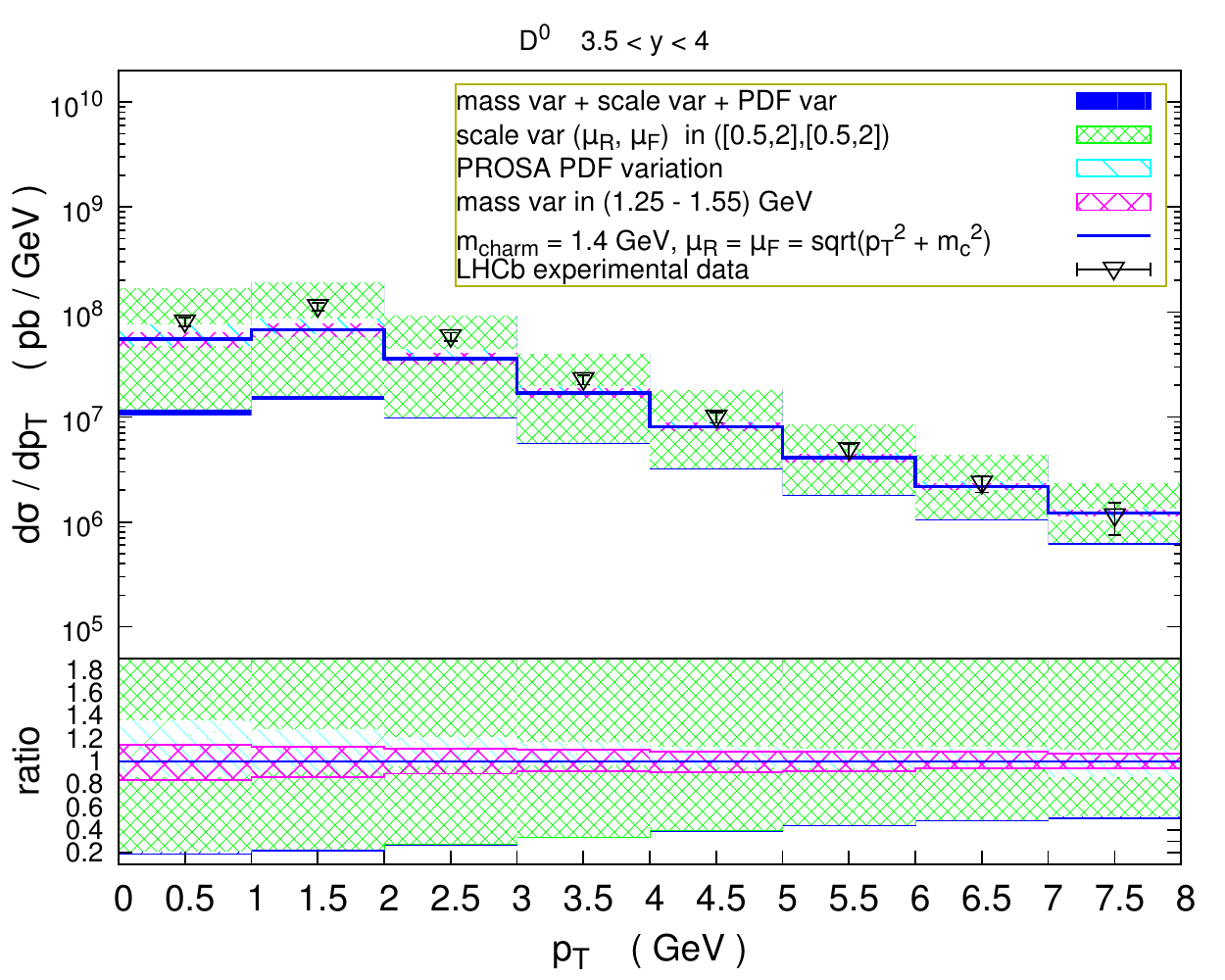}
\includegraphics[width=0.49\textwidth]{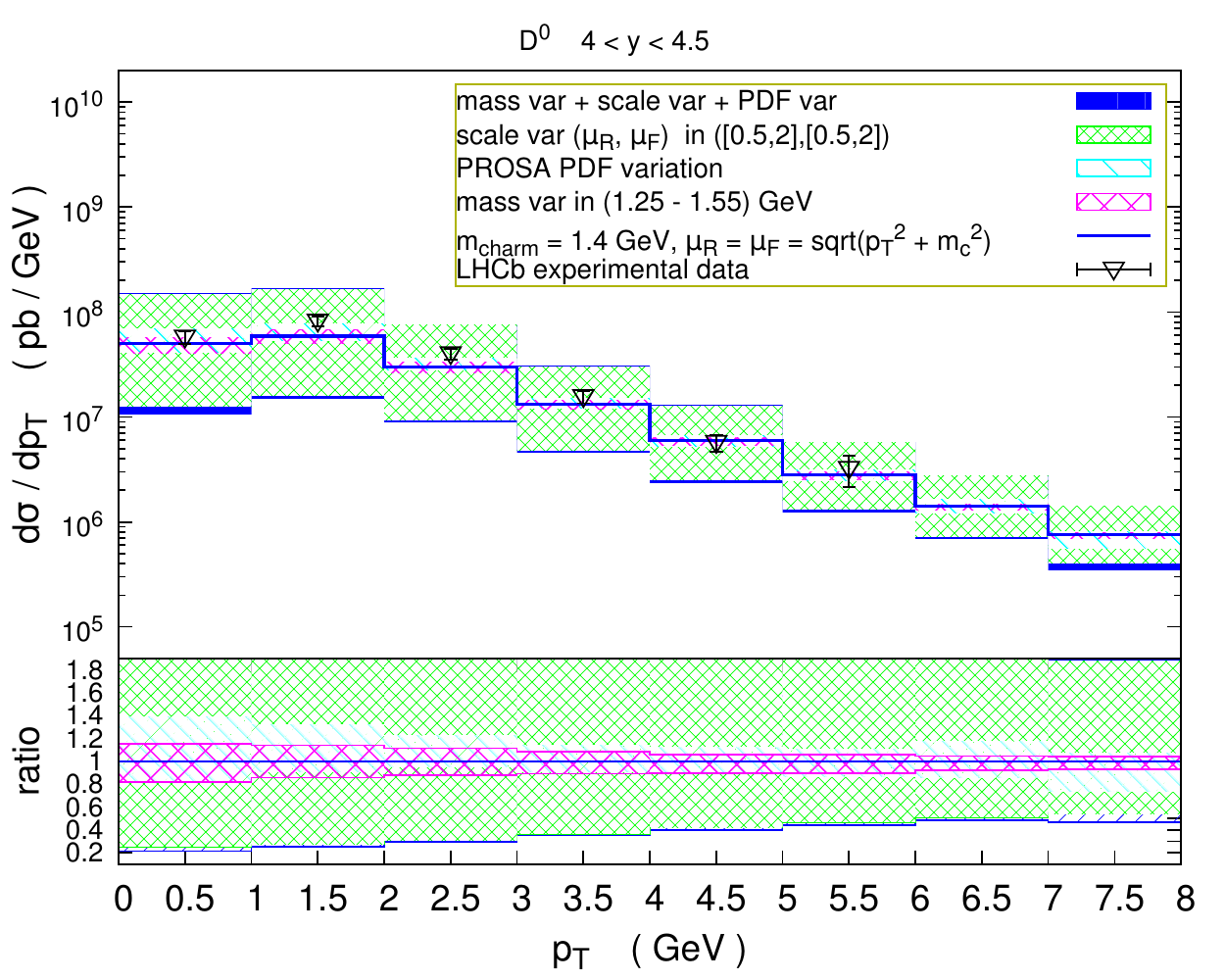}
\end{center}
\vspace*{-5mm}
\caption{\label{fig:d0} 
  Same as Fig.~\ref{fig:dp}, for 
  ($D^0$ + $\bar{D^0}$) hadroproduction at $\sqrt{s}$ = 7 TeV.}
\end{figure}

\begin{figure}
\begin{center}
  \includegraphics[width=0.49\textwidth]{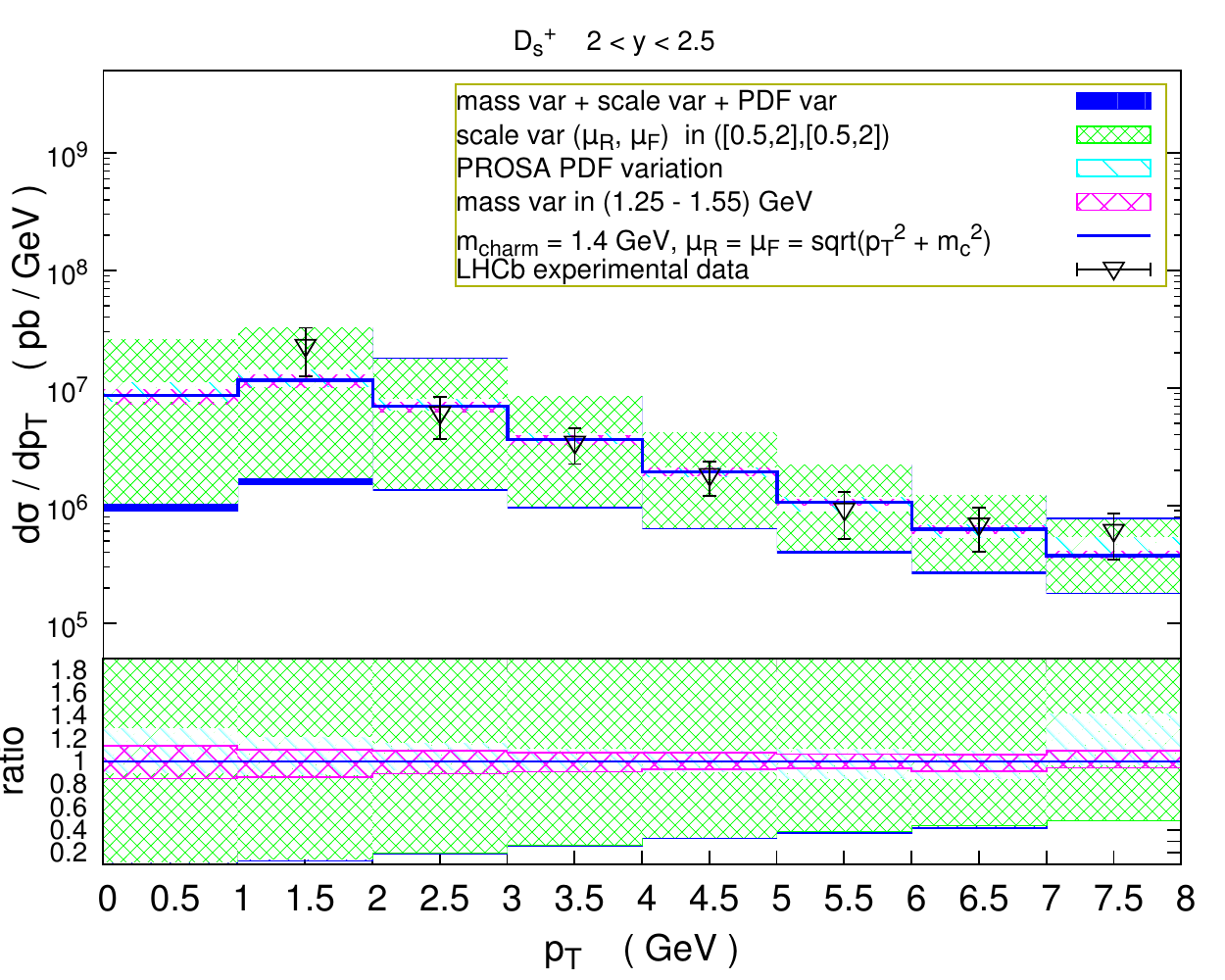}
  \includegraphics[width=0.49\textwidth]{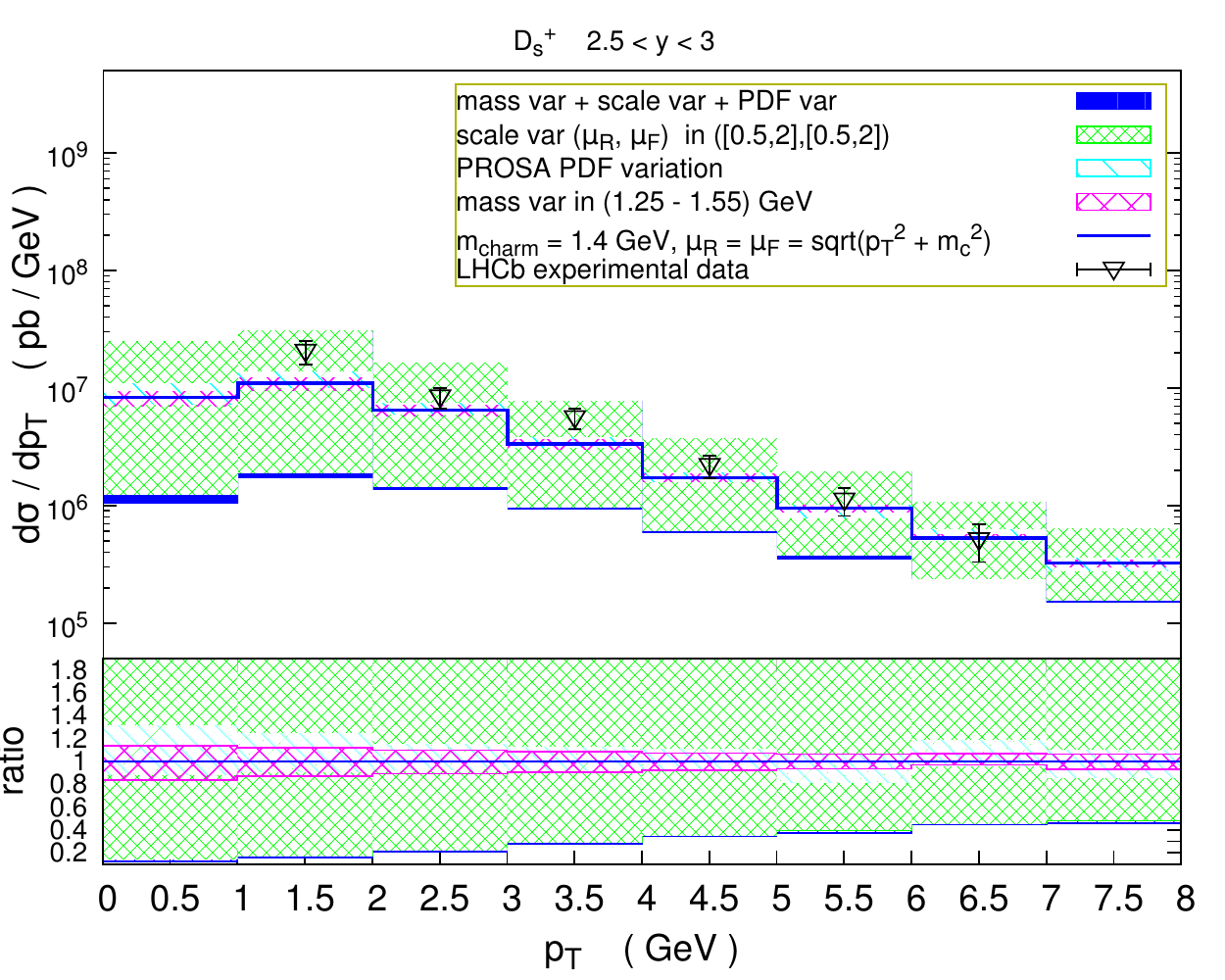}
  \includegraphics[width=0.49\textwidth]{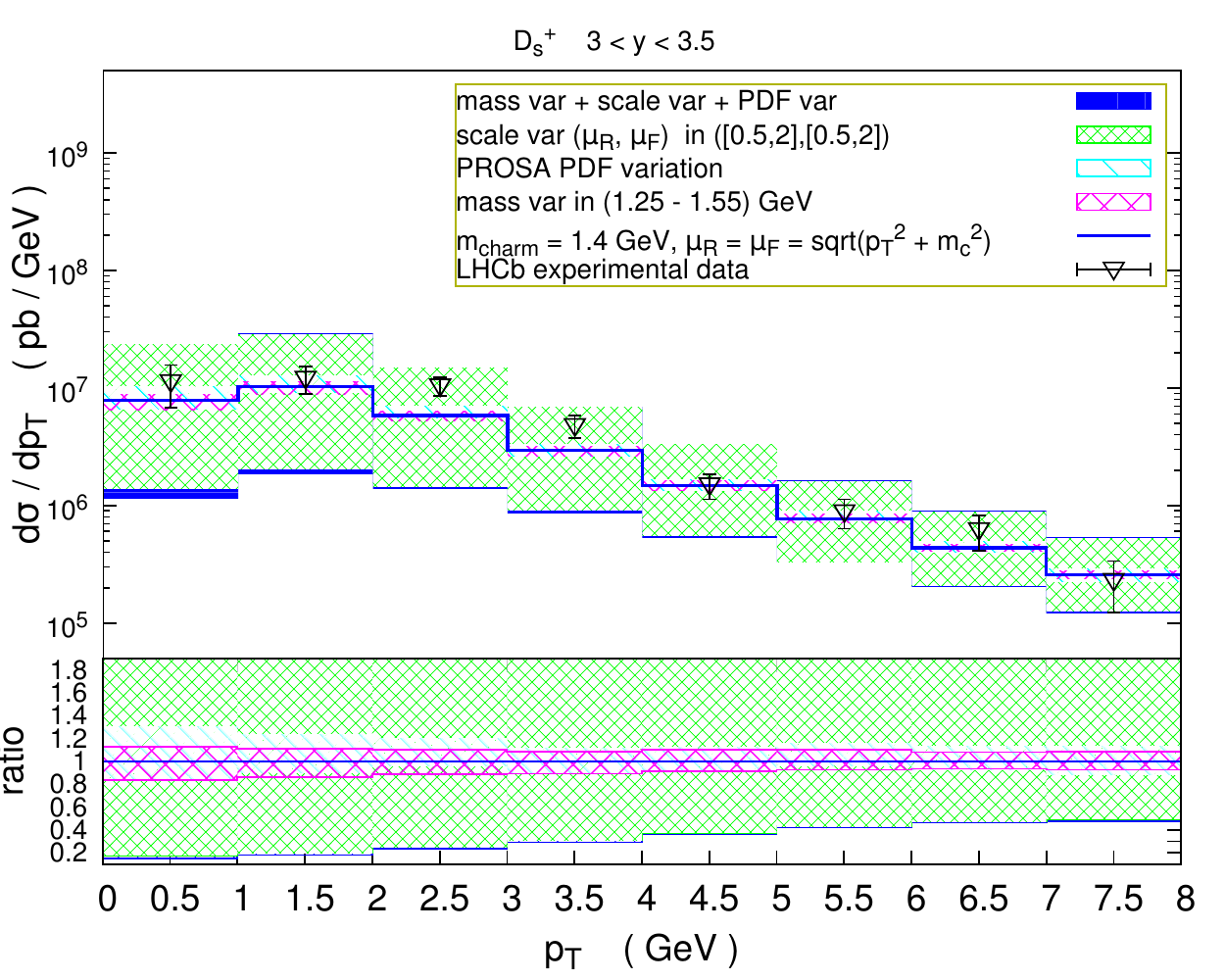}
  \includegraphics[width=0.49\textwidth]{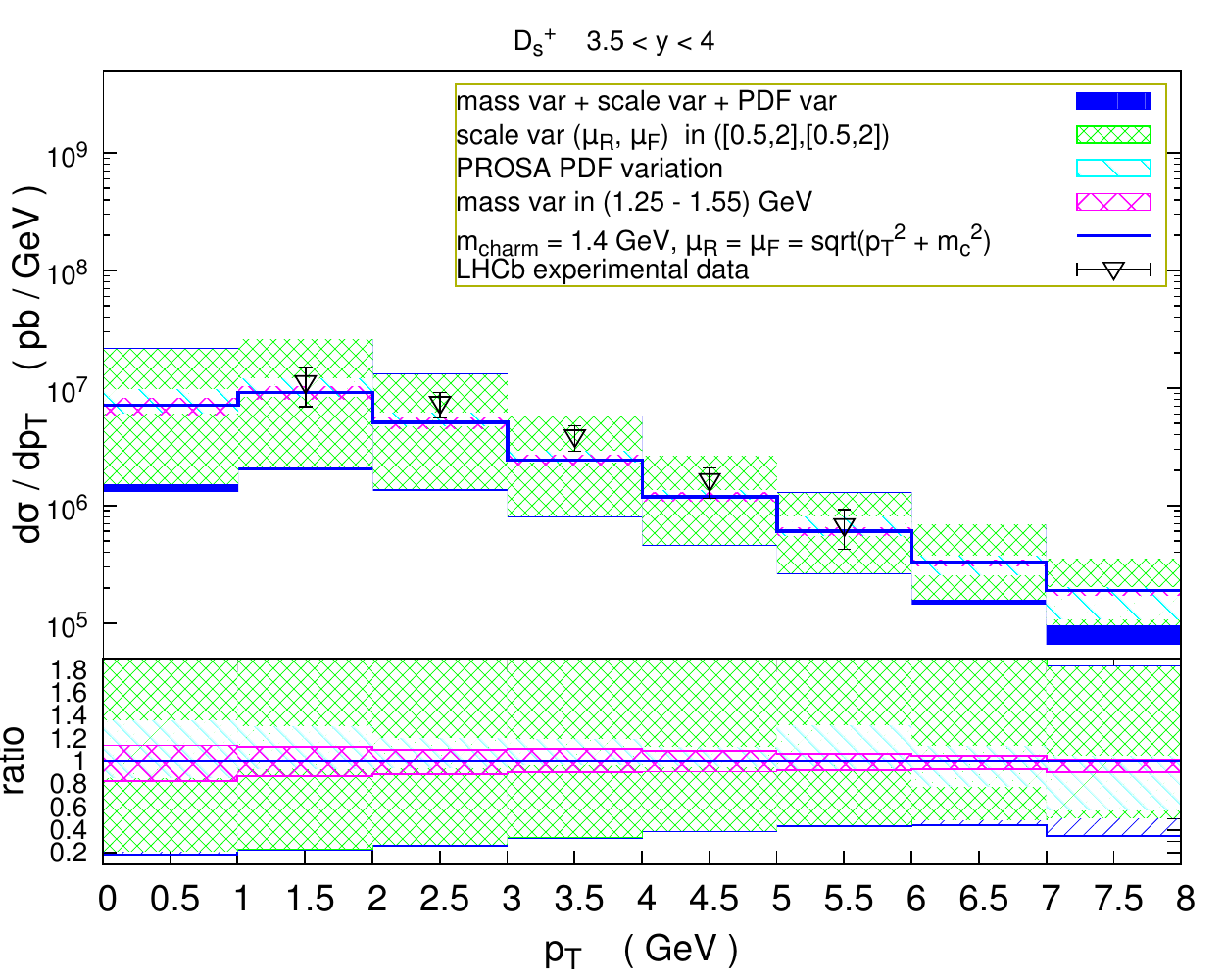}
  \includegraphics[width=0.49\textwidth]{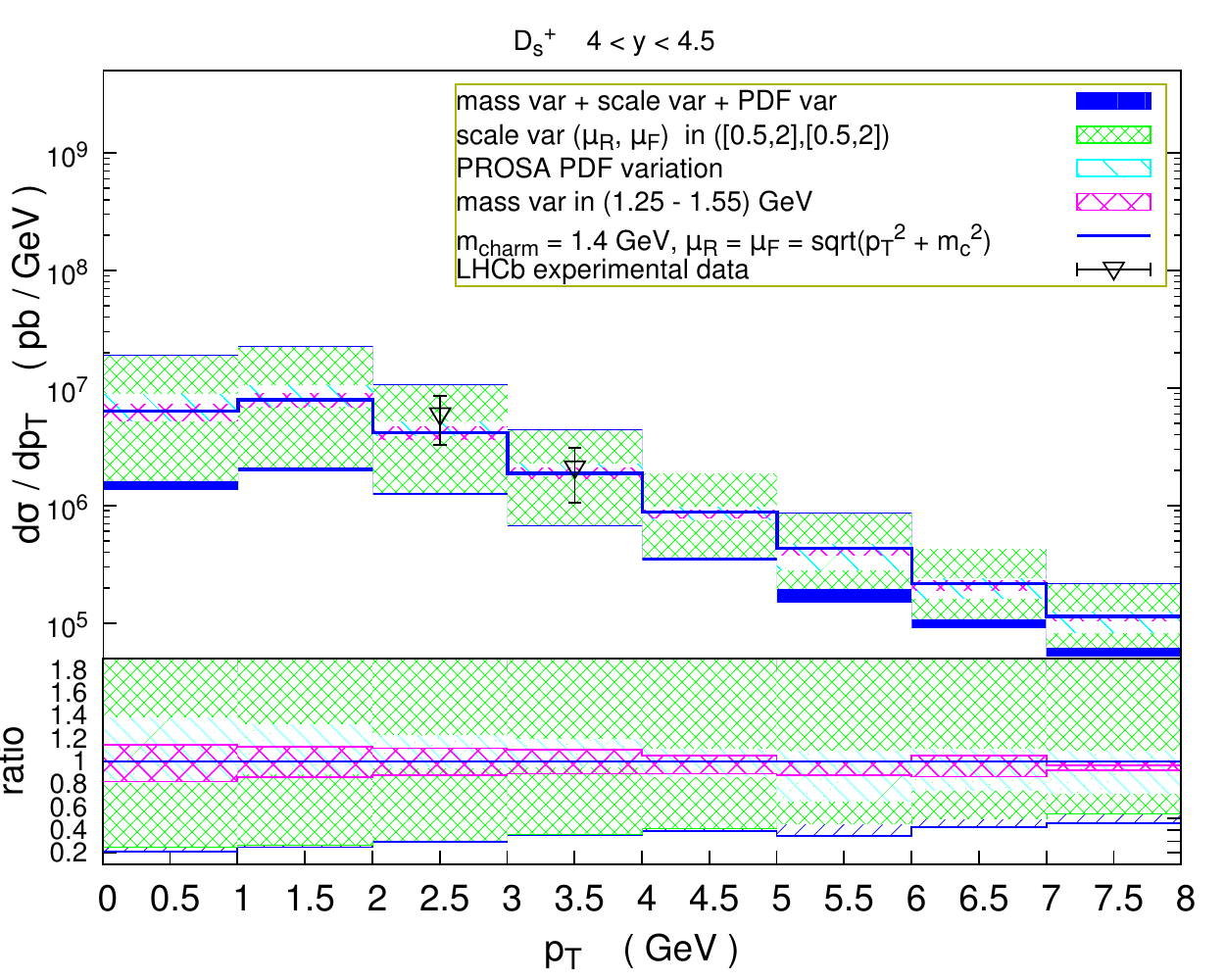}
\end{center}
\vspace*{-5mm}
\caption{\label{fig:sp} 
  Same as Fig.~\ref{fig:dp}, for $D_s^\pm$ hadroproduction at $\sqrt{s}$ = 7 TeV.}
\end{figure}

In this Appendix we provide comparison of theoretical predictions for the hadroproduction of various $D$-mesons 
in our computation as detailed in Section~\ref{sec:theory} to ex\-pe\-ri\-mental data collected by LHCb. 
These refer to absolute values, without making use of any a-posteriori normalization factor.  
\texttt{POWHEGBOX + PYTHIA 6} predictions 
for the spectra of transverse momentum of $D^\pm$, $D^0 + \bar{D}^0$,  $D_s^\pm$ at $\sqrt{s} = 7$~TeV are shown 
in Figs.~\ref{fig:dp}--\ref{fig:sp}, respectively.
Using the PROSA PDFs, $\mu_0=\sqrt{p_{T,\, c}^2 + m_c^2}$ and $m_c = 1.4$~GeV, 
each panel in Figs.~\ref{fig:dp}--\ref{fig:sp} includes 
a tran\-sverse momentum spectrum in the range $p_T \in [0, 8]$~GeV in a different rapidity interval. 

The uncertainties due to scale variation in the range 
($\mu_R$, $\mu_F$) $\in$ \{(1/2, 1/2), (2, 2), (1/2, 1), (1, 1/2 ), (1, 2) (2, 1)\} $\mu_0$, those due to charm mass variation in the range 
$1.25$~GeV $< m_c < 1.55$~GeV and those due to PROSA PDF variation (considering the combination in quadrature of the different sources of variation collected in Table 1) are shown by bands of different style, together with their combination in quadrature. In the lower panels of Figs.~\ref{fig:dp}--\ref{fig:sp}, 
scale, PDF and charm mass uncertainty bands are shown relative to theoretical central predictions.  

\begin{figure}
\begin{center}
\includegraphics[width=0.49\textwidth]{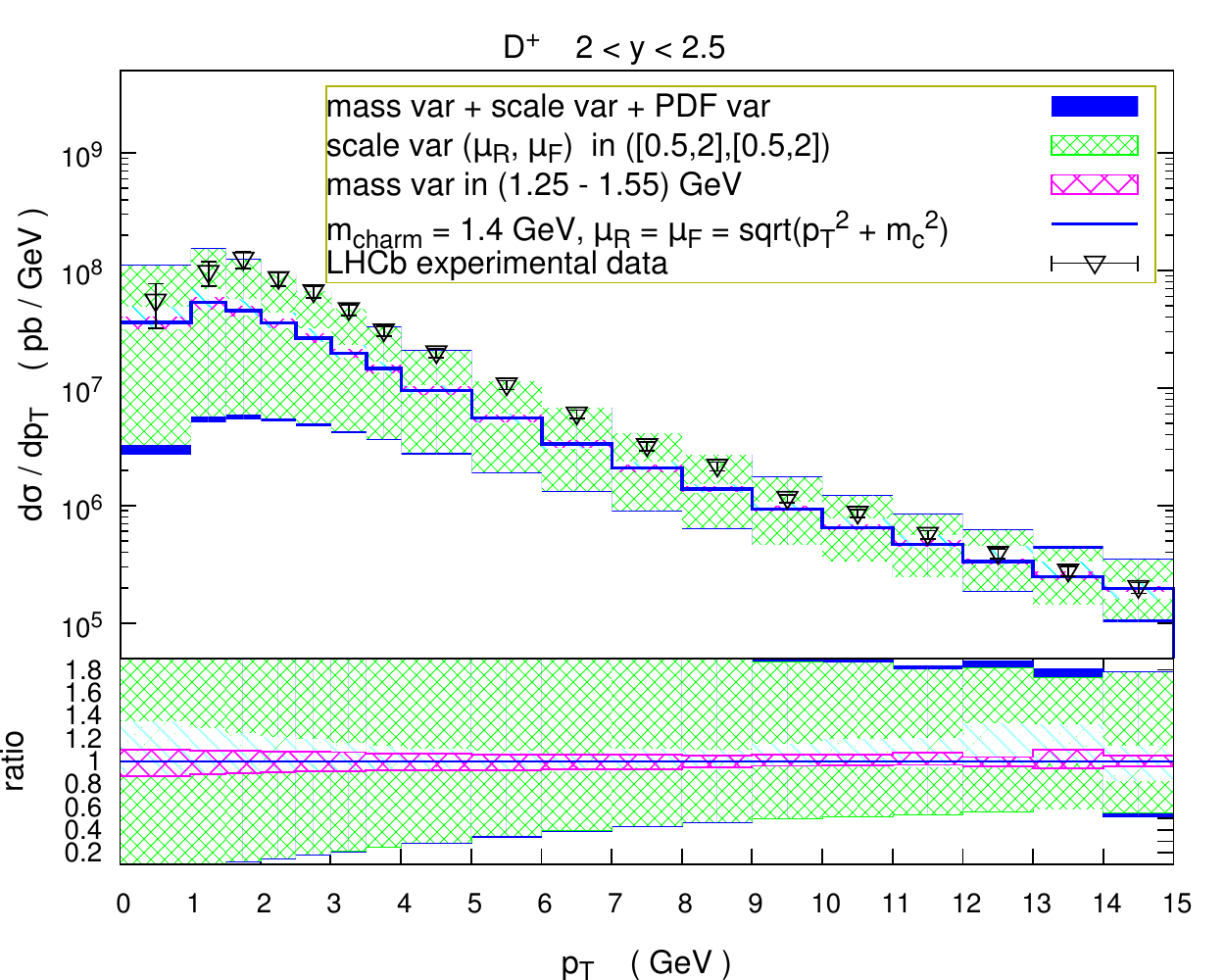}
\includegraphics[width=0.49\textwidth]{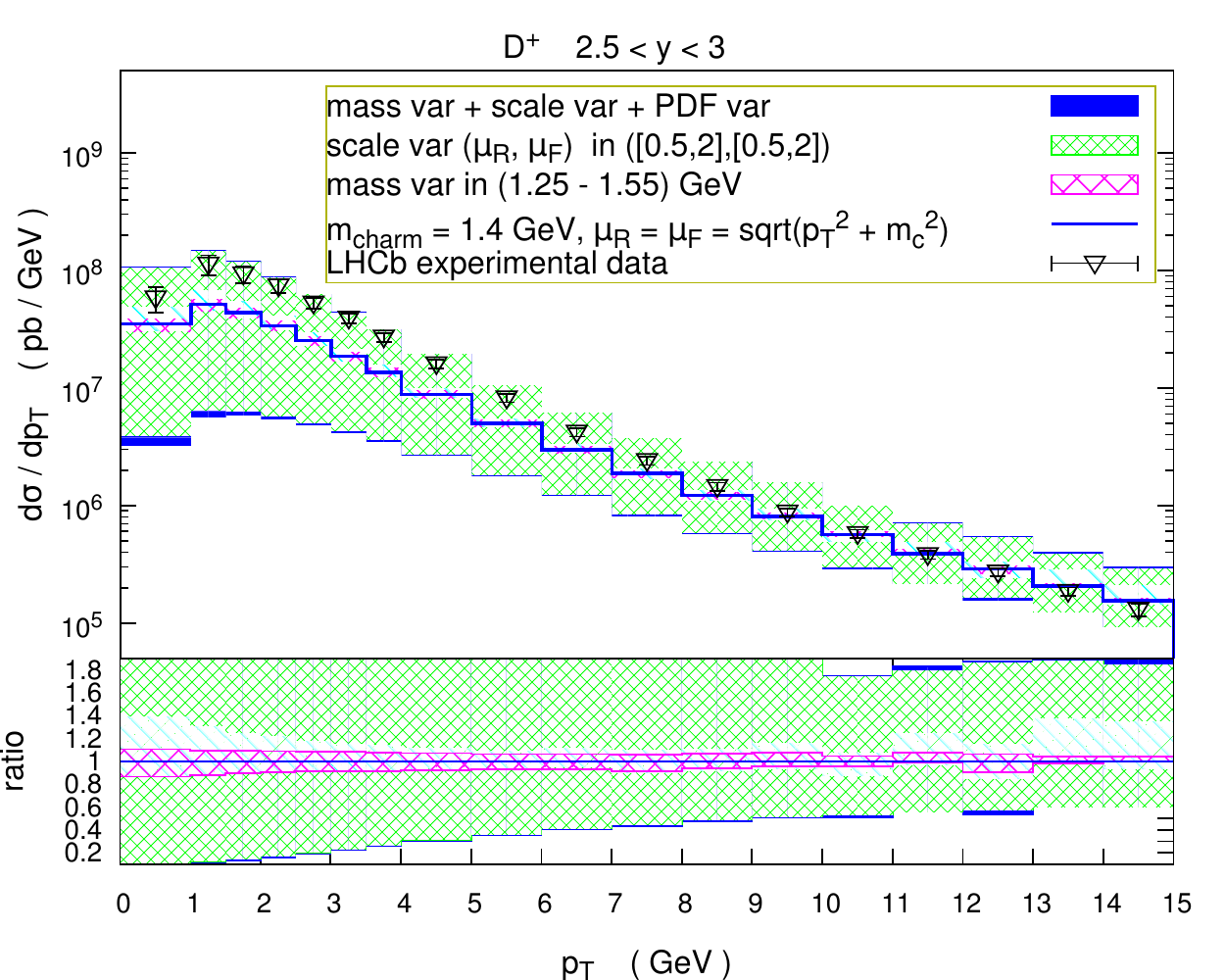}
\includegraphics[width=0.49\textwidth]{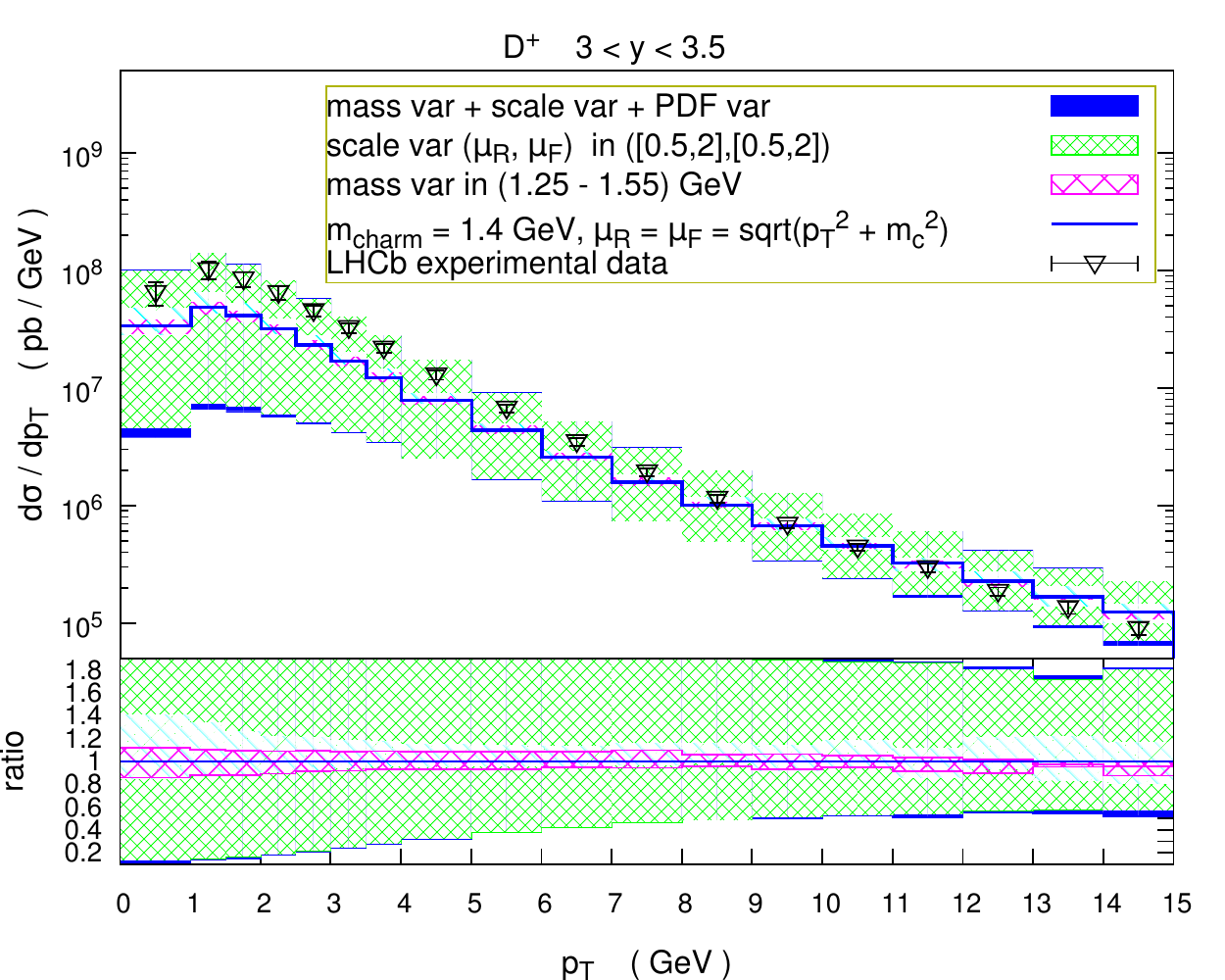}
\includegraphics[width=0.49\textwidth]{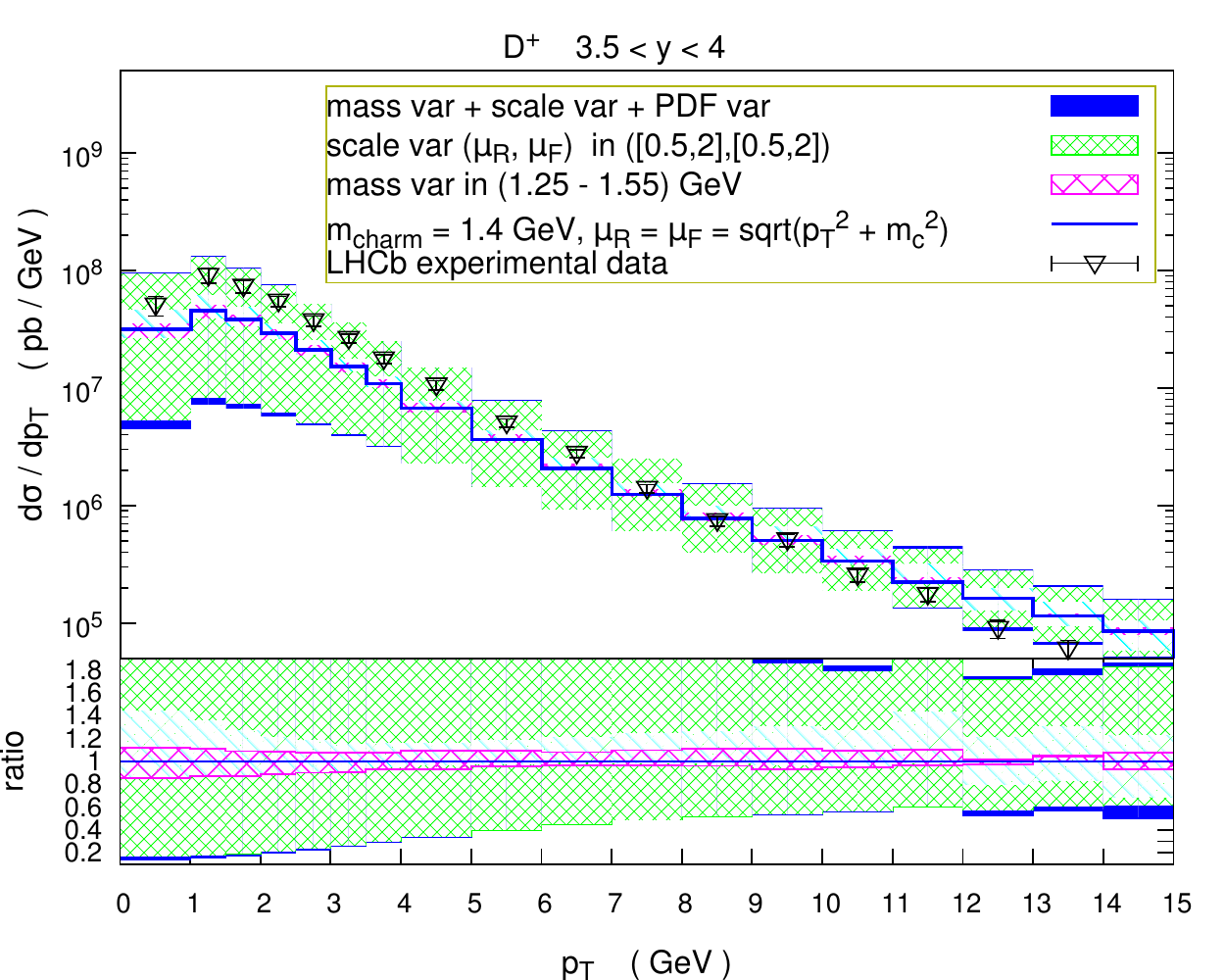}
\includegraphics[width=0.49\textwidth]{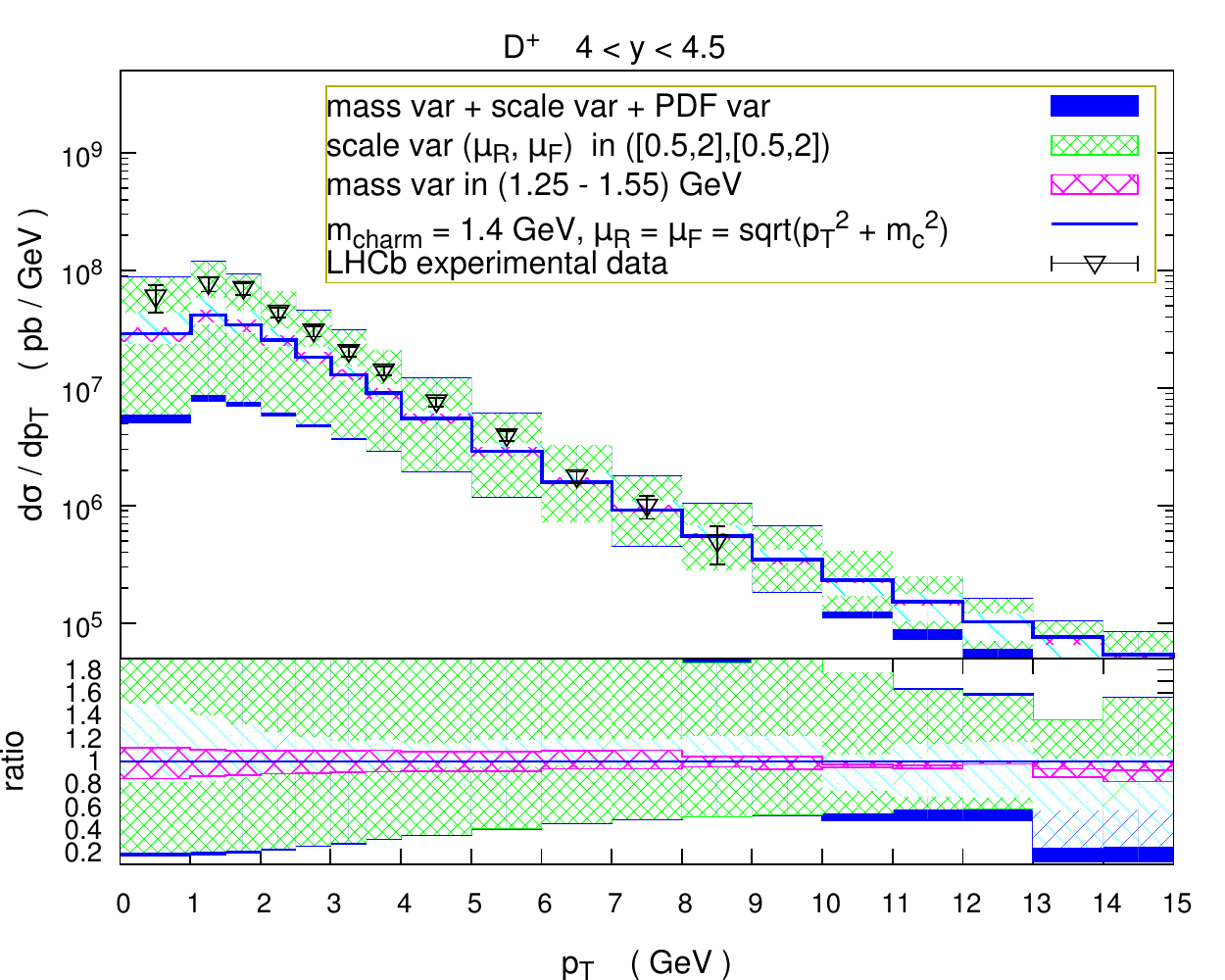}
\end{center}
\vspace*{-5mm}
\caption{\label{fig:dp13} 
Same as Fig.~\ref{fig:dp}, for the LHCb experimental data~\cite{Aaij:2015bpa} on $D^\pm$, hadroproduction at $\sqrt{s} = 13$~TeV. 
} 
\end{figure}

\begin{figure}
\begin{center}
  \includegraphics[width=0.49\textwidth]{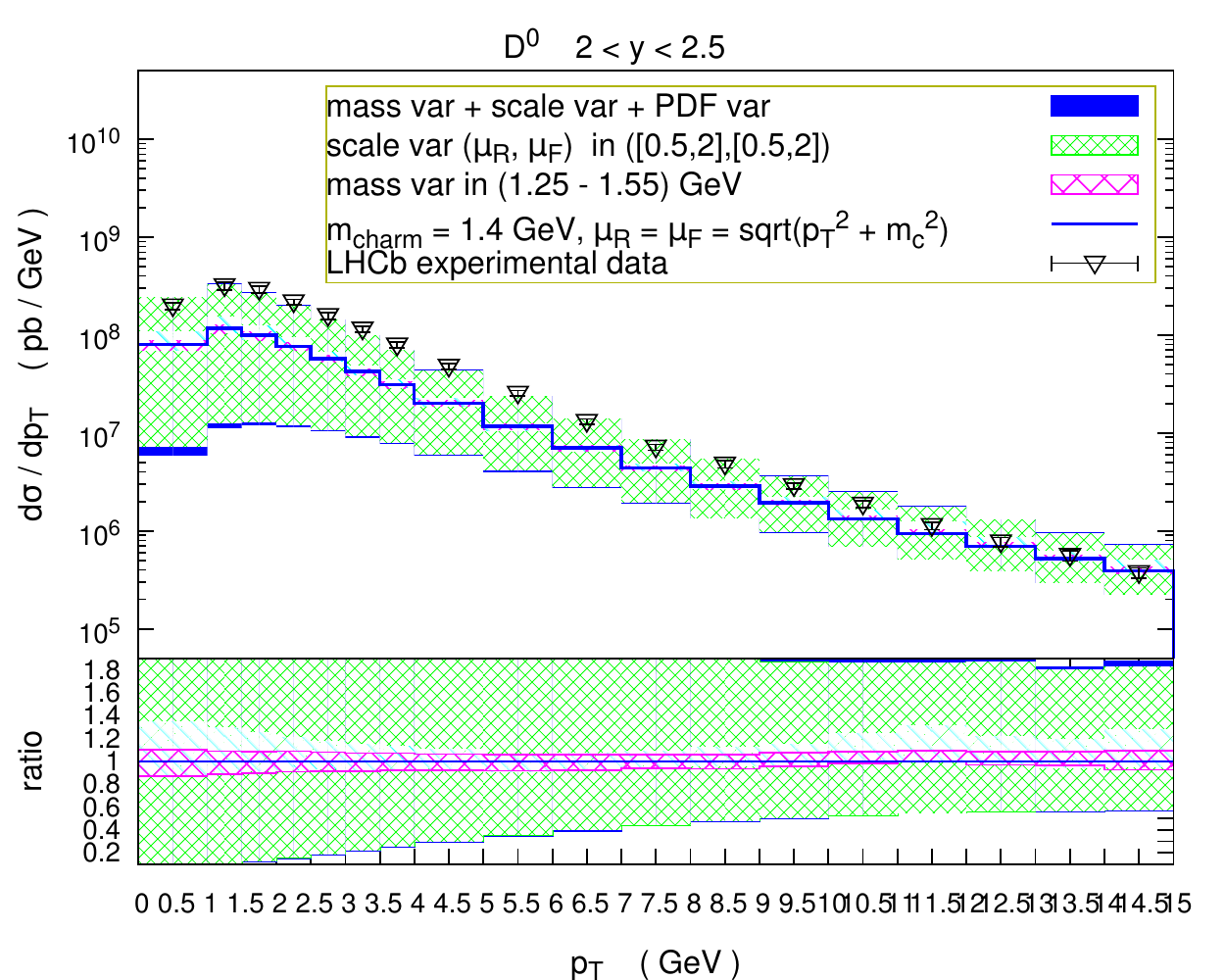}
  \includegraphics[width=0.49\textwidth]{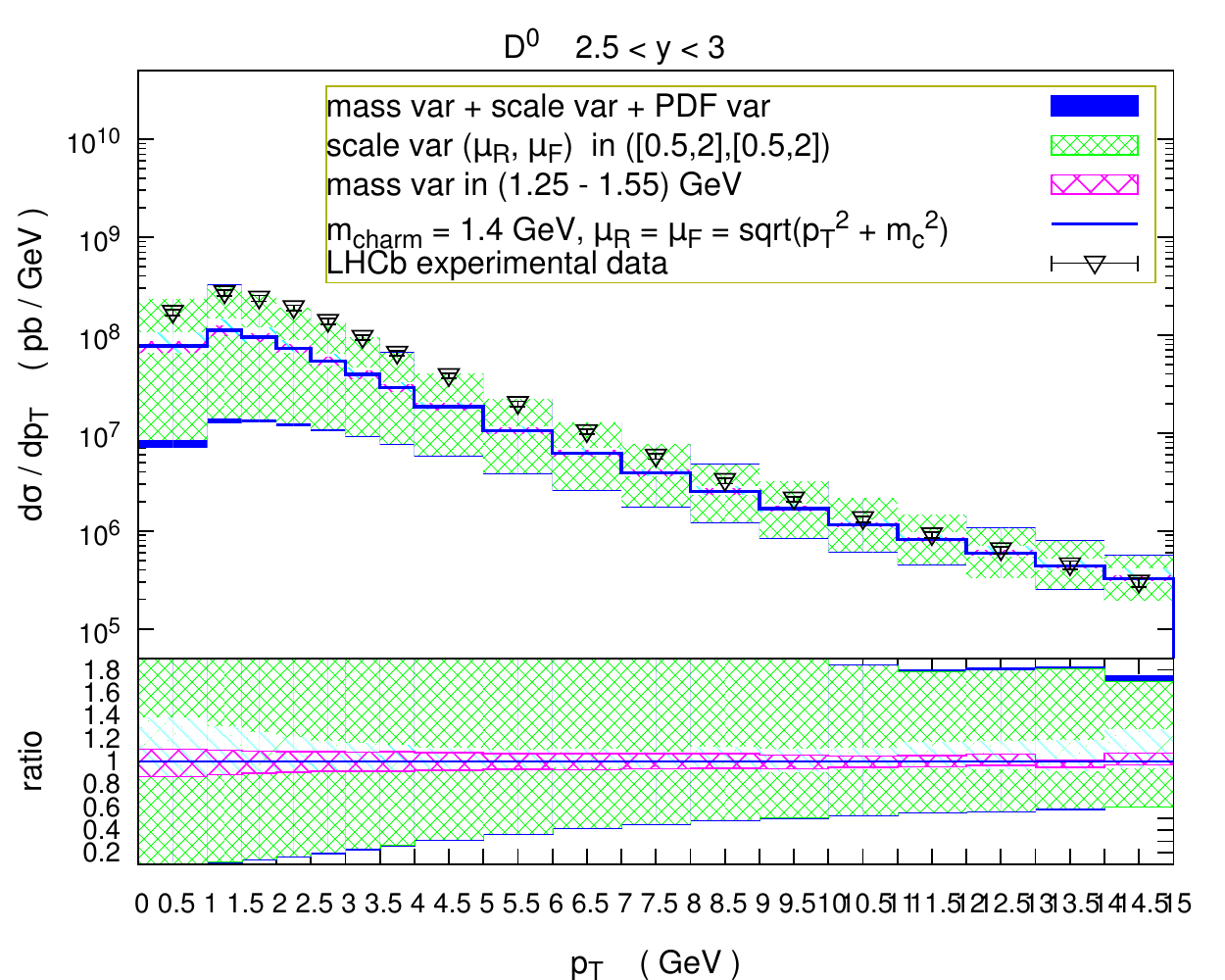}
  \includegraphics[width=0.49\textwidth]{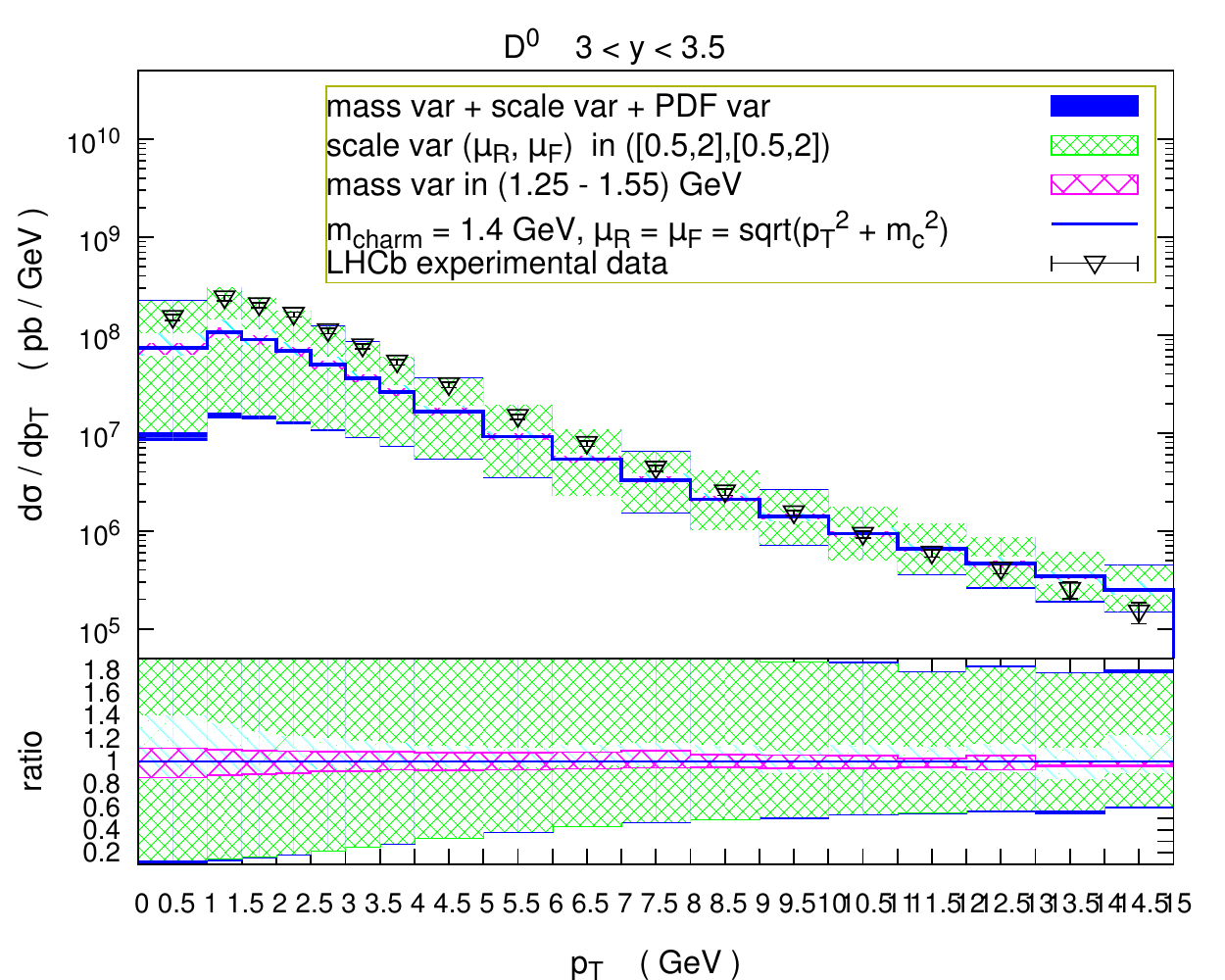}
  \includegraphics[width=0.49\textwidth]{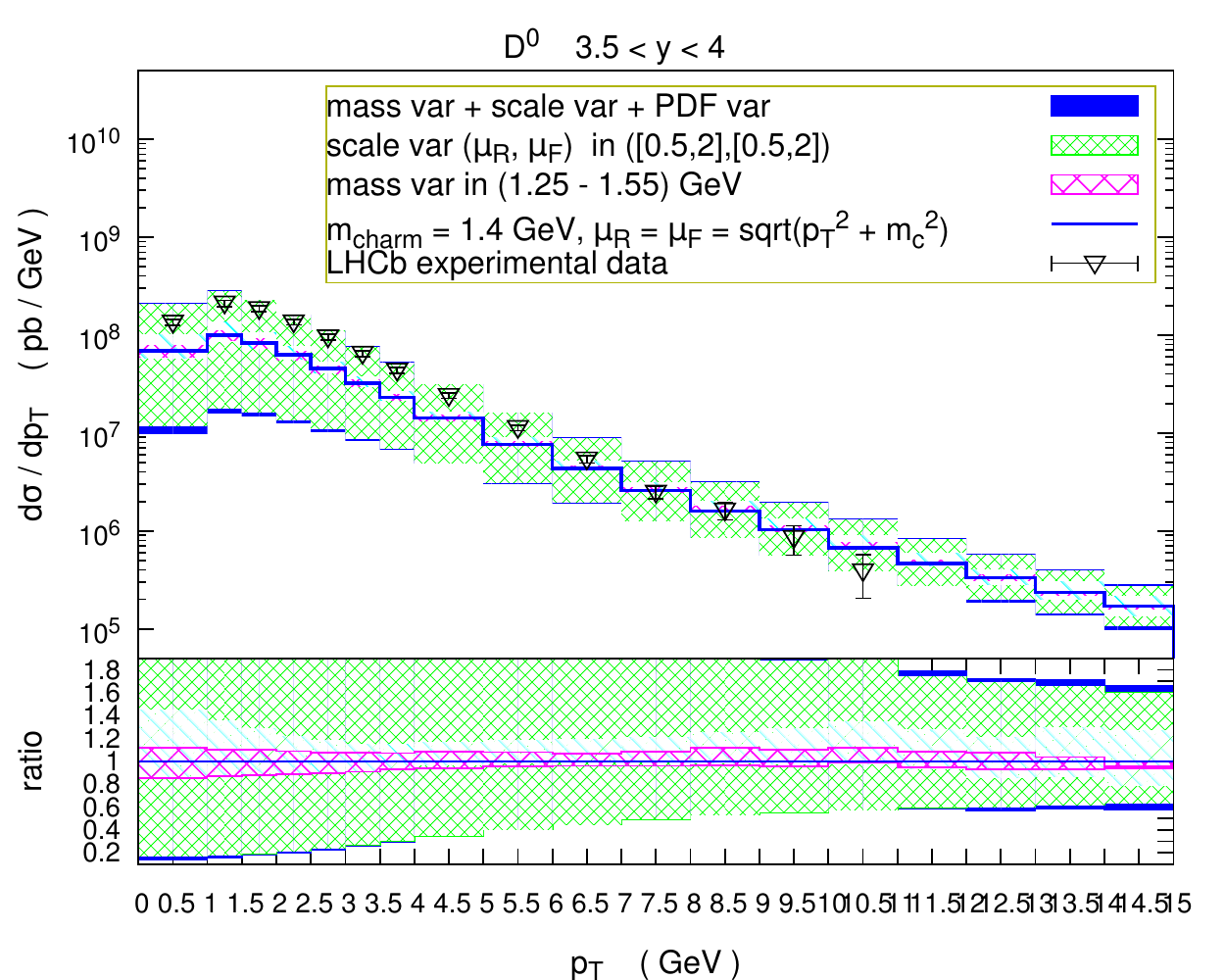}
  \includegraphics[width=0.49\textwidth]{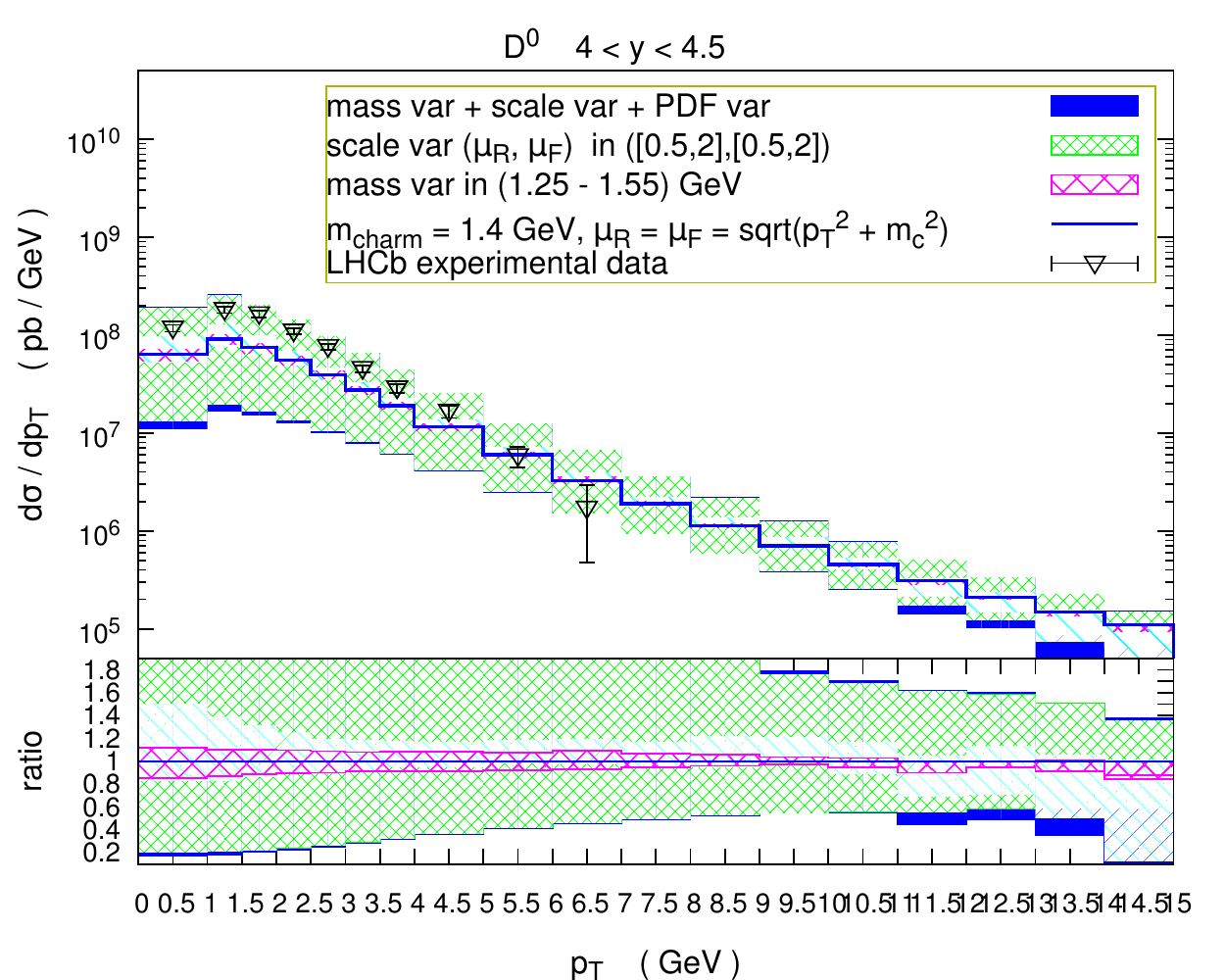}
\end{center}
\vspace*{-5mm}
\caption{\label{fig:d013} 
  Same as Fig.~\ref{fig:dp13}, for ($D^0$ + $\bar{D^0}$) hadroproduction at $\sqrt{s}$ = 13 TeV.}
\end{figure}

\begin{figure}[t]
\begin{center}
  \includegraphics[width=0.49\textwidth]{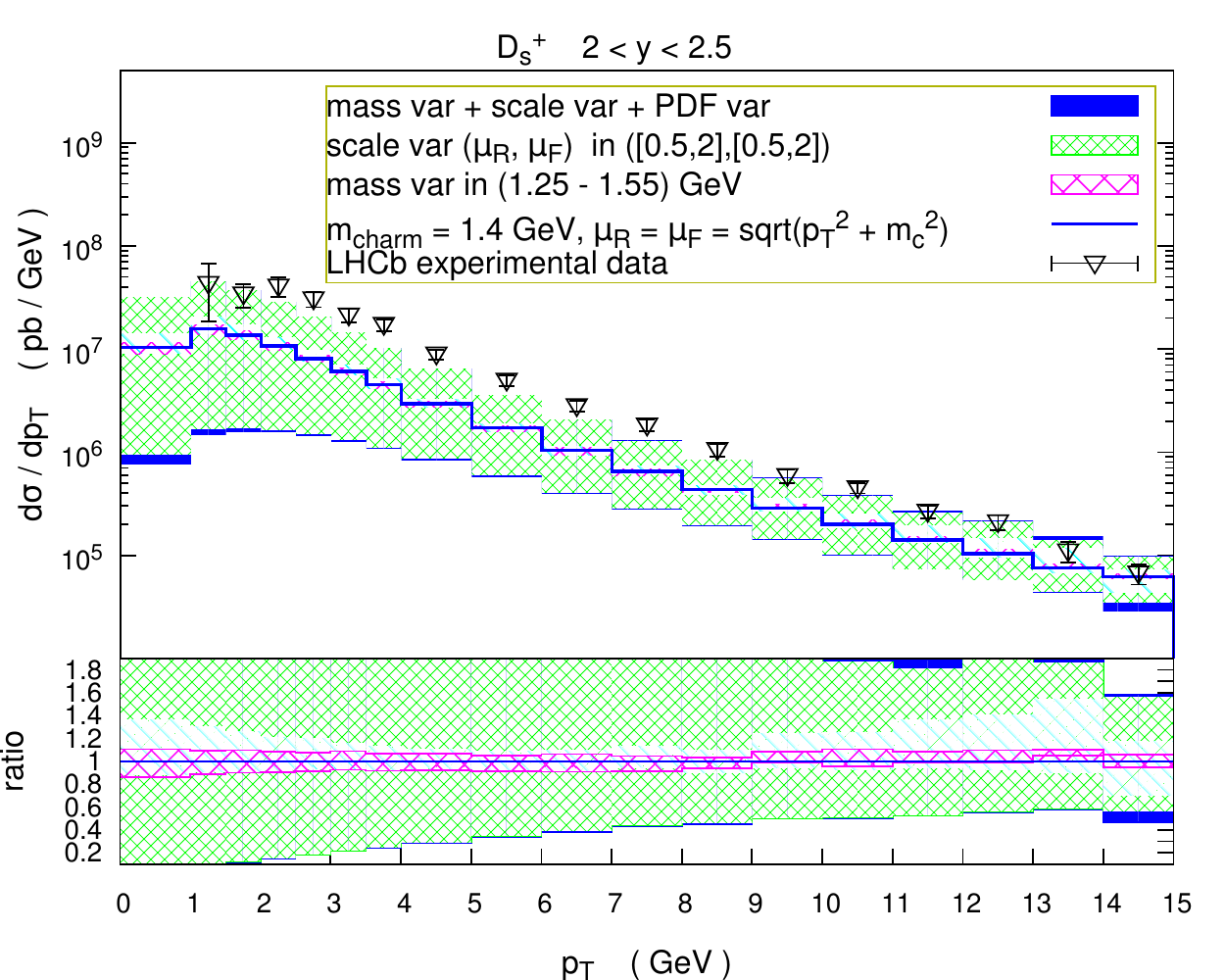}
  \includegraphics[width=0.49\textwidth]{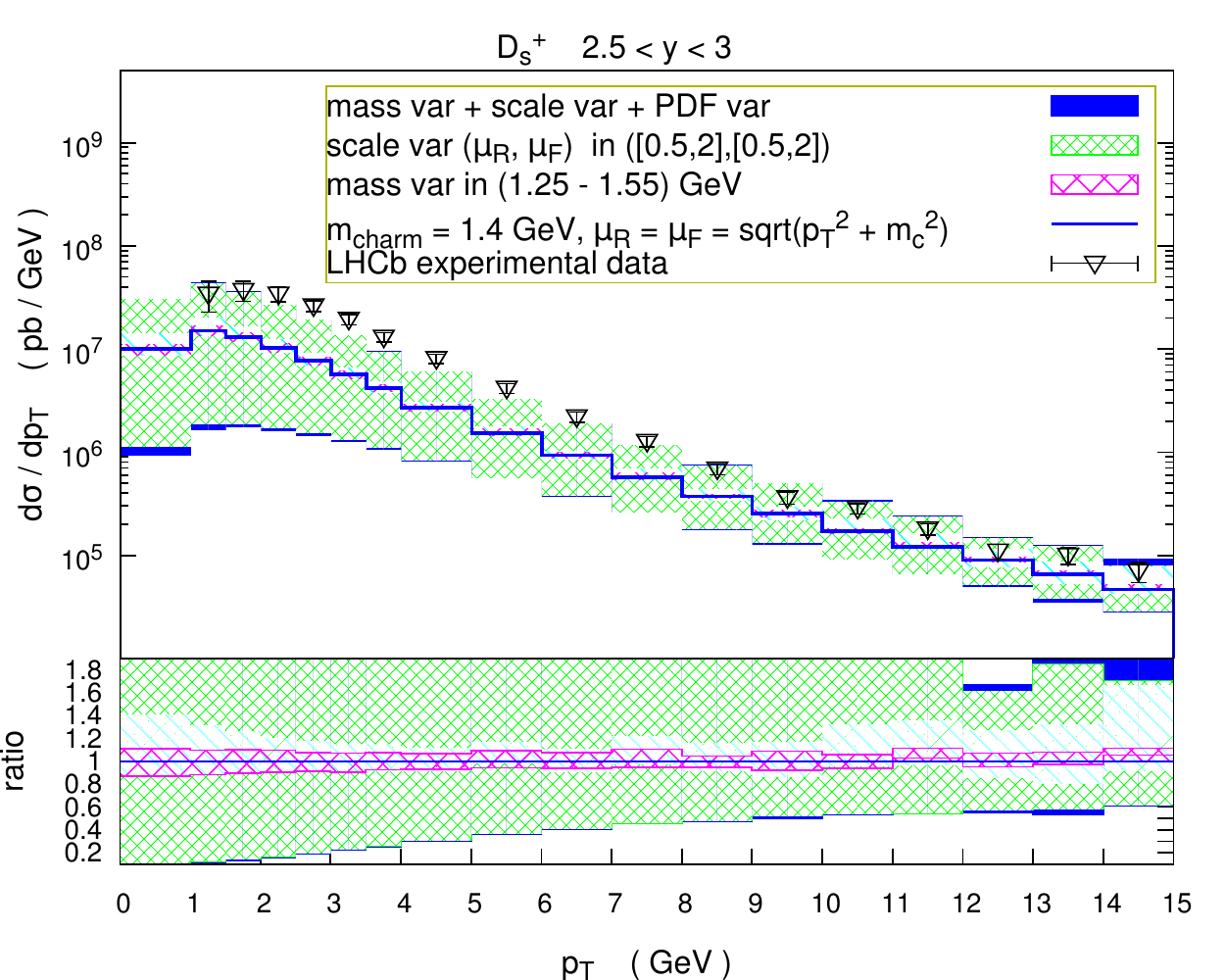}
  \includegraphics[width=0.49\textwidth]{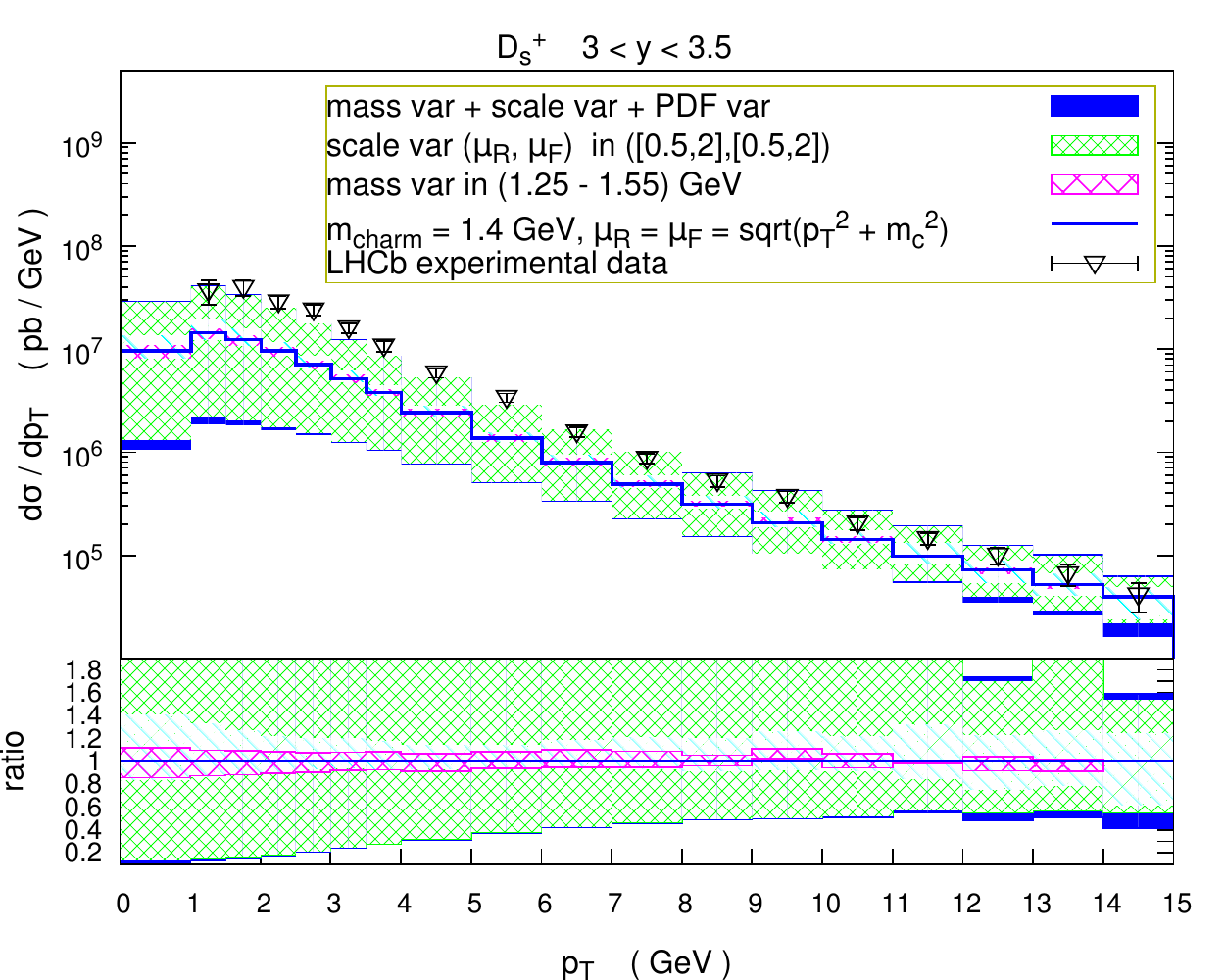}
  \includegraphics[width=0.49\textwidth]{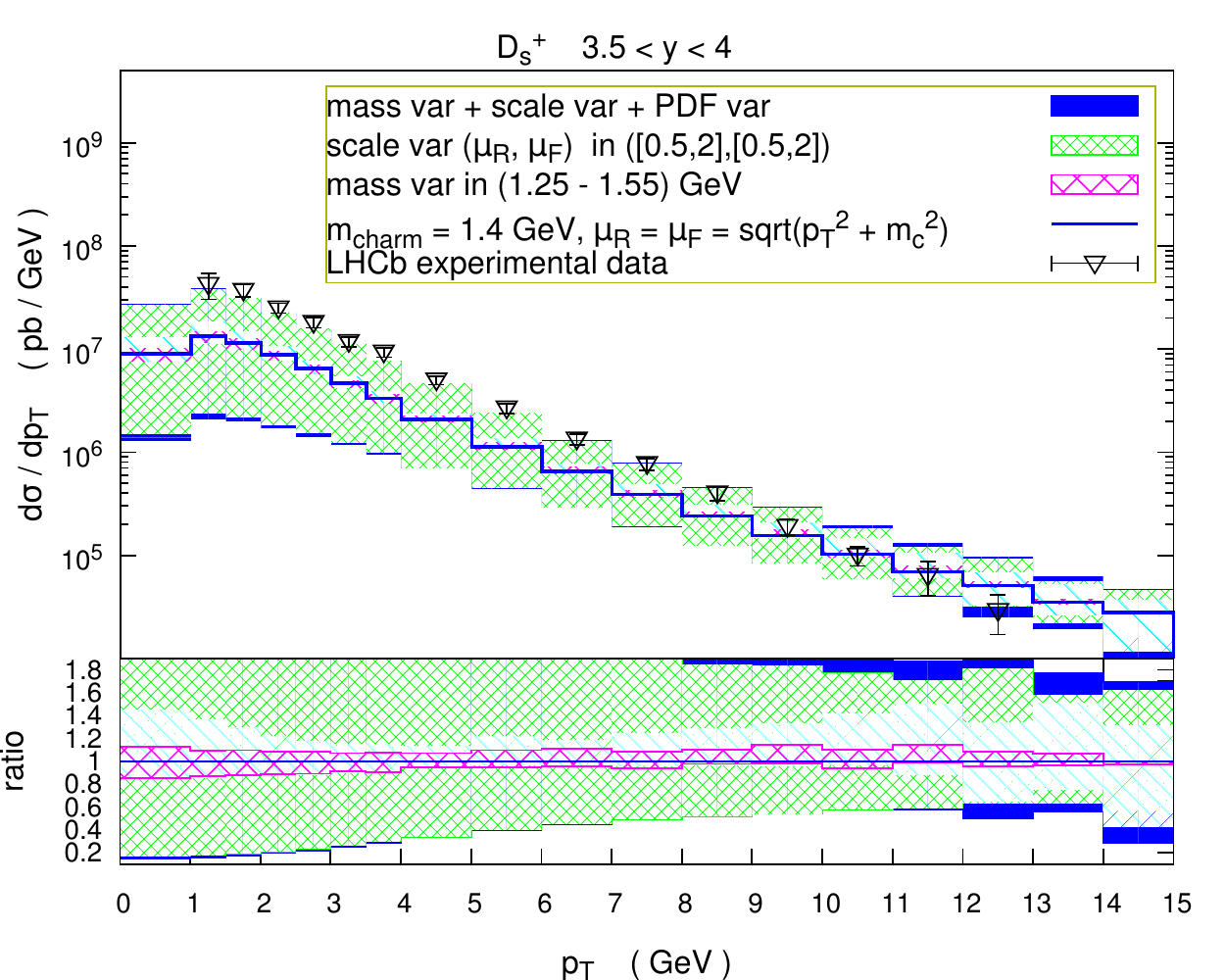}
  \includegraphics[width=0.49\textwidth]{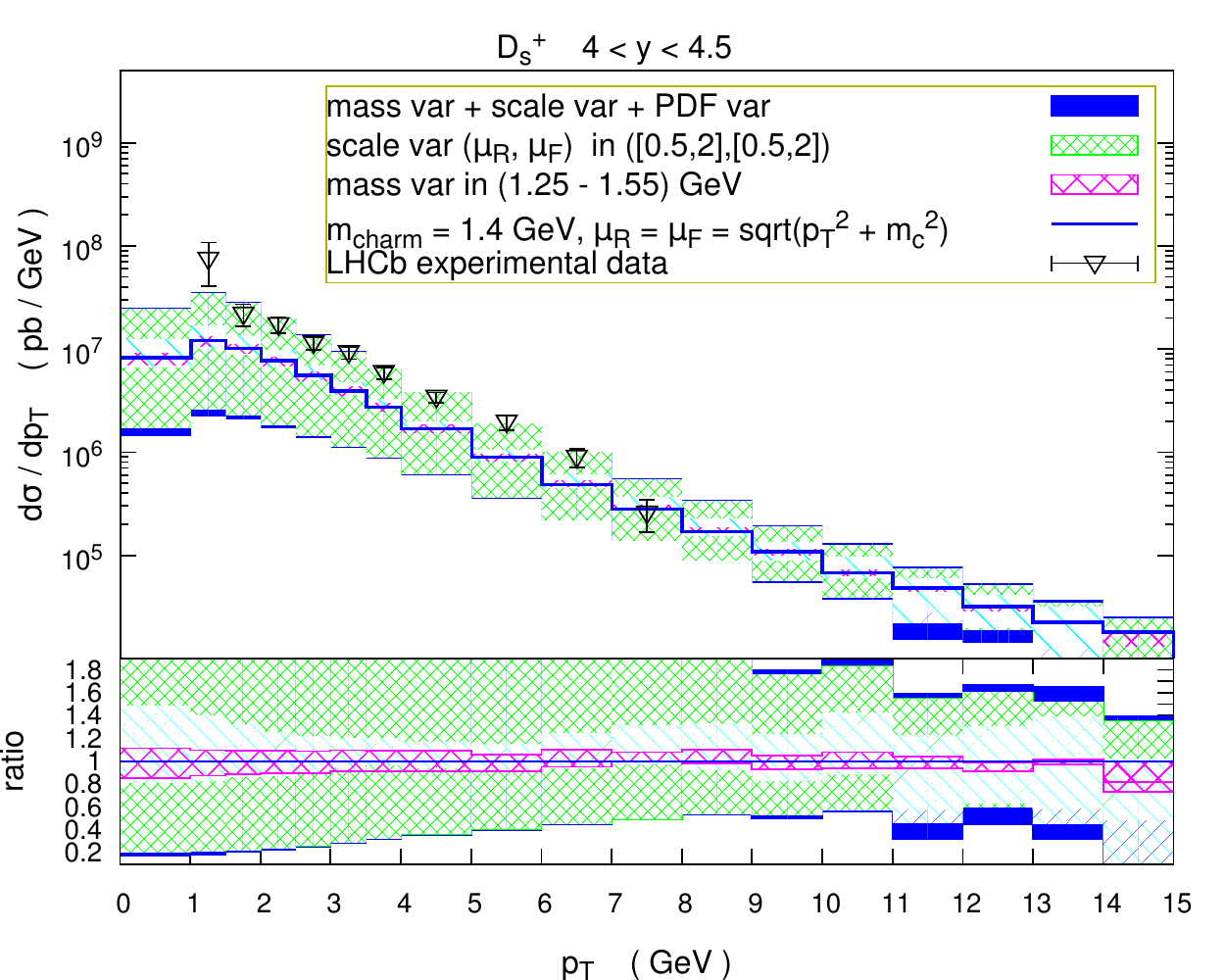}
\end{center}
\vspace*{-5mm}
\caption{\label{fig:sp13} 
  Same as Fig.~\ref{fig:dp13}, for $D_s^\pm$ hadroproduction at $\sqrt{s}$ = 13 TeV.}
\end{figure}

Figs.~\ref{fig:dp13}--\ref{fig:sp13} contain the same study 
for LHCb experimental data~\cite{Aaij:2015bpa} at $\sqrt{s} = 13$~TeV.
Each panel in Figs.~\ref{fig:dp13}--\ref{fig:sp13} for a given rapidity
interval now includes a tran\-sver\-se momentum spectrum in the range $p_T \in [0, 15]$~GeV.

The contribution arising from the feeddown from $D^*$'s and from other excited charmed states is accounted for in the theory predictions for the lowest lying charmed mesons shown in the plots.

While the data at 7 TeV are succesfully described in all rapidity and $p_T$ bins by the QCD computation adopted in this paper, the data at 13 TeV in the low $p_T$ regime turn out to lie above the central theoretical predictions~\footnote{
This remains true even after the update of the ($D^0$ + $\bar{D}^0$) hadroproduction data at 13 TeV, recently published by the LHCb collaboration in the erratum JHEP 09 (2016) 013, which was taken into account in this paper.}. In any case, for ($D^0+\bar{D}^0$) and $D^\pm$, whose contribution actually dominate the total prompt neutrino flux at the neutrino energies explored by IceCube, giving rise to more than 80\% of it, the LHCb experimental data turn out to lie within our scale uncertainty limits. This means that even in case the theoretical description of these data will be improved in the future, the resulting central predictions for prompt neutrino fluxes will likely lie within our present uncertainty bands.

\cleardoublepage
\newpage

\bibliographystyle{JHEP} 
\bibliography{paperprosa}

\end{document}